\documentclass[twocolumn,aps,prx,superscriptaddress,preprintnumbers,bibnotes,nofootinbib]{revtex4-2}

\usepackage[T1]{fontenc}

\usepackage{natbib}
\usepackage{graphicx}
\usepackage{bm}
\usepackage{color}
\usepackage{amsmath}
\usepackage{gensymb} 
\usepackage{physics}
\usepackage{tabularx}
\usepackage{float}
\usepackage{siunitx}
\usepackage{cancel}

\usepackage{changes}
\definechangesauthor[color=blue]{AJP}

\newcommand\Ket[1]{\ensuremath{|{#1}\rangle}}
\newcommand\Bra[1]{\ensuremath{\langle{#1}|}}

\newcommand\unit[2]{\ensuremath{#1~\mathrm{{#2}}}}
\newcommand\Erec{\ensuremath{E_\text{rec}}}

\newcommand{\SLJ}[3]{{\ensuremath{{^{#1}}\mathrm{#2}_{#3}}}}
\newcommand{\TPT}{\SLJ{3}{P}{2}}
\newcommand{\TPO}{\SLJ{3}{P}{1}}
\newcommand{\TPZ}{\SLJ{3}{P}{0}}
\newcommand{\TSO}{\SLJ{3}{S}{1}}
\newcommand{\SSZ}{\SLJ{1}{S}{0}}
\newcommand{\SPO}{\SLJ{1}{P}{1}}

\newcommand{\Isotope}[2]{\ensuremath{^{{#1}}\text{{#2}}}}
\newcommand{\Sr}[1]{\Isotope{{#1}}{Sr}}

\begin{document}

\title{Cavity-enhanced optical lattices for scaling neutral atom quantum technologies to higher qubit numbers}

\author{A. J. Park}
\author{J. Trautmann}
\affiliation{
  Max-Planck-Institut f{\"u}r Quantenoptik,
  Hans-Kopfermann-Stra{\ss}e 1,
  85748 Garching, Germany}
\affiliation{
  Munich Center for Quantum Science and Technology,
  80799 M{\"u}nchen, Germany}

\author{N. \v{S}anti\'c}
\affiliation{
  Max-Planck-Institut f{\"u}r Quantenoptik,
  Hans-Kopfermann-Stra{\ss}e 1,
  85748 Garching, Germany}
\affiliation{
  Munich Center for Quantum Science and Technology,
  80799 M{\"u}nchen, Germany}
\affiliation{
  Institute of Physics,
  Bijeni\v{c}ka cesta 46,
  10000, Zagreb, Croatia}

\author{V. Kl\"{u}sener}
\affiliation{
  Max-Planck-Institut f{\"u}r Quantenoptik,
  Hans-Kopfermann-Stra{\ss}e 1,
  85748 Garching, Germany}
\affiliation{
  Munich Center for Quantum Science and Technology,
  80799 M{\"u}nchen, Germany}

\author{A. Heinz}
\altaffiliation[Current address: ]{
  TOPTICA Photonics,
  Lochhamer Schlag 19,
  82166 Gr{\"a}felfing, Germany}
\affiliation{
  Max-Planck-Institut f{\"u}r Quantenoptik,
  Hans-Kopfermann-Stra{\ss}e 1,
  85748 Garching, Germany}
\affiliation{
  Munich Center for Quantum Science and Technology,
  80799 M{\"u}nchen, Germany}

\author{I. Bloch}
\affiliation{
  Max-Planck-Institut f{\"u}r Quantenoptik,
  Hans-Kopfermann-Stra{\ss}e 1,
  85748 Garching, Germany}
\affiliation{
  Munich Center for Quantum Science and Technology,
  80799 M{\"u}nchen, Germany}
\affiliation{
  Fakult{\"a}t f{\"u}r Physik,
  Ludwig-Maximilians-Universit{\"a}t M{\"u}nchen,
  80799 M{\"u}nchen, Germany}

\author{S. Blatt}
\email{sebastian.blatt@mpq.mpg.de}
\affiliation{
  Max-Planck-Institut f{\"u}r Quantenoptik,
  Hans-Kopfermann-Stra{\ss}e 1,
  85748 Garching, Germany}
\affiliation{
  Munich Center for Quantum Science and Technology,
  80799 M{\"u}nchen, Germany}

\date{\today}

\begin{abstract}

We demonstrate a cavity-based solution to scale up experiments with ultracold atoms in optical lattices by an order of magnitude over state-of-the-art free space lattices.
Our two-dimensional optical lattices are created by power enhancement cavities with large mode waists of \unit{489(8)}{\mu m} and allow us to trap ultracold strontium atoms at a lattice depth of \unit{60}{\mu K} by using only \unit{80}{mW} of input light per cavity axis.
We characterize these lattices using high-resolution clock spectroscopy and resolve carrier transitions between different vibrational levels.
With these spectral features, we locally measure the lattice potential envelope and the sample temperature with a spatial resolution limited only by the optical resolution of the imaging system.
The measured ground-band and trap lifetimes are \unit{18(3)}{s} and \unit{59(2)}{s}, respectively, and the lattice frequency (depth) is long-term stable on the MHz (0.1\%) level.
Our results show that large, deep, and stable two-dimensional cavity-enhanced lattices can be created at any wavelength and can be used to scale up neutral-atom-based quantum simulators, quantum computers, sensors, and optical lattice clocks.
\end{abstract}

\maketitle

\section{Introduction}
\label{sec:intro}

Far-off-resonant optical dipole traps formed by laser light are a foundational technology for modern quantum science~\cite{ashkin97,grimm00}, and are used in quantum simulators~\cite{gross17}, quantum computers~\cite{saffman10,weiss17,browaeys20,morgado21}, optical clocks~\cite{ludlow15}, and matter-wave interferometers~\cite{cronin09}.
When such traps make use of interference between multiple laser beams, they form optical lattices consisting of arrays of nearly identical microtraps that can hold ultracold atoms.
The lattice constant and geometry can be chosen by using different wavelengths and arrangements of the interfering laser beams, while the energy barrier between the traps can be tuned by changing the laser intensity.

Lowering the barriers between the traps lets the atoms tunnel between lattice sites and interact, which is the basis of analog quantum simulations~\cite{georgescu14} of foundational quantum many-body models.
These models are important in condensed matter physics~\cite{esslinger10,gross17}, quantum optics~\cite{devega08,tudela18,krinner18}, quantum chemistry~\cite{arguello19}, and high energy physics~\cite{zohar15,aidelsburger21}.

Raising the barriers between the traps prevents any quantum tunneling or thermally activated hopping of atoms between the traps, creating an array of independent and identical quantum systems.
This parallelism allows simultaneous interrogation and averaging of the signal derived from many identical systems, which is the basis of optical lattice clocks which enable metrology with 18 digits of precision by averaging the signal from thousands of atoms for one hour~\cite{oelker19,bothwell22}.

The system sizes that can be simulated on quantum simulators and the precision of optical lattice clocks are both limited by the number of identical lattice sites that can be used.
Since optical lattices are created by laser beams, the lattice sites are no longer identical on length scales that are comparable to the finite transverse extent of the beams because the lattice potential depth varies.
This inhomogeneity limits scaling atomic quantum technologies to higher numbers of atoms or qubits.

In analog quantum simulators, the depth variation in the optical lattice potential leads to a spatial variation of the tunneling rate, the interaction parameters, and the chemical potential.
Therefore, descriptions of such quantum systems must rely on local-density approximations~\cite{bloch08} and take into account edge effects that can become more important than those in the bulk system.
Analog quantum simulations are often initialized in a low-entropy Mott insulating state, where each lattice site is occupied by a single atom~\cite{bloch08}.
The lattice inhomogeneity limits the size and fidelity of this Mott insulating state, which in turn restricts the size and fidelity of the analog quantum simulation.

Scaling up quantum resources will also improve the stability of state-of-the-art optical lattice clocks.
The fundamental limit to clock stability is the quantum projection noise, which scales $\propto 1/\sqrt{N}$, where $N$ is the number of concurrently interrogated atoms.
In one-dimensional (1D) optical lattice clocks (the most common type), increasing the atom number leads to interaction-induced frequency shifts~\cite{campbell09,lemke11,martin13b}, which lower the clock's accuracy.
These shifts can be reduced by using 2D~\cite{swallows11} or 3D lattices~\cite{akatsuka10,campbell17}, in which case $N$ is limited by the mode area or mode volume of the optical lattices, respectively.
Here, the mode area (volume) refers to the overlap area (volume) of the two (three) orthogonal beams that create the lattices, which is directly proportional to the number of usable lattice sites, and thus $N$.

\begin{figure}
	\centering
	\includegraphics{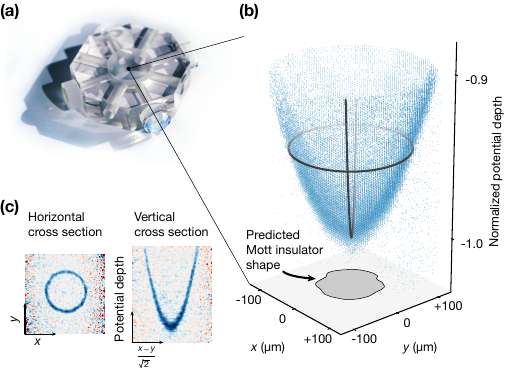}
	\caption{(a) Photograph of the monolithic cavity assembly.
The two-dimensional optical lattice is formed at the intersection of the optical axes of both cavities.
		(b) From the measured potential envelope of the optical lattice, we confirm that each cavity supports a fundamental TEM$_{00}$ mode with $1/e^{2}$ waist of \unit{489(8)}{\mu m} at \unit{914.3}{nm}.
		A Mott insulator of strontium atoms created in an optical lattice with this potential envelope is expected to form in the shaded area, which has a diameter $\sim$\unit{125}{\mu m}.
(c) Cross sections through the experimental data corresponding to horizontal and vertical lines drawn in panel (b).}
	\label{fig:teaser}
\end{figure}

For these reasons, solutions to the lattice inhomogeneity problem have been long sought after.
Towards this goal, methods to improve the homogeneity have been investigated, such as increasing the lattice beam waist~\cite{nicholson12,bothwell22}, the addition of  compensating laser beams~\cite{hart15,mazurenko17,chiu18}, and the projection of box potentials~\cite{gall21,wei21}.
Out of these methods, increasing the lattice beam waist provides the most fundamental improvement, because it is compatible with additional potential-shaping methods.
Moreover, it also provides the most significant improvement: linearly increasing the beam waist quadratically increases the mode area of 2D optical lattices.
However, increasing the beam waist is limited by the available laser power.
A standard solution to limited laser power is to enhance it in an optical cavity, and many research groups have created 1D optical lattices for different purposes~\cite{brusch06,yi11,nicholson12,kulosa15,brown17,yamaguchi19,kawasaki20,bothwell22}.
However, work on 2D~\cite{bowden19,cai20} or 3D cavity lattices~\cite{akatsuka10} is much more scarce, because significant technical challenges must be overcome to ensure stable trapping.

First, the cavity beams must be spatially overlapped precisely to generate optical lattices with large mode area (volume) in 2D (3D).
Obtaining and preserving a good overlap requires long-term mechanical and thermal stability of the whole cavity assembly.
These stability requirements become much harder to fulfill for an increased beam waist, because the mode overlap is sensitive to angular misalignments and this sensitivity scales with the fourth power of the mode waist~\cite{heinz20b,heinz21}.
Second, cavity-enhanced lattices cause increased atomic heating compared to free space lattices by converting laser phase noise and mechanical vibrations into lattice intensity noise.
Therefore, a viable cavity setup for scaling the size of optical lattices must support a large cavity mode size, a high degree of mechanical stability, and a long atomic lifetime.

In this work, we demonstrate the first stable 2D cavity-enhanced optical lattices with a mode area close to \unit{1}{mm^2} that can trap neutral atoms in both high- and low-intensity lattice regimes enabling experiments with both isolated and tunneling atoms.
Our cavity assembly is a monolithic device that contains two independent perpendicular optical cavities that cross at right angles, as shown in the photograph in Fig.~\ref{fig:teaser}(a), leading to a high degree of mechanical and thermal stability.
By precisely optical contacting cavity mirrors to an ultra-low-expansion glass spacer~\cite{heinz21}, we achieve a stable and near-perfect overlap of two orthogonal modes for the first time.
Our new cavity-enhanced 2D lattice setup can adapted to any wavelength of interest, even those for which the available laser power is limited, and our cavity is coated for multiple wavelengths relevant for experiments with ultracold strontium atoms.
We demonstrate the advantage of this flexibility by creating our lattices at an unconventional wavelength of \unit{914.3}{nm} and show more than an order-of-magnitude improvement in system size over free space setups based on the same lasers.

We briefly discuss the design of cavity assembly and its integration into our vacuum system in Section~\ref{sec:setup}.
Then, we characterize the 2D cavity lattices by loading strontium atoms and performing clock spectroscopy at a wavelength where the differential light shift of the clock states is large.
We observe spatially dependent clock excitation with that exhibits carrier and sideband transitions.
The high spectral resolution lets us observe vibrational-mode-specific carrier transitions for the first time.
These new spectral features provide access to the potential and population distribution in each lattice site.
To begin with, we directly map the lattice potential envelope, shown in Fig.~\ref{fig:teaser}(b).
With the measured potential envelope, we quantify the size and homogeneity of the created potential in Section~\ref{sec:fringes}.
In Section~\ref{sec:spectroscopy}, we use the spectrally resolved carriers to measure local temperatures and the polarizability ratio between the two clock states with high precision.
Finally, the monolithic cavity design results in a long atom lifetime in the cavity lattices and an excellent stability of the experimental setup as discussed in Sections~\ref{sec:lifetime} and~\ref{sec:stability}.

Our results create new opportunities for quantum simulations of strongly-coupled light matter interfaces~\cite{devega08,tudela18,krinner18} and quantum chemistry~\cite{arguello19}.
For such simulations, coherent tunneling over hundreds of lattice sites is desirable~\cite{tudela17a} which necessitates a large mode area.
These simulations also require the creation of state-dependent lattices that trap the clock states with a high contrast ratio and long lifetime~\cite{heinz20}, which is challenging due to the particularly limited laser power at the necessary wavelengths.
Our cavity lattices solve both challenges and solve the problem of scaling up optical lattice clocks and neutral-atom quantum computers to tens of thousands of qubits without any changes to the underlying device.

\section{Experimental Setup}
\label{sec:setup}

\subsection{Crossed Cavities}

Any optical lattice experiment faces a trade-off between system size and lattice depth, given by the beam waist and intensity, respectively.
Therefore, limited laser power leads to a technical limit on the system size.
A natural candidate to increase laser intensities is an optical cavity (or resonator).
In a cavity, light circulating between mirrors constructively interferes with incoming light, thus enhancing the laser power circulating within the cavity.
Since the power enhancement factor depends on the mirror reflectivities, whereas a cavity-mode waist depends on the mirrors' radii of curvature, a cavity provides independent control over the circulating power and waist.
This capability lets us achieve deep and large (and thus homogeneous) lattices simultaneously.
Other approaches to improve lattice homogeneity use beam shaping via cylindrical lenses~\cite{wei21} or spatial light modulators~\cite{gall21}.
However, these approaches face laser power limitations much earlier than power-enhancing buildup cavities.

In contrast to existing buildup cavities, we use a monolithic assembly that is both thermally and mechanically stable~\cite{heinz21} to ensure a long-term-stable mode overlap, as shown in Fig.~\ref{fig:teaser}(a).
In brief, we have optically contacted two pairs of mirrors to an octagon-shaped spacer made from ultra-low-expansion glass, forming two cavities that overlap in the center of the spacer.
Our design has no movable parts and is optimized to minimize any potential thermal expansions of the materials that can introduce a relative tilt between the mirrors, which would make the overlap unstable~\cite{heinz20b,heinz21}.
The spacer has bores with different diameters for optical access and high-resolution imaging.
With an interferometric method, we have achieved a near-perfect vertical overlap between the two modes~\cite{heinz20b,heinz21}.

After construction and characterization, the assembly was mounted in a stainless steel vacuum chamber attached to the vacuum system~\cite{snigirev19}.
In Fig.~\ref{fig:setup}(a), we show this science chamber featuring a pair of re-entrant viewports that allow high-resolution imaging of the atomic sample.
The octagon-shaped cavity assembly is mounted to the top viewport in a stress-free manner~\cite{heinz20b}, as sketched in the figure.
After bake-out, we reach a pressure below \unit{3\times 10^{-11}}{mbar} in the science chamber, demonstrating that the cavity assembly is compatible with state-of-the-art ultra high vacuum chambers.

The cavity mirrors are optimized for quantum simulations with strontium atoms and are highly reflective at several selected wavelengths~\cite{heinz21}.
In the remainder of this work, we couple laser light at \unit{914.3}{nm} into the cavities.
At this wavelength, the finesse and the intracavity power enhancement factor are 5025(58) and 1132(13), respectively~\cite{heinz21}.
By coupling a moderate power of $\sim$\unit{80}{mW} into each cavity arm, we create deep lattices with trap frequencies of \unit{116}{kHz}, corresponding to lattice depths of \unit{60}{\mu K}.
We optimize the mode-matching of the input beams to each cavity's fundamental transverse electric field mode (TEM$_{00}$) to $\sim$$99 \%$~\cite{heinz21}.
The laser beams are independently stabilized to the corresponding cavity resonance using acousto-optic modulators (AOMs) as detailed in Appendix~\ref{subsection:experimental_detail}.

\subsection{Atomic sample preparation}

\begin{figure}
  \centering
  \includegraphics{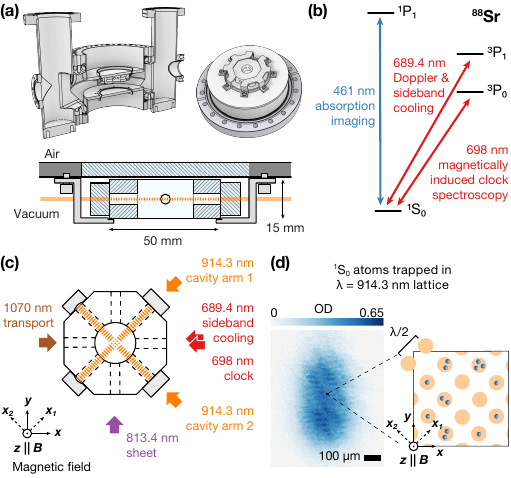}
  \caption{Overview of the experimental setup.
(a) Cross section of the science chamber, which includes a pair of re-entrant viewports (top left).
The cavity assembly rests in a stainless steel cage attached to the top viewport (top right).
A simplified cross section cut of the cavity assembly and its mounting structure is shown at the bottom.
(b) Simplified \Sr{88} energy diagram and optical transitions used in the experiment.
(c) Sketch of the laboratory and lattice coordinate frames and the relevant laser beams.
\Sr{88} atoms in the \SSZ{}($g$) state are transported into the assembly and are loaded into a trap created by the \unit{914.3}{nm} cavity lattice beams and a light sheet at \unit{813.4}{nm}.
The laser beam at \unit{689.4}{nm} is used for direct sideband cooling, and the laser beam at \unit{698}{nm} drives the clock transition induced by a bias magnetic field $B$ along $z$.
(d) {\it{In-situ}} absorption image of $g$ atoms in the combined trap of the lattices and light sheet (left).
On the right is an illustration showing the atoms occupying the lattice sites.}
  \label{fig:setup}
\end{figure}

To benchmark the performance of our optical lattice setup, we load strontium atoms in the vibrational ground states of the deep cavity lattices.
We begin by preparing \Sr{88} atoms using a robust and rapid magneto-optical trap~\cite{snigirev19} that operates on the narrow \SSZ{}-\TPO{} transition, as shown in Fig.~\ref{fig:setup}(b).
Subsequently, we transport the atoms into the center of the cavity assembly by moving the focus of an optical dipole trap beam at \unit{1070}{nm}~\cite{leonard14}.
The beam propagates along the $x$ axis as shown in Fig.~\ref{fig:setup}(c), has a 1/$e^{2}$ beam waist of \unit{50}{\mu m}, and induces a trap depth of $k_B \times$ \unit{45}{\mu K} where $k_B$ is the Boltzmann constant.
Here and in the following, the trap depth is understood as the potential energy difference between the center of the trap and the gravity-induced saddle point.

During transport, atoms spread axially over a few mm due to the weak axial confinement of the transport beam, and the temperature of the atomic cloud rises to $\sim$\unit{7}{\mu K}.
After transport, we perform narrow-line Doppler cooling in a crossed dipole trap.
The crossed dipole trap is created by overlapping the 1070-nm-transport beam and a light sheet at \unit{813.4}{nm}, as sketched in Fig.~\ref{fig:setup}(c).
The elliptical light sheet has a 1/$e^{2}$ beam waist of \unit{400}{\mu m} (\unit{13}{\mu m}) along the $x$ ($z$) axis with a trap depth of \unit{5}{\mu K}, corresponding to trap frequencies of (20, 5, 500) Hz along the ($x$, $y$, $z$) axes.
Subsequently, the transport beam is turned off, and we let the atomic cloud expand in the light sheet.

To adiabatically load the atoms into the cavity lattices sketched in Fig.~\ref{fig:setup}(c), we linearly ramp up the intensity of the cavity beams to a lattice depth (frequency) of \unit{60}{\mu K} (\unit{116}{kHz}).
Here, the lattice depth refers to a horizontal modulation depth assuming an infinitely extended 1D lattice where the lattice trap frequency $\nu_t$ and modulation depth $V$ are related by $\nu_t/\nu_\text{rec}=2\sqrt{V/h\nu_\text{rec}}$, where $\nu_{\text{rec}}=h/2M\lambda^{2}$ is the lattice recoil frequency for an atom of mass $M$, and $\lambda$ is the lattice wavelength.
At this point, the atoms are trapped in the potential created by the sheet and cavity beams.
The cavity beams by themselves would produce deep lattices horizontally, but produce a relatively weak dipole trap vertically with a trap frequency of $\sim$\unit{50}{Hz}.
Therefore, we achieve a tighter confinement along $z$ by having the light sheet intersect the lattice beams approximately at the minimum of the potential created by the lattice and gravity, as shown in the vertical potential in Fig.~\ref{fig:clock}(a).

After loading the atoms into the lattices, we cool the atoms to the vibrational ground band using sideband cooling on the \SSZ{}-\TPO{} transition~\cite{leibfried03,ido03}, where the \unit{689.4}{nm} cooling beam propagates horizontally at 45$^{\circ}$ to the lattice axes, as sketched in Fig.~\ref{fig:setup}(c).
Then, we ramp down the lattice power after sideband cooling to drop the atoms that are trapped only by the lattices but not by the light sheet, and we ramp the lattices back up.
Subsequently, we measure in-trap density profiles with {\it{in-situ}} absorption imaging along $z$~\cite{snigirev19}.
In Fig.~\ref{fig:setup}(d), we show a typical absorption image.
We use a large field of view and low resolution imaging with \unit{5.40(8)}{\mu m} per pixel to image the large atomic distribution and do not resolve the lattice structure.
Based on the optical density, we expect an average atom number of $\sim$1 per lattice site at the center of the trap.

\subsection{\Sr{88} clock excitation}
\label{section:clock}

\begin{figure}[t]
  \centering
  \includegraphics{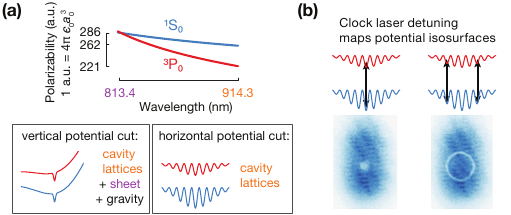}
  \caption{Spatially dependent clock spectroscopy in non-magic lattices.
(a) Dynamic dipole polarizabilities of $g$ and $e$ states (top).
At the light sheet wavelength of \unit{813.4}{nm}, $g$ and $e$ have the same polarizability ({\it{magic}} condition).
In contrast, at the cavity lattice wavelength of \unit{914.3}{nm}, $e$ experiences a polarizability reduced by \unit{15}{\%} compared to $g$ ({\it{non-magic}} condition).
For this reason, the light sheet creates identical traps for $g$ and $e$, while the cavity beams create a slightly stronger potential for $g$ than for $e$.
Vertical and horizontal cuts through the trapping potential created by the light sheet and cavity lattice are sketched below (polarizability difference between $g$ and $e$ is exaggerated for a clear illustration).
(b) {\it{In-situ}} absorption images after probing the clock transition at different detunings indicated by the black arrows.
The differential ac-Stark shifts of the non-magic cavity lattices allow local addressing, leading to ring structures that map the lattice envelope.}
  \label{fig:clock}
\end{figure}

We perform high-resolution spectroscopy on the ultra-narrow clock transition between \SSZ{} ($g$) and \TPZ{} ($e$) of strontium, as shown in Fig.~\ref{fig:setup}(b).
Unlike in the fermionic isotope \Sr{87}, where the clock transition is weakly electric-dipole-allowed~\cite{boyd07b} with a linewidth of \unit{1.35(3)}{mHz}~\cite{muniz21}, accessing the transition of the bosonic isotope \Sr{88} requires an external magnetic field~\cite{taichenachev06}.
Despite this requirement, we use \Sr{88}, because of its high natural abundance and simple electronic structure, which leads to simpler spectroscopic features.

To perform clock spectroscopy, we merge a \unit{698}{nm} clock probe beam into the same optical path as the \unit{689.4}{nm} sideband cooling beam, as sketched in Fig.~\ref{fig:setup}(c).
We apply clock laser light for $\sim$\unit{600}{ms} and a bias magnetic field of $\sim$\unit{45}{G} parallel to $z$, unless specified otherwise.
The clock probe beam has a $1/e^{2}$ waist of $\sim$\unit{285}{\mu m} and a power of \unit{21}{mW}, where the waist was calibrated as in Ref.~\cite{heinz20}.

The cavity lattices are created at a wavelength of \unit{914.332}{nm}.
At this wavelength, the differential polarizability of the clock states $\alpha_g-\alpha_e$ is $\simeq 0.15\,\alpha_g$, as evident from the polarizability plots of $g$ and $e$ shown in Fig.~\ref{fig:clock}(a).
Here, $\alpha_k$ specifies the polarizability of a state $k$.
The differential light shift is proportional to the light intensity and the differential polarizability.
As a result, the clock transition frequency shifts, and the magnitude of the shift varies as a function of lattice intensity.
Lattices in which the two clock states experience different light shifts are called {\it{non-magic}}.
In contrast, we intentionally make the light sheet operate at the {\it{magic}} wavelength of \unit{813.4}{nm}, such that it does not shift the transition.
Therefore, the local clock shift only originates from the cavity beams.
The trapping potential for $g$ and $e$ created by the combination of cavity beams, light sheet, and gravity is illustrated in Fig.~\ref{fig:clock}(a).
In the vertical potential cut, we see a dimple created by the \unit{813.4}{nm} sheet, and its trap depth is identical for both $g$ and $e$.
The horizontal potential cut is dominated by the cavity beams, and $e$ experiences a weaker lattice depth than $g$.
Therefore, we see that only the cavity lattices determine the differential light shift in the horizontal plane.

We model the cavity light intensity as the sum of two orthogonal TEM$_{00}$ Gaussian beams with $1/e^{2}$ waist $w$.
We have assumed that the waist of both cavity modes is the same because the two cavities are constructed in the same way~\cite{heinz21}.
We also assume that the waist stays constant over the area of our interest, which is valid because of the long Rayleigh length of the beams, $z_R\sim$ \unit{80}{cm}.
When the clock laser frequency is tuned close to the maximum differential ac-Stark shift, we excite $g$ atoms in the center as illustrated in Fig.~\ref{fig:clock}(b).
In contrast, when the laser is red detuned from the maximum, we excite $g$ to $e$ in an equipotential region.
This region takes the shape of a ring, reflecting the spatial cross section of the light intensity.
Taking such cross sections at different detunings enables us to map out the lattice trap envelope created by the cavity beams.

\section{Characterizing the Lattice Envelope}
\label{sec:fringes}

\begin{figure}[t]
  \centering
  \includegraphics{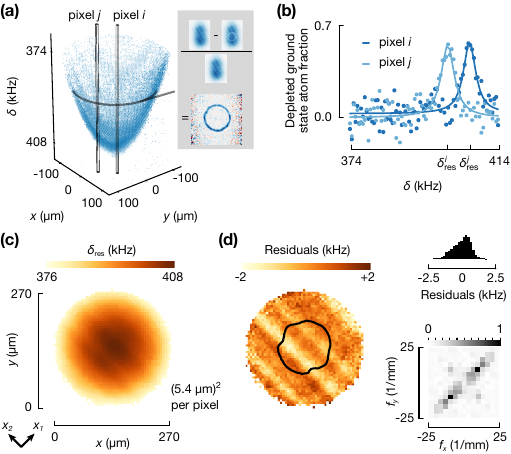}
  \caption{Lattice envelope characterization.
(a) Envelope of the lattice potential reconstructed by stacking the images of depleted $g$ fractions obtained at different clock laser detunings $\delta$ (left).
To obtain these images, we take an image with (OD$_\text{clk}$) and without (OD$_\text{bg}$) clock excitations.
For each detuning, we calculate the depleted $g$ fraction (OD$_\text{bg}$-OD$_\text{clk}$)/OD$_{\text{bg}}$ (right).
(b) For every pixel, we plot the depleted $g$ fraction as a function of detuning $\delta$ and extract the peak frequency $\delta_{\text{res}}$ with a Lorentzian fit (solid lines).
(c) Resulting image of $\delta_{\text{res}}$ for every pixel in panel (a).
The cavity axes are labelled as $x_1$ and $x_2$.
(d) When fitting the potential envelope predicted by ideal TEM$_{00}$ modes to the data in Panel (c), we find fit residuals showing fringe patterns coaligned with the lattice axes.
A histogram (peak-normalized spatial Fourier transform) of the residuals is shown in the top (bottom) right.
The black contour line on the image of residuals shows the expected Mott-insulator shape, based on the envelope data in Panel (c).
The resulting shape is shown as the shaded region in Fig.~\ref{fig:teaser}(b).}
  \label{fig:mapping}
\end{figure}

We use the measured equipotential surfaces to characterize the waist and homogeneity of the potential.
At each clock laser detuning $\delta$, we take two absorption images, one without clock excitation, OD$_{\text{bg}}$, and one with clock excitation, OD$_{\text{clk}}$.
From these two images, we extract a normalized difference image, (OD$_{\text{bg}}$-OD$_{\text{clk}}$)/OD$_{\text{bg}}$, and reconstruct the potential map as illustrated in Fig.~\ref{fig:mapping}(a).
This post-processed image reflects the fraction of $g$ atoms that have been {\it{depleted}} by the clock excitation.
Here, the {\it{depleted atoms}} include both those atoms that are still in $e$ at the time of imaging and those that have been lost from the trap after the excitation due to inelastic excited state collisions~\cite{lisdat09,bishof11}.
The details of the excitation process are described in Appendix~\ref{subsection:experimental_detail}.
We use this post-processed image representing the depleted ground state fraction for further analysis to eliminate possible systematic errors originating from the initial density distribution.

From a series of post-processed images taken at different detunings $\delta$, we determine each pixel's resonant detuning $\delta_\text{res}$, which is proportional to the lattice envelope averaged over each pixel.
Example traces of a pixel's fractional depleted $g$ atoms are shown in Fig.~\ref{fig:mapping}(b).
Each trace is individually fitted to a Lorentzian lineshape to extract $\delta_\text{res}$ for its pixel.
The distribution of the reduced $\chi^{2}$ of all the fits is centered around 0.9, and the statistical error on $\delta_\text{res}$ from the fits is $\sim$\unit{100}{Hz}.
The image of $\delta_\text{res}$ maps the shape of the potential and is shown in Fig.~\ref{fig:mapping}(c).
The variation of the potential depth across the whole image is $\sim$\unit{10}{\%} of the total ac-Stark shift, since we only load atoms into the central lattice region.

To quantify the waist and deviation of the measured $\delta_{\text{res}}$ from the expected values, we fit the image of $\delta_{\text{res}}$ to a fit function that models the potential given by the superposition of two orthogonal TEM$_{00}$ cavity modes.
The details of the fit function and fit parameters are described in Appendix~\ref{subsection:fit_function_derivation}.
We perform a weighted least squares fit and obtain a cavity mode waist of \unit{489(8)}{\mu m}, where the uncertainty arises mostly from the uncertainty in the image system magnification.
Although the fit captures the global Gaussian shape well, the residuals reveal that there are additional fringes shown in Fig.~\ref{fig:mapping}(d).
Since the peak-to-peak amplitude of the most dominant fringe ($\sim$\unit{3}{kHz}) is an order of magnitude larger than the error on the $\delta_\text{res}$ estimates ($\sim$\unit{100}{Hz}), the fringes are well resolved.
The statistical uncertainties show that our method can resolve structures as small as \unit{300}{ppm} of the total ac-Stark shift.
Due to the additional inhomogeneous fringes, the reduced $\chi^{2}$ of the lattice envelope fit to the $\delta_{\text{res}}$ data is $\sim$5, and the histogram of residuals is asymmetric, as shown at the top right of Fig.~\ref{fig:mapping}(d).

We observe that the inhomogeneous fringes are well aligned with the cavity axes, and that they are more (less) pronounced parallel to the optical axis of cavity arm 2 (arm 1), corresponding to lattice coordinate $x_2$ ($x_1$), as defined in Fig.~\ref{fig:setup}(c).
We characterize the fringe spacings using the peak-normalized 2D Fourier transform of the zero-padded fit residuals.
The result is shown at the bottom right of Fig.~\ref{fig:mapping}(d) and reveals a factor of three larger fringe amplitude along coordinate $x_{2}$ compared to $x_{1}$, peaking at a fringe wavelength of $\sim$\unit{65}{\mu m}.
Despite the different magnitudes, the Fourier transform shows similar spatial frequency components along both axes.
This similarity in the frequency components strongly suggests that a common mechanism causes the fringes along both axes.

The presence of these fringes is surprising, since the mode-cleaning effect of the cavities is expected to lead to a clean potential.
We consider three possible scenarios in which the light circulating within the cavities can exhibit such fringes: (1) scattering from defects or dust particles on the cavity mirrors, or contributions of higher transverse modes due to (2) mode mixing from imperfect cavity mirror surfaces, or (3) imperfect input coupling~\cite{heinz21}.

Based on the fringe wavelengths that we observe, we conclude that a dust particle or scattering center would have to be present $\sim$\unit{350}{\mu m} displaced from the mirror's center.
However, our estimates show that such a scattering center would have to scatter tens of percent of the circulating power into a solid angle of 2$\pi$ to explain the fringe amplitude that we observe.
Given that the fringe amplitude differs by an order of magnitude between each cavity, scattering such a large amount of light should have resulted in a very different finesse for both cavities.
Our \emph{in-situ} measurements show that the finesse of both cavities agrees within \unit{10}{\%}.

If we associate the fringes with a contribution of other transverse modes, the observed \unit{65}{\mu m} fringe spacing requires admixture of TEM$_{mn}$ modes with $m>100$, where $m$ counts the number of nodes in the lattice plane.
The beam profile associated with such a mode would have a much larger rms diameter than the \unit{489}{\mu m} waist of the TEM$_{00}$ mode that we observe.
The mirror profile at such large transverse extents is not spherical anymore, because it enters the transition into the flat annulus used for optical contacting~\cite{heinz21,heinz20b}, which might explain the irregular fringe patterns in Fig.~\ref{fig:mapping}(d).
However, we observe fringes with a peak-to-peak amplitude at the 1\% level on top of the expected TEM$_{00}$ profile.
This amplitude can only be explained if the cavity enhances the higher order mode and produces a stable interference pattern between both fundamental and high-order mode.
In particular, it is highly improbable that the fringes could be explained by interference between the \unit{40}{MHz} modulation sidebands used to stabilize the laser to the cavity, because any such interference pattern averages out on the time scales relevant for the atomic motion.
An explanation that might be consistent with our observations is an accidental mode degeneracy of the fundamental with a very high-order transverse mode, such that interference between both modes enhances the fringe contrast.

Another potential source of fringes could be artifacts from superradiant scattering of atoms in the cavity lattices or diffraction of the imaging beam from the periodic atom distribution.
Determining whether the fringes are present in the cavity lattices or whether they are artifacts of the detection method requires further investigation using a combination of improved beam shaping and alternative detection methods such as site-resolved fluorescence imaging.
For the remainder of this work, we assume that the fringes are present in the potential and show that even under this worst-case assumption, the cavity lattices outperform all other existing solutions for scaling up 2D optical lattices.

To our knowledge, we have created the largest far-off resonant 2D optical lattices for trapping ultracold atoms.
Our cavity mode waists are more than five times larger than what can be created using the most powerful laser available at this wavelength while preserving the lattice depth, resulting in more than an order of magnitude improvement on the number of available lattice sites.
From the measured intensity profile, we estimate an achievable Mott insulator size for the fermionic isotope \Sr{87}, which has more suitable scattering properties than the bosonic isotope \Sr{88}.
In two dimensions, the interaction energy $U$ required to form a Mott insulator is approximately 8$t$, where $t$ is the tunneling rate~\cite{esslinger10}.
For a fixed scattering length, the lattice depth can be tuned to satisfy the condition mentioned above.
At a typical depth of $\sim$10 $h\nu_{\text{rec}}$, a Mott insulator forms within the region where the energy shift due to the lattice envelope is smaller than the interaction energy between two atoms.
We find that the peak-to-peak amplitude of the fringes is three times smaller than the interaction energy, which is $\sim$\unit{700}{Hz} for \Sr{87}.
Therefore, we expect a homogeneous Mott insulator extending up to the boundary set by the interaction energy.
The fringes cause small distortions of the Mott insulator shape, as shown in Fig.~\ref{fig:mapping}(d).
However, our results show that the size would not significantly differ from the ideal size created with perfectly homogeneous lattices.

From these estimates which are detailed in Appendix~\ref{subsection:mott_size}, we expect that the Mott insulator state will occupy a region with a diameter of $D\simeq$ \unit{125}{\mu m} at a wavelength $\lambda=$ \unit{914.3}{nm}.
This diameter corresponds to $N\simeq\pi(D/\lambda)^{2}\simeq 6\times10^{4}$ lattice sites.
The area of the region does not vary much as a function of wavelength $\lambda$, although the number of sites changes quadratically due to the change in lattice spacing.
For this reason, our cavity assembly offers a solution to create large Mott insulators at any wavelength of interest~\cite{heinz20,campbell17} supported by the cavity mirrors.

\section{Local clock spectroscopy in non-magic lattices}
\label{sec:spectroscopy}

In the previous Section, the discussion focused on driving the most dominant carrier transition between the lowest vibrational states of the $g$ and $e$ lattices.
The carrier spectrum discussed in Fig.~\ref{fig:mapping}(b) was modeled with a Lorentzian function.
However, the spectrum can become more complex when transitions between higher vibrational states are considered.
We now make use of the high spectral resolution of the clock laser to resolve spectral transitions between such higher vibrational states.
This new capability enables us to precisely determine the polarizability ratio of the clock states without having to calibrate the lattice intensities.
This ratio is an important quantity that determines the magnitude of the differential light shift and can be used to calibrate state-of-the-art atomic structure calculations~\cite{heinz20}.
Moreover, we find that we can use this method to locally measure temperature with a spatial resolution only limited by the imaging optics.

To understand the spectra we measure, we first consider the sideband spectrum in a deep optical lattice, where tunneling is suppressed.
In this case, the spectrum resembles the spectrum of a harmonically trapped ion~\cite{leibfried03}, and a trapped atom occupies a discrete vibrational level $n$.
Assuming an infinitely extended 1D lattice without any radial confinement and considering the quartic distortion by the sinusoidal lattice potential, the vibrational energy spectrum is~\mbox{\cite{blatt09}}
\begin{equation}
E_n/h=\nu_t(n+1/2)-\frac{\nu_{\text{rec}}}{2}(n^{2}+n+1),
\label{eq:vibrational_energy}
\end{equation}
where $\nu_t$ is the on-site lattice trap frequency.
With this expression, we can explain all the spectral transitions observed in atoms trapped in a deep non-magic optical lattice.

A typical spectrum consists of three different types of transitions: carrier transitions that maintain the vibrational state, and red and blue sideband transitions that respectively remove and add a motional quantum.
In a magic wavelength lattice, the transition frequencies of all carrier transitions are degenerate.
From Eqn.~(\ref{eq:vibrational_energy}), we find that the first red and blue sideband transitions are detuned from the carrier by $\nu_t-\nu_{\text{rec}}$.
Typically, the sideband transitions are highly suppressed compared to the carriers transitions by the Lamb-Dicke factor~\cite{leibfried03}.
The red sideband is even further suppressed in a laser-cooled sample due to a large ground vibrational state population.
In this work, we focus on the carrier and the first blue sideband transitions.

\begin{figure*}[t]
  \centering
  \includegraphics{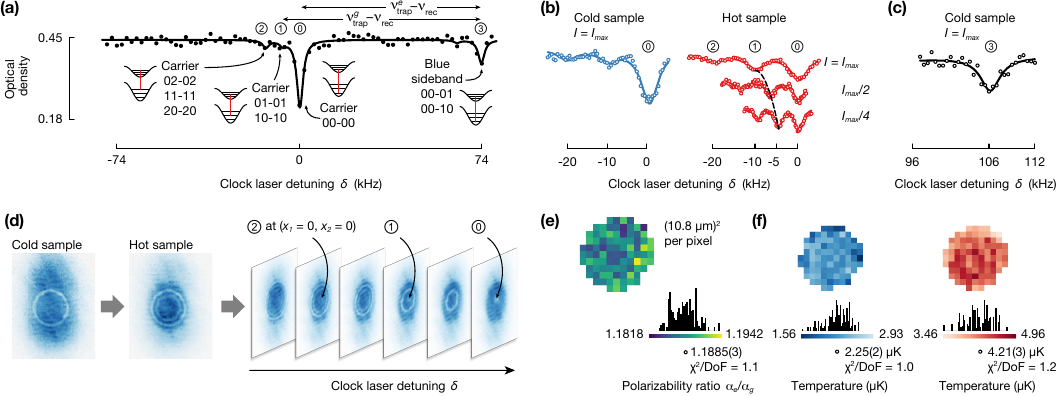}
  \caption{Local clock spectroscopy of carrier and sideband transitions in non-magic lattices.
(a) A sparse clock sideband spectrum in a 2D non-magic optical lattice.
The solid line is a fit to a multi-peak Lorentzian function, in which the peak locations were fixed at the expected detunings.
(b) A more finely resolved spectrum covering a detuning range of the carrier transitions for the samples with two different temperatures, one colder (left) and the other, hotter (right).
To maximize the splitting, we increased $\nu_t$ to $\sim$\unit{109}{kHz} from $\sim$\unit{77}{kHz} (at which (a) was taken).
For the hotter samples, we vertically stack three spectra taken at different lattice intensities $I$, at which $I_\text{max}$ results in $\nu_t \simeq\unit{109}{kHz}$.
The dotted line is a guide to the eye for easier visualization of the carrier-splitting-frequency as a function of the lattice intensity.
(c) A blue-sideband spectrum taken at $\nu_t\simeq\unit{109}{kHz}$.
(d) Comparison of the absorption images of the cold and hot samples after the clock excitation at a fixed detuning (left) and a series of images of the hot sample as a function detuning.
(e) Per-pixel polarizability ratio and weighted histogram.
(f) Temperature maps of cold (left) and hot (right) samples with weighted histograms.}
  \label{fig:spectroscopy}
\end{figure*}

The spectrum becomes more complex in a non-magic lattice, where the trap potential is state-dependent.
Thus, the trap frequencies of $g$ and $e$ lattices are different, and we use $\nu^{k}_{t}$ to denote the trap frequency of state $k$.
In this case, the carrier transitions are no longer degenerate, but are split by $\nu^{e}_t-\nu^{g}_t$ according to Eqn.~(\ref{eq:vibrational_energy}).
Moreover, the first blue sideband is detuned by $\nu^{e}_{t}-\nu_{\text{rec}}$ from the carrier transition of the lowest vibrational state.

In 2D, the vibrational levels are labelled by two independent vibrational numbers $n_1$ and $n_2$, each corresponding to a vibrational band of one of the lattices.
Since the two lattices are orthogonal and do not interfere, the energy spectrum is given by $E_{n_{1}}+E_{n_{2}}$.
For simplicity, we consider the case when both lattices have an equal intensity (or depth), $I=I_{1}=I_{2}$.
In this case, the carrier transitions split according to the total vibrational number $n_T=n_1+n_2$ of the states involved.
Therefore, each carrier transition is ($n_{T}$+1)-fold degenerate.
An example of a coarse scan over the clock excitation spectrum for $\nu^{e}_t\simeq\unit{77}{kHz}$ in our non-magic 2D lattices is shown in Fig.~\ref{fig:spectroscopy}(a).

We work in a resolved carrier regime that has not been previously explored.
The frequency splitting between the two neighboring carrier transitions is $(\nu^{g}_t-\nu^{e}_t)\propto\sqrt{I}(\sqrt{\alpha_e}-\sqrt{\alpha_g})$.
To maximize the splitting, we increase $\nu^{e}_t$ to $\sim$\unit{109}{kHz}, and take a high-resolution spectrum, zooming into the carrier transitions as shown in the left part of Fig.~\ref{fig:spectroscopy}(b).
We observe up to three different carrier transitions, of which the one from the lowest vibrational state is the most blue-detuned.
The amplitude of the three peaks becomes more comparable when we intentionally heat the sample by applying a beam resonant with the \SSZ{}-\TPO{} transition as shown in the right part of Fig.~\ref{fig:spectroscopy}(b).
As we decrease the lattice depth linearly, we observe that the splitting reduces quadratically as expected.
Similar to the carrier transitions, the first blue sideband transitions split as well.
However, here we focus on the first blue sideband of the lowest vibrational state, which is shown in Fig.~\ref{fig:spectroscopy}(c).

The splitting of the carrier transitions is clearly visible in the ground state images as well.
As described in Section~\ref{sec:fringes}, we excite atoms in a ring shape, reflecting the equipotential surfaces of the cavity lattice envelope.
When we increase the population of the higher vibrational states by heating the sample, we see additional smaller rings appearing in Fig.~\ref{fig:spectroscopy}(d).
Each ring results from driving the carrier transitions from different vibrational states, which are resonant at different locations.
Similar to what we have seen in Section~\ref{sec:fringes}, all three rings move inward as the detuning increases due to the spatially dependent ac-Stark shift.
The dominant carrier transition, which involves the lowest vibrational states, arrives at the center last because it is the most blue-detuned transition.
Moreover, the spacing between two neighboring rings increases as the rings approach the center since the potential becomes flatter.

With the resolved carrier and blue-sideband spectrum, we first extract the polarizability ratio $\alpha_g/\alpha_e$, which is one of the parameters that determine the magnitude of the differential ac-Stark shift.
Since $\alpha_k\propto (\nu^{k}_t)^{2}$, the polarizability ratio $\alpha_g/\alpha_e=(\nu^{g}_t/\nu^{e}_t)^{2}$.
To measure $\nu^{e}_t$ and $\nu^{g}_t$, we use an analysis method similar to the one used in Section~\ref{sec:fringes}.
For every two-by-two averaged pixel, we determine the frequency difference between the lowest carrier and first blue sideband peaks, which is $\nu^{e}_t-\nu_{\text{rec}}$.
To measure $\nu^{g}_t$, we heat the sample to better observe the different carrier peaks.
For each spectrum of the averaged pixel, we fit a three peak Lorentzian function with the frequency difference between the peaks constrained to be the same.
From the fit, we determine the frequency splitting $\Delta \delta$ between the carriers for each averaged pixel, and combine this value with $\nu^{e}_t$ to obtain $\nu^{g}_t$ for every pixel, $\nu^{g}_t=\nu^{e}_t+\Delta \delta$.
The error bars of the parameter estimates from the fits are rescaled according to the reduced $\chi^2$ of the fits to compensate for the non-Gaussian noise of the absorption images.

In Fig.~\ref{fig:spectroscopy}(e), we show the polarizability ratio estimated from the pixel-to-pixel $\nu_{g}$ and $\nu_{e}$ maps.
The weighted mean of the ratio $\alpha_g/\alpha_e$ = 1.1885 $\pm$ (3$\times$10$^{-4}$)$_{\text{stat}}$ $\pm$ (1$\times$10$^{-3}$)$_{\text{sys}}$, which is in good agreement with the theory described in Appendix ~\ref{subsection:polarizablity_theory}.
The systematic uncertainty arises from the experimental drifts between the hot and cold data sets that are used to extract $\nu_{e}$ and $\Delta \delta$, respectively.
This uncertainty can be greatly reduced by further minimizing the elapsed time between the data sets.
The variance of the ratio across the sample can be explained by the variance of each pixel because the reduced $\chi^{2}$ is 1.14.
Therefore, we conclude that we do not observe a systematic variation of the ratio across the sample.
Our method provides improved robustness compared to a similar method explored in Ref.~\cite{mcdonald15} because we can make use of the resolved carrier spectrum combined with spectral imaging.

Finally, we extract the local temperatures of the sample using the carrier spectrum.
In the temperature regime that we are considering, the vibrational populations are Boltzmann-distributed.
In this case, the temperature $T$ can be estimated by measuring the relative population $p_0/p_1$ of the first two non-degenerate levels and the energy spacing between them, using $k_BT=h(\nu^{g}_t-\nu_{\text{rec}})/$ln($2p_0/p_1$).
We estimate $\nu^{g}_t-\nu_{\text{rec}}$ for each pixel using the same method as described above.
Next, we use a hotter sample and determine the peak locations and amplitudes by fitting a three-peak Lorentzian function with equal frequency difference between the peaks to each averaged pixel, and we compare the amplitudes of the two most blue-detuned carrier peaks to estimate $p_1/p_0$.
For the colder sample, we repeat the same procedure but keep the peak locations fixed to those determined from the hotter sample.

The extracted temperature maps and weighted histograms are shown in Fig.~\ref{fig:spectroscopy}(f).
We clearly observe a temperature difference between the cold and hot samples at \unit{2.25(2)}{\mu K} and \unit{4.21(3)}{\mu K}, respectively.
The temperature variation across the samples are within the temperature uncertainty of each pixel because the reduced $\chi^{2}$ is 1.01 and 1.2 for cold and hot samples, respectively.
The temperatures of the cold sample correspond to $\sim$80\% vibrational ground state fraction.

Our local thermometry assumes that the induced atom losses during the clock excitation (discussed in Appendix~\ref{subsection:experimental_detail}) do not influence the temperature estimates.
This assumption is corroborated by repeating the temperature measurements with shorter clock excitation durations and negligible loss, where we did not observe temperature differences.
The thermometry technique based on the carrier spectrum has been proven to be more precise than methods based on time-of-flight or sideband spectrum due to its high signal-to-noise ratio~\cite{mcdonald15,han18}.
We have improved the technique's robustness by resolving the carriers and extended it to probe local temperatures.
When combined with higher numerical aperture imaging, this technique will enable spectroscopic measurements of motional band populations with single-site resolution.

\section{Lifetime}
\label{sec:lifetime}

\begin{figure}
  \centering
  \includegraphics{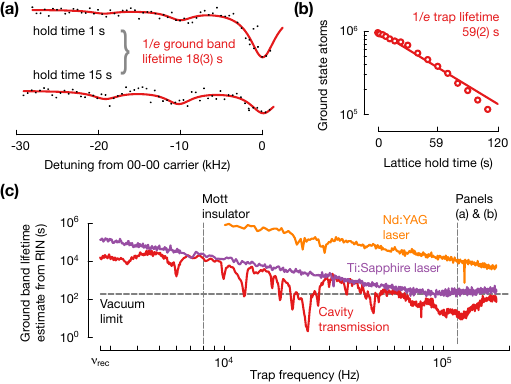}
  \caption{Lifetime in the cavity lattices.
(a) Resolved carrier spectra in the cavity lattices at a trap depth (frequency) of 457 \Erec (\unit{116}{kHz}).
The spectrum on top (bottom) was taken after holding atoms for \unit{1}{s} (\unit{15}{s}) in the lattices.
We extract a $1/e$ lattice ground band lifetime of $\sim$\unit{18(3)}{s}.
(b) Number of $g$ atoms as a function of lattice hold time.
 We extract an overall $1/e$ trap lifetime of \unit{59(2)}{s}.
(c) Estimated ground band lifetime as a function of lattice depth.
The estimation was performed by rescaling the relative intensity noise (RIN) of a Ti:Sapphire laser transmitted through cavity 2.
For reference, we show the same estimates obtained from the RIN of the cavity input light and the RIN of a highly intensity-stable Nd:YAG laser.}
  \label{fig:lifetime}
\end{figure}

In optical traps, heating induces excitation to higher motional bands, leading to motional state decoherence and subsequent atom loss.
To characterize the heating sources in our setup, we measure the lifetimes of the ground band population and the overall lifetime of atoms trapped in the cavity lattices.

We characterize the ground band lifetime using the resolved carrier spectrum technique presented in the previous Section.
This technique offers a new way to probe the motional ground band populations in the high-lattice-depth regime with high signal-to-noise.
In Fig.~\ref{fig:lifetime}(a), we take two carrier spectra after holding the atoms in the lattices at a modulation depth (trap frequency) of 457 $E_{\text{rec}}$ (\unit{116}{kHz}) for \unit{1}{s} and for \unit{15}{s}.
By comparing the populations in the ground band at these times and assuming an exponential heating rate, we extract a ground band lifetime of \unit{18(3)}{s}, which is comparable to state-of-the-art free space lattice experiments~\cite{blatt15} at similar depths in units of the recoil energy.

In addition to the ground band lifetime, we also measure the overall trap lifetime at the same lattice modulation depth.
The overall trap lifetime serves as a good benchmark to compare with other setups where the ground band lifetime is not accessible.
In Fig.~\ref{fig:lifetime}(b), we show the number of $g$ atoms trapped in the cavity lattices and light sheet as a function of the trap hold time, and we extract a trap $1/e$ lifetime of \unit{59(2)}{s}.

The heating mechanisms in optical traps include collisions with background gas, incoherent scattering of trap light, and laser-noise-induced heating~\cite{blatt15}.
Based on the longest trap lifetime we have measured, we project a vacuum-limited lifetime $>$\unit{180}{s}.
The expected lifetime due to incoherent light scattering is also more than two orders of magnitude longer than the observed ground-band lifetime, leaving laser noise as the main source of heating.
Moreover, we observe that the lifetime changes depending on the parameters of the laser's intensity and frequency stabilization control loops.

Laser-noise-induced heating arises due to laser beam intensity and pointing fluctuations.
In deep optical lattices, where each lattice site can be approximated as a harmonic trap, the laser intensity (pointing) noise power spectral density at $2\nu_t$ ($\nu_t$) causes parametric heating~\cite{savard97} that results in transitions between lattice bands that are two (one) motional quanta apart.
In traps enhanced by optical cavities, we expect the intensity fluctuations to dominate for two reasons.
First, heating from pointing fluctuations is strongly suppressed due to the resonator's mechanical stability ~\cite{mosk01}.
Second, locking a laser to a cavity resonance converts laser frequency noise into amplitude noise, increasing the latter beyond that in free-space optical lattices~\cite{lodewyck10}.
Thus, we focus on the relative intensity noise (RIN) of the laser transmitted by the cavities.
Combining the RIN and ground band lifetime measurements, we estimate expected lifetimes for different lattice depths as shown in Fig.~\ref{fig:lifetime}(c).
Here, we use the parametric heating rate $\propto \nu^{2}_t S(2\nu_t)$, where $S(2\nu_t)$ is the power spectral density of the fractional intensity noise at $2\nu_t$, to scale the lifetimes to the measurement shown in Fig~\ref{fig:lifetime}(a), based on the model in Ref.~\cite{blatt15}.
However, based on the rescaled RIN of the transmission, we conclude that the ground band lifetime will be vacuum-limited in most regions, even at wavelengths where very-low-noise non-planar-ring-oscillator lasers~\cite{harry10} are not available.

\section{Long-term Stability}
\label{sec:stability}

\begin{figure}
  \centering
  \includegraphics{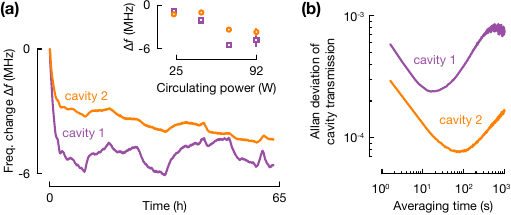}
  \caption{Stability of the cavity lattices (a) Cavity resonance frequency change $\Delta f$ as a function of time for both cavities. At $t=0$, we couple the laser beams into the cavities, which create a circulating power of $\sim$\unit{92}{W} per cavity. We see a rapid initial decrease of \unit{6}{MHz}, which corresponds to an expansion of the cavity length by \unit{1}{nm}. Then, $\Delta$$f$ settles and fluctuates with a peak-to-peak amplitude of $\sim$\unit{2}{MHz}. Inset: $\Delta$$f$ as a function of circulating power for caivty 1 (squares) and 2 (circles). Here, $\Delta$$f$ is extracted by fitting the trace of the first hour to an exponential function. (b) Fractional intensity variation of the cavity transmission plotted as a function of time.}
  \label{fig:stability}
\end{figure}

Finally, we characterize the long-term frequency and intensity stability of the lattice beams.
In our setup, the lattice laser is stabilized to a resonant frequency of the cavity, which slowly changes as the cavity shrinks or expands.
Since we do not actively stabilize the cavity length, any length change directly influences the laser frequency.
Although we actively stabilize the input beam power before coupling it into the cavity, we do not additionally stabilize the power of the transmitted light.
Therefore, the beam power inside the cavity is susceptible to mechanical drifts of the incoupling optical components and the performance of the input intensity servo.
In this Section, we quantify the frequency and intensity drifts of our setup and discuss the consequences for optical lattice clocks and quantum simulators using our cavity lattices.

To estimate the laser frequency drift due to the cavity length change, we continuously measure two parameters: (i) the laser frequency by beating it with a femtosecond optical frequency comb and (ii) the frequency of the double-pass AOM used to stabilize the laser to the cavity resonance frequency (Appendix~\ref{subsection:experimental_detail}).
By subtracting the two numbers, we obtain the cavity's resonance frequency drift independently of the laser frequency drift.
In Fig.~\ref{fig:stability}(a), we plot the change of the resonance frequencies for both cavities as a function of time, starting with the moment we couple light into the cavities.
We use a circulating power of $\sim$\unit{92}{W} in each cavity, matching the conditions of the measurements in Sections~\ref{sec:fringes},~\ref{sec:spectroscopy}, and~\ref{sec:lifetime}.
Within the first hour, we observe a rapid decrease in the resonance frequency of $\sim$\unit{6}{MHz}.
Subsequently, the resonance frequencies settle but fluctuate with a peak-to-peak amplitude of $\sim$\unit{2}{MHz}.

The decrease by \unit{6}{MHz} corresponds to a cavity expansion of \unit{1}{nm} compared to the nominal cavity length of \unit{50}{mm}.
When repeating this measurement at different power levels, we see that the expansion reduces proportionally (see inset).
We conclude that the cavity mirrors scatter and absorb part of the circulating light due to the cavity mirror losses, which are $\sim$\unit{100}{ppm} based on our finesse and transmission measurements.
These losses deposit heat on the cavity which therefore expands.
We attribute the fluctuations on the long time scales to slow environmental temperature changes, which could be further reduced by stabilizing the temperature of the vacuum chamber.

We use the measured resonance stability to estimate a lower bound of the accuracy of an optical lattice clock based on our cavities.
Compared to cavities that are tunable in length~\cite{mosk01,akatsuka10,letargat13,schiller12,bowden19}, we will have to overcome the obstacle of constructing a fully monolithic cavity with resonances as close to the magic wavelength as possible.
This problem could be solved by adapting our optical contacting methods~\cite{heinz20b} to tune the resonance frequency with an accuracy of \unit{10}{MHz}.
The cavity resonance frequency can be further fine-tuned by placing it in a temperature-controlled enclosure that is also required to create a well-defined blackbody-radiation environment~\cite{nicholson15,bowden19}.
To estimate the clock inaccuracy due to detuning from the magic wavelength condition, we assume a lattice depth of 100 \Erec~\cite{lemonde05}, which will lead to a cavity frequency variation of $\sim$\unit{1.3}{MHz} caused by coupling light into the cavity.
With these assumptions, we obtain a clock frequency variation of \unit{2}{mHz}~\cite{shi15}, corresponding to a fractional clock accuracy of $\approx 5 \times 10^{-18}$.
We believe that the largest contribution to the cavity frequency variation is the mirror loss, which can be reduced by more than an order of magnitude when using mirrors with $<$\unit{10}{ppm} loss.
This reduction would improve the frequency stability by an order of magnitude, assuming that reducing the mirror loss would proportionally reduce the cavity frequency variation.
The necessary temperature control to minimize the black body shift uncertainty reduces the frequency fluctuations caused by environmental changes to the same level.
Other systematic effects such as charge-buildup on dielectric surfaces and a non-uniform blackbody radiation background due to the cavity spacer can be addressed using existing techniques.
Charge buildup can be reduced by placing electrodes on the inside of the central cavity bore to compensate for stray electric fields and to evaluate the stray field's magnitude~\cite{bowden19}.
Blackbody radiation shifts can be suppressed by providing a temperature-controlled environment~\cite{nicholson15,bowden19} at low temperatures~\cite{ushijima15,schymik21}.
With these improvements, we project a possible clock accuracy below $10^{-18}$, which would let state-of-the-art 2D or 3D optical lattice clocks make use of the scaling advantage provided by our cavity lattices.

Finally, we characterize the long-term stability of the cavity lattice depth by measuring the cavity transmission.
The Allan deviation of the transmitted power as a function of the averaging time is shown in Fig.~\ref{fig:stability}(b).
We observe a fractional instability below $10^{-3}$ for typical $\sim$\unit{20}{s} cycle times of quantum gas microscope experiments.
Furthermore, the transmission of cavity 2 is more stable than the one of cavity 1, which is consistent with the lower input power sensitivity observed in Fig.~\ref{fig:stability}(a).
By measuring the transmission and implementing a slow feedback loop on the cavity input powers, we could preserve the stability over many experimental runs which would result in long-term-stable quantum simulation parameters.
The fractional stabilities of the tunneling rate $t$ and the interaction energy $U$ roughly scale as $(3/4)(\sigma_{V}/V)$, where $\sigma_{V}/V$ is the fractional uncertainty of the lattice depth $V$.
Therefore, we expect that it is feasible to achieve a long-term stability of $10^{-3}$ for $t$ and $U$.

\section{Conclusion}
\label{sec:conclusion}

We have presented a new cavity-based experimental platform for scaling quantum simulators, quantum computers, and clocks based on neutral atoms trapped in optical lattices.
Our lattices increase the number of available lattice sites by more than an order of magnitude compared to state-of-the-art free-space lattices~\cite{blatt15}.
Currently, most far-off-resonant optical lattices are created using high-power Nd:YAG lasers.
Here, our solution opens new opportunities to create large and deep lattices at any desired wavelength supported by the cavity mirror coatings.

As a demonstration, we loaded strontium atoms into two-dimensional optical lattices generated at \unit{914.332}{nm}, which is well-adapted for narrow-line laser cooling of strontium atoms.
The lattice laser beams have waists of \unit{489(8)}{\mu m} and operate at a circulating power of \unit{92}{W}.
This circulating power is more than an order of magnitude larger than commercially available laser power at this wavelength.
Despite the large beam waist, atoms are trapped in lattices as deep as 457 \Erec, corresponding to trap frequencies of \unit{116}{kHz}.
In these non-magic optical lattices, we perform high-resolution clock spectroscopy.
Extending the work of Refs.~\cite{shibata14,marti18}, we show a highly sensitive method to reconstruct the lattice intensity envelope from the local clock shift.
The statistical uncertainty of our reconstruction method shows that intensity deviations as small as \unit{300}{ppm} of the peak intensity can be resolved.
From the reconstructed intensity map, we estimate the size and shape of a future Mott insulator state and conclude that the state will consist of $6\times10^{4}$ atoms.
This atom number is more than an order of magnitude larger than in state-of-art 2D optical lattices generated from free space laser beams~\cite{blatt15}.

The combination of high resolution laser spectroscopy and deep non-magic lattices allows us to resolve different motional carrier transitions of the lattices for the first time.
We use this capability to locally measure the sample temperature with high spatial resolution.
The resolved carrier spectrum also provides a method to directly measure the ground-band lifetime.
We observe ground-band and lattice lifetimes of \unit{18(3)}{s} and \unit{59(2)}{s} respectively, and a long-term lattice frequency (depth) stability on the MHz (0.1\%) level.
Our results demonstrate that there are no disadvantages of cavity-based far-off-resonant optical lattices compared to free space, while allowing the creation of deep and large optical lattices at wavelengths where the available laser power is limited.
These cavity lattices create new opportunities for analog and digital quantum simulation, including controlled collisional phase gates~\cite{daley08,daley11,daley11b}, quantum simulations of light-matter interfaces~\cite{devega08,tudela17a,tudela17b,tudela18} and quantum chemistry~\cite{arguello19}.
Moreover, the strong reduction in harmonic confinement will reduce the finite size effects for any optical lattice quantum simulator and will reduce the experimental time required for measuring quantum many-body correlations.

The cavity lattices can also be used to improve the precision of lattice-based atom interferometers and optical lattice clocks by providing more identical particles to reduce the quantum projection noise.
Our compact and stable cavity design will enable near-future applications of optical atomic clocks that require hands-off operation outside of laboratories such as in satellites and airplanes~\cite{koller17,grotti18,wolf09,origlia18}.
Finally, neutral atom arrays in optical tweezers interacting via Rydberg states have become a promising candidate for quantum computing~\cite{saffman10,weiss17,browaeys20,morgado21}, but current array sizes have been limited to $\sim$400 sites, partly due to the high laser power required to create larger arrays~\cite{wang20}.
With our optical cavity lattices, these neutral-atom quantum computers can be scaled to tens of thousands of qubits.

\begin{acknowledgments}
  We thank M.~Safronova for providing the matrix elements required for the polarizability ratio calculation and for stimulating discussions, and D.~Yankelev for critical reading of the manuscript.
  This work was supported by funding from the European Union (PASQuanS Grant No.
817482).
  A.\,J.\,P. was supported by a fellowship from the Natural Sciences and Engineering Research Council of Canada (NSERC), funding ref. no. 517029, N.\,\v{S}.
was supported by a Marie Sk\l{}odowska-Curie individual fellowship, grant agreement no. 844161, and V.\,K. was supported by a Hector Fellow Academy fellowship.
\end{acknowledgments}

\appendix

\section{Experimental details}
\label{subsection:experimental_detail}

\subsection{Crossed cavity locks}
The cavity lattice beams at \unit{914.332}{nm} are generated from a Ti:Sapphire laser.
The output of the laser is divided into four paths: a path for locking the laser's frequency to a pre-stabilization cavity, two paths for coupling into each cavity arm of the crossed cavity assembly, and an optical heterodyne beat setup with a femtosecond optical frequency comb.
Locking the laser's frequency to a pre-stabilization cavity with piezo-tunable length allows us to keep the laser's frequency close to the resonance frequencies of the crossed cavities.

The length of the pre-stabilization cavity is actively stabilized by locking one of the resonance frequencies to the clock laser frequency with a Pound–Drever–Hall (PDH) lock.
The clock laser itself is frequency stabilized to an ultra-stable reference cavity with a finesse of $\sim$280000.
Once the length of the pre-stabilization cavity is stabilized, we lock the Ti:Sapphire laser's frequency to another of its resonances via the PDH technique as well.
For this purpose, the error signal is fed back to a piezo attached to the Ti:Sapphire laser's bow-tie cavity mirrors.

The two beams that couple into the cavities are each frequency-shifted with separate double-pass acousto-optic modulators (AOMs) and then fiber-coupled to the optical setup for coupling into the cavities.
The AOMs tune the frequency of each beam to the resonant frequency for each cavity.
To maintain the desired frequency for each beam, we again use the PDH technique, and use the error signal to control the radio-frequency (rf) signal driving the AOMs.
The two cavity are separated in frequency by $\sim$\unit{60}{MHz}/$\nu_\mathrm{FSR}$, with free spectral range $\nu_\mathrm{FSR} = c/2L = \unit{3}{GHz}$, cavity length $L=$ \unit{50}{mm}, and speed of light $c$.

During the experiments, we actively stabilize the intensity of the laser beams coupled to the cavity.
For this stabilization, we split off $\sim$$1 \%$ of each beam's power and send it onto a photodiode.
The photodiode signal is compared to a DC signal to create an error signal which is then fed into a proportional-integral (PI) controller that actively controls the amplitude of the rf signals of the double pass AOMs.
With this locking scheme, we control the intensity sent to the cavities.
The input cavity beams can also be shut off rapidly by turning off the rf power.
Disabling the frequency and intensity locks of the two cavity beams does not affect the frequency stability of the Ti:Sapphire laser, since its frequency is pre-stabilized.

\begin{figure}
 \centering
  \includegraphics{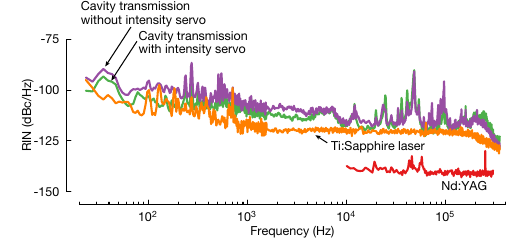}
  \caption{Characterization of laser noise.
Relative intensity noise of the Ti:Sapphire laser (orange), transmission of cavity 2 with (green) and without (purple) intensity servo, compared with the relative intensity noise of a low-noise Nd:YAG laser (red).}
\label{fig:rin_spectra}
\end{figure}

To characterize the laser noise, we measure the relative intensity noise (RIN) of the laser under three different conditions: (1) when the laser frequency is locked only to the pre-stabilization cavity, (2) as in (1) but with additional frequency stabilization to the crossed cavities, and (3) the same as (2) but with intensity stabilization, where conditions (2) and (3) are measured after the transmission through the crossed cavity.
In Fig.~\ref{fig:rin_spectra}, we show such measurements for cavity 2.
For comparison, we also measure the RIN of a commercial low-noise Nd:YAG laser.
We use these measurements to estimate the ground-band lifetimes at different trap frequencies in Section~\ref{sec:lifetime}.

Finally, an optical heterodyne beat of the Ti:Sapphire laser and the optical frequency comb is used to monitor the absolute frequency of the laser.
By simultaneously measuring this frequency and the frequency of the double-pass AOMs that are used for locking, we can determine the absolute frequencies of both cavity modes used to trap the atoms.
We use this method to measure the long-term stability of our experimental setup in Section~\ref{sec:stability}.

\subsection{Clock excitation}

\begin{figure}
 \centering
  \includegraphics{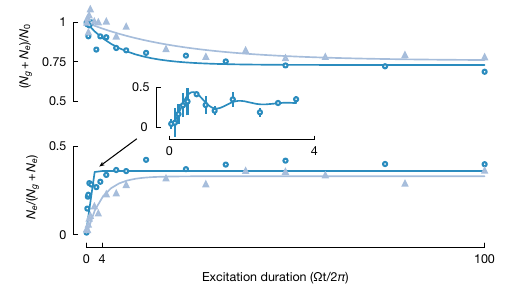}
  \caption{Clock excitation dynamics.
Total atom number $N_g+N_e$ normalized to the initial total atom number $N_0$ (top) and excited fraction (bottom) as a function of clock excitation duration.
The inset is a zoomed-in portion of the initial excitation dynamics, fit with an exponentially decaying sine.
The measurements are repeated an external magnetic field at $\sim$\unit{45}{G} (triangles) and $\sim$\unit{225}{G} (circles).
The data are fit with exponential functions as a guide to the eye.}
  \label{fig:clock_dynamics}
\end{figure}

In our experiments, the clock excitation dynamics are susceptible to decoherence mechanisms due to elastic $e$-$g$ collisions and fast inelastic $e$-$e$ collisions in \Sr{88}~\cite{lisdat09}.
The elastic collisions reset the coherence between $g$ and $e$, but the population stays constant, while inelastic collisions cause atom loss.

To distinguish between the two effects, we use a detection scheme that can image both $g$ and $e$ atoms separately.
To image the in-trap density of $g$ atoms, we use absorption imaging on the \SSZ{}-\SPO{} transition~\cite{snigirev19} as explained in the main text.
To image $e$ atoms, we remove $g$ atoms by applying light resonant with the \SSZ{}-\SPO{} transition and repump $e$ back to $g$ by applying \unit{679}{nm} and \unit{707}{nm} laser light resonant with the \TPZ{}-\TSO{} and \TPT{}-\TSO{} transitions, respectively~\cite{lisdat09}.

Using the above method, we excite atoms in the center of the lattices and take $g$ and $e$ images at different clock excitation durations.
From these measurements, we study how the total atom number, {\it{i.e.}} $N_g+N_e$, and the excited state fraction, {\it{i.e.}} $N_e/(N_g+N_e)$, evolve as a function of the clock excitation duration, where $N_s$ specifies the number of atoms in state $s$.
Since we work with non-magic lattices and an imaging resolution of \unit{5.40(8)}{\mu m}, the excited state fraction derived from each pixel is averaged over many different clock laser detunings.
For this reason, we study the clock excitation dynamics at the center of the lattices, where the lattice envelope is flattest.
The results obtained from averaging the central four pixels are plotted in Fig.~\ref{fig:clock_dynamics}, where we have repeated the measurement at two different magnetic fields, \unit{45}{G} and \unit{225}{G}, respectively.
Following Ref.~\cite{taichenachev06}, the different magnetic field values proportionally scale the Rabi frequency $\Omega$~\cite{taichenachev06}.
Therefore, we have rescaled the clock laser duration according to the strength of the magnetic field.

We observe a decay of the total number of atoms as shown in the bottom of Fig.~\ref{fig:clock_dynamics}, which is expected due to the inelastic $e$-$e$ collisions~\cite{lisdat09}.
At the center of the atomic cloud, we estimate $\sim$1 atom per lattice site on average from the in-situ images.
Assuming a Poisson distribution, we expect that $\sim$\unit{40}{\%} of the populated lattice sites are occupied by more than one atom.
Therefore, the clock spectroscopy in our setup is susceptible to atom loss.
Here, this loss is advantageous in characterizing the potential, because it enhances the signal-to-noise ratio of the ground state depletion images used for the technique.
From the data shown in Fig.~\ref{fig:clock_dynamics}, we expect that the \unit{40}{\%} of the atoms lost from the trap and \unit{60}{\%} of the $e$ atoms create the depleted images used in Section~\ref{sec:fringes}.

From the dynamics of the excited state fractions shown in Fig.~\ref{fig:clock_dynamics} (top), we observe an exponential rise to a steady-state value, which resembles strongly dephased Rabi dynamics.
At a high magnetic field of \unit{225}{G} at which we expect $\Omega\sim$ 2$\pi\times$\unit{500}{Hz}, we observe clear Rabi oscillations that quickly dephase as shown in the inset of Fig.~\ref{fig:clock_dynamics} (top).
In addition to elastic collisions, there are several other mechanisms that can cause dephasing in our experiments, such as misalignment of the clock probe beam~\cite{blatt09}, the clock laser linewidth, the clock and lattice laser intensity noise, and the effect of averaging several pixels where each pixel contains contributions from many clock laser detunings.
Among these possibilities, our estimates show that the lattice intensity fluctuations are the most dominant dephasing mechanism.
At a differential ac-Stark shift of $\sim$\unit{400}{kHz}, a fractional lattice intensity stability of $\sim$10$^{-3}$ causes inhomogeneity in $\delta$ of \unit{400}{Hz}, which is already on the order of $\Omega/2\pi$ at \unit{225}{G}.
Our setup is particularly susceptible to lattice intensity noise because the techniques described in the main text rely on a large differential ac-Stark shift.
Such susceptibility to the lattice intensity noise can be removed by creating the lattices at the magic wavelength.

\section{Mott-insulator size estimate}
\label{subsection:mott_size}

\begin{figure}
  \centering
  \includegraphics{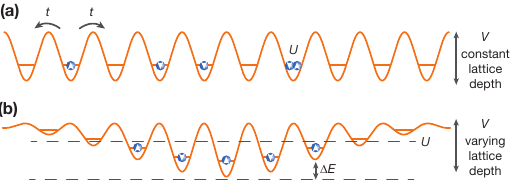}
  \caption{Fermionic atoms in an optical lattice.
(a) Ultracold fermionic atoms trapped in infinitely extended optical lattices of depth $V$ can tunnel between sties at rate $t$.
Two fermions of opposite spin on the same site interact with interaction energy $U$.
(b) Finite-sized optical lattices vary quadratically in depth causing a site-dependent energy offset $\Delta E$, typically referred as harmonic confinement or a lattice envelope.}
  \label{fig:optical_lattices}
\end{figure}

One of the most common initial states for quantum simulations is a Mott insulator, a low entropy initial state in which a single atom occupies each lattice site.
Since this is the starting state, the achievable Mott insulator size determines the size and thus the complexity of the simulation.
In this Section, we show how this Mott insulator size depends on the waists of the beams that create the optical lattices.

The most important energy scales of quantum simulation in optical lattices are $t$ and $U$, which respectively characterize the tunneling rate of atoms between neighboring sites and the interaction energy between two atoms on the same site as shown in Fig.~\ref{fig:optical_lattices}(a).
In two dimensions, a fermionic spin-1/2 Mott-insulator forms when the interaction energy $U$ is approximately the same as the ground bandwidth $8t$~\cite{esslinger10}, filling each lattice site with exactly one atom.

However, since the lattice is not perfectly flat, there is an additional site-specific energy offset $\Delta E$ as shown in Fig.~\ref{fig:optical_lattices}(b).
This offset leads to an additional constraint, $\Delta E < U$, which limits the size of the Mott insulator.
To estimate the system size, we first estimate at which lattice depth the constraint $U=8t$ is satisfied~\cite{esslinger10}.
Following Ref.~\cite{bloch08},

\begin{align}
U=\sqrt{8/\pi}kaE_{\text{rec}}(V/E_\text{rec})^{3/4},
\label{eq:J}
\end{align}

\begin{align}
t\sim\frac{4}{\sqrt{\pi}}E_\text{rec}\bigg(\frac{V}{E_{\text{rec}}}\bigg)^{3/4}\text{exp}\bigg[-2\bigg(\frac{V}{E_\text{rec}}\bigg)^{1/2}\bigg],
\label{eq:U}
\end{align}
where $k=2\pi/\lambda$ and $a$ are the wave-vector and the scattering length, respectively.
The lattice potential can be approximated as
\begin{align}
V(x_1,x_2)\sim V[e^{-2x^{2}_1/w^{2}}\text{cos}^{2}(kx_1)+e^{-2x^{2}_2/w^{2}}\text{cos}^{2}(kx_2)].
\end{align}
Using the expressions above, a Mott insulator forms when $V=E_{\text{rec}} [\text{ln}(8\sqrt{2}/ka)]^{2}/4$.
Let us consider the fermionic isotope \Sr{87} which has $a=96a_0$, where $a_0$ is the Bohr radius, and a nuclear spin $I=9/2$.
When using nuclear-spin-polarized atoms in a mixture of $\ket{g,m_I=\pm 9/2}$ in optical lattices created at a wavelength of \unit{914}{nm}, the transition to a Mott insulator is expected to occur at $V\sim 8 E_\text{rec}$.
At this depth, the interaction energy $U$ is $\sim$\unit{728}{Hz}.
We can compare this energy $U$ with the site-specific $\Delta E$ to estimate the size of the Mott insulator.
The site-specific energy $\Delta E$ scales with the harmonic confinement of the lattice, $M\omega^{2}r^{2}/2$ where $r^{2}=x_{1}^{2}+x_{2}^{2}$ and $\omega=\sqrt{4V/Mw^{2}}$ is the radial trap frequency.
Here, $M$ is the mass of a \Sr{87} atom, and $w$ is the $1/e^{2}$ waist of the beams that create the lattices.

Setting $\Delta E=U$, we find the radius of the Mott insulator $r_{\text{Mott}}=w\sqrt{U/2V}$.
Considering the waist of our cavity modes $w$ = \unit{489(8)}{\mu m}, we expect a radius of $\sim$\unit{60}{\mu m}.
The area defined by this radius corresponds to the region where the lattice depth is within $96-97$\% of the maximum.
The total atom number in this region $N_{\text{atoms}}=\pi r^{2}/(\lambda/2)^{2}$ which is 4$\times10^{4}$ to 11$\times10^{4}$ depending on the wavelength, which ranges from 1064 to \unit{689}{nm}.
Performing the same estimates using a waist of \unit{80}{\mu m} as used in Ref.~\cite{blatt15}, we obtain an atom number ranging from 1$\times10^{3}$ to 3$\times 10^{3}$ for the same wavelength range.

\section{Fit function derivation}
\label{subsection:fit_function_derivation}

In this Section, we derive the fit function used in Section~\ref{sec:fringes}, describing the spatially dependent detuning that is resonant with the clock transition.

The energy shift of each clock state can be decomposed into two parts: a shift induced by the ac-Stark effect and the zero-point vibrational energy in each lattice site.
The former is given by $-\alpha_k I(x_1,x_2)/2\epsilon_0 c$, where $\epsilon_0$ is the vacuum permittivity, $I(x_1,x_2)$ is the total lattice intensity from the both lattices, and $k$ labels the state $e$ or $g$.
In orthogonal 2D optical lattices where the two lattices have an identical beam waist $w$, we can write the lattice envelope as $I(x_1,x_2)=I_0[e^{-2(x_1-x^{0}_1)^{2}/w^{2}}+(1+\varepsilon)e^{-2(x_2-x^{0}_2)^{2}/w^{2}}]$, where $\varepsilon$ specifies the intensity balance between the two lattices, $I_0$ is the peak intensity, and $x^{0}_j$ specifies the position of the lattice intensity maximum.
The second cause of the energy shift is the zero-point vibrational energy of each state.
Assuming that the atoms occupy the vibrational ground state, the zero-point energy experienced by each state is given by $h\nu^{k}_t(x_j)/2$ per lattice axis, where $\nu^{k}_t(x_j)$ specifies the on-site lattice trap frequency that state $k$ experiences along the lattice axis $x_j$.
The lattice trap frequency also depends on polarizability and light intensity.
Following the above definitions, $\nu_t^{k}(x_1)=2\sqrt{\nu_{\text{rec}}\alpha_k I(x_1)/(2c\epsilon_0 h)}$ and $\nu_t^{k}(x_2)=2\sqrt{\nu_{\text{rec}}(1+\varepsilon)\alpha_k I(x_2)/(2c\epsilon_0 h)}$.

The resonant condition occurs when the detuning with respect to the free space resonance matches the additional shifts,
\begin{align}
\delta_{\text{res}}&=\frac{1}{2\epsilon_0 c h}(\alpha_g - \alpha_e) I(x_1,x_2) \nonumber \\
& \qquad +\sqrt{\frac{\nu_{\text{rec}}I(x_1)}{2c\epsilon_0 h}}(\sqrt{\alpha_e}-\sqrt{\alpha_g})\nonumber  \\
& \qquad +\sqrt{\frac{\nu_{\text{rec}}(1+\varepsilon)I(x_2)}{2c\epsilon_0 h}}(\sqrt{\alpha_e}-\sqrt{\alpha_g}).
\label{eq:energy_difference}
\end{align}

In our experiments, we spectroscopically measure the peak trap frequency of $g$ or $e$ to calibrate the peak lattice intensity, $I_0$, and the polarizability ratio, $\alpha_g/\alpha_e$.
We rewrite the above expression with respect to these quantities, where the peak trap frequency is $\nu_t^k=2\sqrt{\nu_{\text{rec}}\alpha_k I_0/(2c\epsilon_0 h)}$.
Then, we can rewrite $\delta_{\text{res}}$ as
\begin{align}
\delta_{\text{res}}&=\frac{1}{\nu_\text{rec}}\bigg(\frac{\nu^{e}_t}{2}\bigg)^{2}\bigg(\frac{\alpha_g}{\alpha_e}-1\bigg) \bigg[e^{-2(x_1-x^{0}_1)^{2}/w^{2}}\nonumber\\
& \qquad \qquad \qquad \qquad \qquad +(1+\varepsilon)e^{-2(x_2-x^{0}_2)^{2}/w^{2}}\bigg] \nonumber \\
& +\frac{\nu^{e}_t(e^{-(x_1-x^{0}_1)^{2}/w^{2}})}{2}\bigg(1-\sqrt{\frac{\alpha_g}{\alpha_e}}\bigg) \nonumber \\
& +\frac{\nu^{e}_t\sqrt{1+\varepsilon}(e^{-(x_2-x^{0}_2)^{2}/w^{2}})}{2}\bigg(1-\sqrt{\frac{\alpha_g}{\alpha_e}}\bigg),
\label{eq:fit_function}
\end{align}
where we have chosen the excited state trap frequency $\nu^{e}_t$ rather than the ground state trap frequency for convenience.

We fit the data described in Section~\ref{sec:fringes} to the fit function shown in Eqn.~(\ref{eq:fit_function}) with an additional frequency offset $f_0$.
The fitted parameters are $x^{0}_1,x^{0}_2,f_0,w,$ and $\varepsilon$.
We obtain a cavity mode waist of \unit{489(8)}{\mu m} with a reduced $\chi^{2}$ of  $\sim 5$, as discussed in the main text.

\section{Clock state polarizabilities}
\label{subsection:polarizablity_theory}

The polarizability of an electronic state is determined by contributions from the core and valence electrons.
The core part of the polarizability can be calculated in the single-electron approximation, including random-phase approximation corrections~\cite{safronova99}.
The valence part of the atomic state $i$ is given by the sum of contributions over all electric-dipole coupled states $k$.
This part can be decomposed further into scalar, vector, and tensor parts.
For the clock states of \Sr{88} where both states have $J=0$, only the scalar part contributes~\cite{heinz20}, and the polarizability can be calculated according to~\cite{safronova15}
\begin{equation}
\alpha_{k} = \dfrac{2}{3 \hbar} \sum_{l} \frac{ \omega_{kl}}{\omega_{kl}^{2} - \omega^{2}} | \Bra{l} D \Ket{k} |^2.
\label{eq:polarizability}
\end{equation}
Here $\Bra{l} D \Ket{k} $ is the reduced dipole matrix element between the clock state $k$ and state $l$.
We use $\omega_{kl}$ to denote the corresponding transition frequency.
The valence part also depends on the frequency $\omega$ of the light field interacting with the atom.

\begin{table}[H]
  \centering
  \setlength{\tabcolsep}{6pt}%
\begin{tabularx}{0.95\linewidth}{lllS}
  \\
  State $l$ & $\Delta E$ & $\Bra{l} D \Ket{k}$ & $\alpha_k$ \\
   & (cm$^{-1}$) & ($e a_0$) & \multicolumn{1}{c}{($4\pi\epsilon_0 a_0^3$)} \\
  \hline
  \\
      & & $k = 5s^{2}\,\SSZ{}$ & \\
  \\
  $5s5p$\SLJ{3}{P}{1} & 14504 &  0.1510 & 0.53\\
  $5s5p$\SLJ{1}{P}{1} & 21698 &  5.248 & 248.98\\
  $5s6p$\SLJ{3}{P}{1} & 33868 &  0.034 & 0.01\\
  $5s6p$\SLJ{1}{P}{1} & 34098 &  0.282 & 0.38\\
  Other & & & 5.8 \\
  Core & & & 5.3 \\
  Total & & & 261.0$\pm$1.2\\
  \\
      & & $k = 5s5p\,\TPZ{}$ & \\
  \\
  $5s4d$\SLJ{3}{D}{1} & 3842 &  2.671 &  -38.25\\
  $5s6s$\SLJ{3}{S}{1} & 14721 &  1.968 & 85.92\\
  $5s5d$\SLJ{3}{D}{1} & 20689 &  2.450 & 58.91\\
  $5p^{2}$\SLJ{3}{P}{1} & 21083 &  2.605 & 64.44\\
  $5s7s$\SLJ{3}{S}{1} & 23107 &  0.515 & 2.16\\
  Other & & & 42.07 \\
  Core & & & 5.55 \\
  Total & & & 220.8$\pm$2.3 \\
\end{tabularx}
\caption{Contributions to scalar polarizability $\alpha_{k}$ of the $g$ ($e$) clock state $5s^{2}\,\SSZ{}$ ($5s5p\,\TPZ{}$) at \unit{914.332}{nm}.
The transition energies $\Delta E$~\cite{nistasd} are listed in  $\textrm{cm}^{-1}$ and the reduced electric-dipole matrix elements $\Bra{l} D \Ket{k}$ and polarizability contributions~\cite{safronova13,safronovapriv} are shown in atomic units, where $e$ ($a_0$) is the electron charge (Bohr radius).
Here, \textit{Other} refers to contributions from states which are not listed explicitly and \textit{Core} refers to the core polarizability.
The uncertainties for individual polarizability contributions result from propagating uncertainties in the matrix elements~\cite{safronovapriv} and lead to the quoted uncertainty for the total polarizability.}
\label{tab:polarizability}
\end{table}

We calculate the polarizability of the $g$ ($e$) clock state $5s^2\,\SSZ{}$ ($5s5p \,\TPZ{}$) at \unit{914.332}{nm} in Tab.~\ref{tab:polarizability}.
From the total polarizabilities, we extract $\alpha_{g}/\alpha_{e} = 1.18 \pm 0.01$.


\begin{thebibliography}{84}%
\makeatletter
\providecommand \@ifxundefined [1]{%
 \@ifx{#1\undefined}
}%
\providecommand \@ifnum [1]{%
 \ifnum #1\expandafter \@firstoftwo
 \else \expandafter \@secondoftwo
 \fi
}%
\providecommand \@ifx [1]{%
 \ifx #1\expandafter \@firstoftwo
 \else \expandafter \@secondoftwo
 \fi
}%
\providecommand \natexlab [1]{#1}%
\providecommand \enquote  [1]{``#1''}%
\providecommand \bibnamefont  [1]{#1}%
\providecommand \bibfnamefont [1]{#1}%
\providecommand \citenamefont [1]{#1}%
\providecommand \href@noop [0]{\@secondoftwo}%
\providecommand \href [0]{\begingroup \@sanitize@url \@href}%
\providecommand \@href[1]{\@@startlink{#1}\@@href}%
\providecommand \@@href[1]{\endgroup#1\@@endlink}%
\providecommand \@sanitize@url [0]{\catcode `\\12\catcode `\$12\catcode
  `\&12\catcode `\#12\catcode `\^12\catcode `\_12\catcode `\%12\relax}%
\providecommand \@@startlink[1]{}%
\providecommand \@@endlink[0]{}%
\providecommand \url  [0]{\begingroup\@sanitize@url \@url }%
\providecommand \@url [1]{\endgroup\@href {#1}{\urlprefix }}%
\providecommand \urlprefix  [0]{URL }%
\providecommand \Eprint [0]{\href }%
\providecommand \doibase [0]{https://doi.org/}%
\providecommand \selectlanguage [0]{\@gobble}%
\providecommand \bibinfo  [0]{\@secondoftwo}%
\providecommand \bibfield  [0]{\@secondoftwo}%
\providecommand \translation [1]{[#1]}%
\providecommand \BibitemOpen [0]{}%
\providecommand \bibitemStop [0]{}%
\providecommand \bibitemNoStop [0]{.\EOS\space}%
\providecommand \EOS [0]{\spacefactor3000\relax}%
\providecommand \BibitemShut  [1]{\csname bibitem#1\endcsname}%
\let\auto@bib@innerbib\@empty
\bibitem [{\citenamefont {Ashkin}(1997)}]{ashkin97}%
  \BibitemOpen
  \bibfield  {author} {\bibinfo {author} {\bibfnamefont {A.}~\bibnamefont
  {Ashkin}},\ }\bibfield  {title} {\bibinfo {title} {Optical trapping and
  manipulation of neutral particles using lasers},\ }\href
  {https://doi.org/10.1073/pnas.94.10.4853} {\bibfield  {journal} {\bibinfo
  {journal} {Proc. Natl. Acad. Sci. U.S.A.}\ }\textbf {\bibinfo {volume}
  {94}},\ \bibinfo {pages} {4853} (\bibinfo {year} {1997})}\BibitemShut
  {NoStop}%
\bibitem [{\citenamefont {Grimm}\ \emph {et~al.}(2000)\citenamefont {Grimm},
  \citenamefont {Weidem{\"u}ller},\ and\ \citenamefont
  {Ovchinnikov}}]{grimm00}%
  \BibitemOpen
  \bibfield  {author} {\bibinfo {author} {\bibfnamefont {R.}~\bibnamefont
  {Grimm}}, \bibinfo {author} {\bibfnamefont {M.}~\bibnamefont
  {Weidem{\"u}ller}},\ and\ \bibinfo {author} {\bibfnamefont {Y.~B.}\
  \bibnamefont {Ovchinnikov}},\ }\bibfield  {title} {\bibinfo {title} {Optical
  dipole traps for neutral atoms},\ }\href
  {https://doi.org/10.1016/S1049-250X(08)60186-X} {\bibfield  {journal}
  {\bibinfo  {journal} {Adv. At. Mol. Opt. Phys.}\ }\textbf {\bibinfo {volume}
  {42}},\ \bibinfo {pages} {95} (\bibinfo {year} {2000})}\BibitemShut {NoStop}%
\bibitem [{\citenamefont {Gross}\ and\ \citenamefont {Bloch}(2017)}]{gross17}%
  \BibitemOpen
  \bibfield  {author} {\bibinfo {author} {\bibfnamefont {C.}~\bibnamefont
  {Gross}}\ and\ \bibinfo {author} {\bibfnamefont {I.}~\bibnamefont {Bloch}},\
  }\bibfield  {title} {\bibinfo {title} {Quantum simulations with ultracold
  atoms in optical lattices},\ }\href {https://doi.org/10.1126/science.aal3837}
  {\bibfield  {journal} {\bibinfo  {journal} {Science}\ }\textbf {\bibinfo
  {volume} {357}},\ \bibinfo {pages} {995} (\bibinfo {year}
  {2017})}\BibitemShut {NoStop}%
\bibitem [{\citenamefont {Saffman}\ \emph {et~al.}(2010)\citenamefont
  {Saffman}, \citenamefont {Walker},\ and\ \citenamefont
  {M\o{}lmer}}]{saffman10}%
  \BibitemOpen
  \bibfield  {author} {\bibinfo {author} {\bibfnamefont {M.}~\bibnamefont
  {Saffman}}, \bibinfo {author} {\bibfnamefont {T.~G.}\ \bibnamefont
  {Walker}},\ and\ \bibinfo {author} {\bibfnamefont {K.}~\bibnamefont
  {M\o{}lmer}},\ }\bibfield  {title} {\bibinfo {title} {Quantum information
  with {Rydberg} atoms},\ }\href {https://doi.org/10.1103/RevModPhys.82.2313}
  {\bibfield  {journal} {\bibinfo  {journal} {Rev. Mod. Phys.}\ }\textbf
  {\bibinfo {volume} {82}},\ \bibinfo {pages} {2313} (\bibinfo {year}
  {2010})}\BibitemShut {NoStop}%
\bibitem [{\citenamefont {Weiss}\ and\ \citenamefont
  {Saffman}(2017)}]{weiss17}%
  \BibitemOpen
  \bibfield  {author} {\bibinfo {author} {\bibfnamefont {D.~S.}\ \bibnamefont
  {Weiss}}\ and\ \bibinfo {author} {\bibfnamefont {M.}~\bibnamefont
  {Saffman}},\ }\bibfield  {title} {\bibinfo {title} {Quantum computing with
  neutral atoms},\ }\href {https://doi.org/10.1063/PT.3.3626} {\bibfield
  {journal} {\bibinfo  {journal} {Physics Today}\ }\textbf {\bibinfo {volume}
  {70}},\ \bibinfo {pages} {44} (\bibinfo {year} {2017})}\BibitemShut {NoStop}%
\bibitem [{\citenamefont {Browaeys}\ and\ \citenamefont
  {Lahaye}(2020)}]{browaeys20}%
  \BibitemOpen
  \bibfield  {author} {\bibinfo {author} {\bibfnamefont {A.}~\bibnamefont
  {Browaeys}}\ and\ \bibinfo {author} {\bibfnamefont {T.}~\bibnamefont
  {Lahaye}},\ }\bibfield  {title} {\bibinfo {title} {Many-body physics with
  individually controlled {Rydberg} atoms},\ }\href
  {https://doi.org/10.1038/s41567-019-0733-z} {\bibfield  {journal} {\bibinfo
  {journal} {Nature Physics}\ }\textbf {\bibinfo {volume} {16}},\ \bibinfo
  {pages} {132} (\bibinfo {year} {2020})}\BibitemShut {NoStop}%
\bibitem [{\citenamefont {Morgado}\ and\ \citenamefont
  {Whitlock}(2021)}]{morgado21}%
  \BibitemOpen
  \bibfield  {author} {\bibinfo {author} {\bibfnamefont {M.}~\bibnamefont
  {Morgado}}\ and\ \bibinfo {author} {\bibfnamefont {S.}~\bibnamefont
  {Whitlock}},\ }\bibfield  {title} {\bibinfo {title} {Quantum simulation and
  computing with {Rydberg}-interacting qubits},\ }\href
  {https://doi.org/10.1116/5.0036562} {\bibfield  {journal} {\bibinfo
  {journal} {AVS Quantum Sci.}\ }\textbf {\bibinfo {volume} {3}},\ \bibinfo
  {pages} {023501} (\bibinfo {year} {2021})}\BibitemShut {NoStop}%
\bibitem [{\citenamefont {Ludlow}\ \emph {et~al.}(2015)\citenamefont {Ludlow},
  \citenamefont {Boyd}, \citenamefont {Ye}, \citenamefont {Peik},\ and\
  \citenamefont {Schmidt}}]{ludlow15}%
  \BibitemOpen
  \bibfield  {author} {\bibinfo {author} {\bibfnamefont {A.~D.}\ \bibnamefont
  {Ludlow}}, \bibinfo {author} {\bibfnamefont {M.~M.}\ \bibnamefont {Boyd}},
  \bibinfo {author} {\bibfnamefont {J.}~\bibnamefont {Ye}}, \bibinfo {author}
  {\bibfnamefont {E.}~\bibnamefont {Peik}},\ and\ \bibinfo {author}
  {\bibfnamefont {P.~O.}\ \bibnamefont {Schmidt}},\ }\bibfield  {title}
  {\bibinfo {title} {Optical atomic clocks},\ }\href
  {https://doi.org/10.1103/RevModPhys.87.637} {\bibfield  {journal} {\bibinfo
  {journal} {Rev. Mod. Phys.}\ }\textbf {\bibinfo {volume} {87}},\ \bibinfo
  {pages} {637} (\bibinfo {year} {2015})}\BibitemShut {NoStop}%
\bibitem [{\citenamefont {Cronin}\ \emph {et~al.}(2009)\citenamefont {Cronin},
  \citenamefont {Schmiedmayer},\ and\ \citenamefont {Pritchard}}]{cronin09}%
  \BibitemOpen
  \bibfield  {author} {\bibinfo {author} {\bibfnamefont {A.~D.}\ \bibnamefont
  {Cronin}}, \bibinfo {author} {\bibfnamefont {J.}~\bibnamefont
  {Schmiedmayer}},\ and\ \bibinfo {author} {\bibfnamefont {D.~E.}\ \bibnamefont
  {Pritchard}},\ }\bibfield  {title} {\bibinfo {title} {Optics and
  interferometry with atoms and molecules},\ }\href
  {https://doi.org/10.1103/RevModPhys.81.1051} {\bibfield  {journal} {\bibinfo
  {journal} {Rev. Mod. Phys.}\ }\textbf {\bibinfo {volume} {81}},\ \bibinfo
  {pages} {1051} (\bibinfo {year} {2009})}\BibitemShut {NoStop}%
\bibitem [{\citenamefont {Georgescu}\ \emph {et~al.}(2014)\citenamefont
  {Georgescu}, \citenamefont {Ashhab},\ and\ \citenamefont
  {Nori}}]{georgescu14}%
  \BibitemOpen
  \bibfield  {author} {\bibinfo {author} {\bibfnamefont {I.~M.}\ \bibnamefont
  {Georgescu}}, \bibinfo {author} {\bibfnamefont {S.}~\bibnamefont {Ashhab}},\
  and\ \bibinfo {author} {\bibfnamefont {F.}~\bibnamefont {Nori}},\ }\bibfield
  {title} {\bibinfo {title} {Quantum simulation},\ }\href
  {https://doi.org/10.1103/RevModPhys.86.153} {\bibfield  {journal} {\bibinfo
  {journal} {Rev. Mod. Phys.}\ }\textbf {\bibinfo {volume} {86}},\ \bibinfo
  {pages} {153} (\bibinfo {year} {2014})}\BibitemShut {NoStop}%
\bibitem [{\citenamefont {Esslinger}(2010)}]{esslinger10}%
  \BibitemOpen
  \bibfield  {author} {\bibinfo {author} {\bibfnamefont {T.}~\bibnamefont
  {Esslinger}},\ }\bibfield  {title} {\bibinfo {title} {{Fermi-Hubbard} physics
  with atoms in an optical lattice},\ }\href
  {https://doi.org/10.1146/annurev-conmatphys-070909-104059} {\bibfield
  {journal} {\bibinfo  {journal} {Annual Review of Condensed Matter Physics}\
  }\textbf {\bibinfo {volume} {1}},\ \bibinfo {pages} {129} (\bibinfo {year}
  {2010})}\BibitemShut {NoStop}%
\bibitem [{\citenamefont {de~Vega}\ \emph {et~al.}(2008)\citenamefont
  {de~Vega}, \citenamefont {Porras},\ and\ \citenamefont {Cirac}}]{devega08}%
  \BibitemOpen
  \bibfield  {author} {\bibinfo {author} {\bibfnamefont {I.}~\bibnamefont
  {de~Vega}}, \bibinfo {author} {\bibfnamefont {D.}~\bibnamefont {Porras}},\
  and\ \bibinfo {author} {\bibfnamefont {J.~I.}\ \bibnamefont {Cirac}},\
  }\bibfield  {title} {\bibinfo {title} {Matter-wave emission in optical
  lattices: Single particle and collective effects},\ }\href
  {https://doi.org/10.1103/PhysRevLett.101.260404} {\bibfield  {journal}
  {\bibinfo  {journal} {Phys. Rev. Lett.}\ }\textbf {\bibinfo {volume} {101}},\
  \bibinfo {pages} {260404} (\bibinfo {year} {2008})}\BibitemShut {NoStop}%
\bibitem [{\citenamefont {Gonz{\'a}lez-Tudela}\ and\ \citenamefont
  {Cirac}(2018)}]{tudela18}%
  \BibitemOpen
  \bibfield  {author} {\bibinfo {author} {\bibfnamefont {A.}~\bibnamefont
  {Gonz{\'a}lez-Tudela}}\ and\ \bibinfo {author} {\bibfnamefont {J.~I.}\
  \bibnamefont {Cirac}},\ }\bibfield  {title} {\bibinfo {title}
  {Non-{M}arkovian quantum optics with three-dimensional state-dependent
  optical lattices},\ }\href {https://doi.org/10.22331/q-2018-10-01-97}
  {\bibfield  {journal} {\bibinfo  {journal} {Quantum}\ }\textbf {\bibinfo
  {volume} {2}},\ \bibinfo {pages} {97} (\bibinfo {year} {2018})}\BibitemShut
  {NoStop}%
\bibitem [{\citenamefont {Krinner}\ \emph {et~al.}(2018)\citenamefont
  {Krinner}, \citenamefont {Stewart}, \citenamefont {Pazmi{\~n}o},
  \citenamefont {Kwon},\ and\ \citenamefont {Schneble}}]{krinner18}%
  \BibitemOpen
  \bibfield  {author} {\bibinfo {author} {\bibfnamefont {L.}~\bibnamefont
  {Krinner}}, \bibinfo {author} {\bibfnamefont {M.}~\bibnamefont {Stewart}},
  \bibinfo {author} {\bibfnamefont {A.}~\bibnamefont {Pazmi{\~n}o}}, \bibinfo
  {author} {\bibfnamefont {J.}~\bibnamefont {Kwon}},\ and\ \bibinfo {author}
  {\bibfnamefont {D.}~\bibnamefont {Schneble}},\ }\bibfield  {title} {\bibinfo
  {title} {Spontaneous emission of matter waves from a tunable open quantum
  system},\ }\href {https://doi.org/10.1038/s41586-018-0348-z} {\bibfield
  {journal} {\bibinfo  {journal} {Nature}\ }\textbf {\bibinfo {volume} {559}},\
  \bibinfo {pages} {589} (\bibinfo {year} {2018})}\BibitemShut {NoStop}%
\bibitem [{\citenamefont {Arg{\"u}ello-Luengo}\ \emph
  {et~al.}(2019)\citenamefont {Arg{\"u}ello-Luengo}, \citenamefont
  {Gonz{\'a}lez-Tudela}, \citenamefont {Shi}, \citenamefont {Zoller},\ and\
  \citenamefont {Cirac}}]{arguello19}%
  \BibitemOpen
  \bibfield  {author} {\bibinfo {author} {\bibfnamefont {J.}~\bibnamefont
  {Arg{\"u}ello-Luengo}}, \bibinfo {author} {\bibfnamefont {A.}~\bibnamefont
  {Gonz{\'a}lez-Tudela}}, \bibinfo {author} {\bibfnamefont {T.}~\bibnamefont
  {Shi}}, \bibinfo {author} {\bibfnamefont {P.}~\bibnamefont {Zoller}},\ and\
  \bibinfo {author} {\bibfnamefont {J.~I.}\ \bibnamefont {Cirac}},\ }\bibfield
  {title} {\bibinfo {title} {Analogue quantum chemistry simulation},\ }\href
  {https://doi.org/10.1038/s41586-019-1614-4} {\bibfield  {journal} {\bibinfo
  {journal} {Nature}\ }\textbf {\bibinfo {volume} {574}},\ \bibinfo {pages}
  {215} (\bibinfo {year} {2019})}\BibitemShut {NoStop}%
\bibitem [{\citenamefont {Zohar}\ \emph {et~al.}(2015)\citenamefont {Zohar},
  \citenamefont {Cirac},\ and\ \citenamefont {Reznik}}]{zohar15}%
  \BibitemOpen
  \bibfield  {author} {\bibinfo {author} {\bibfnamefont {E.}~\bibnamefont
  {Zohar}}, \bibinfo {author} {\bibfnamefont {J.~I.}\ \bibnamefont {Cirac}},\
  and\ \bibinfo {author} {\bibfnamefont {B.}~\bibnamefont {Reznik}},\
  }\bibfield  {title} {\bibinfo {title} {Quantum simulations of lattice gauge
  theories using ultracold atoms in optical lattices},\ }\href
  {https://doi.org/10.1088/0034-4885/79/1/014401} {\bibfield  {journal}
  {\bibinfo  {journal} {Reports on Progress in Physics}\ }\textbf {\bibinfo
  {volume} {79}},\ \bibinfo {pages} {014401} (\bibinfo {year}
  {2015})}\BibitemShut {NoStop}%
\bibitem [{\citenamefont {Aidelsburger}\ \emph {et~al.}(2021)\citenamefont
  {Aidelsburger}, \citenamefont {Barbiero}, \citenamefont {Bermudez},
  \citenamefont {Chanda}, \citenamefont {Dauphin}, \citenamefont
  {Gonz{\'a}lez-Cuadra}, \citenamefont {Grzybowski}, \citenamefont {Hands},
  \citenamefont {Jendrzejewski}, \citenamefont {J{\"u}nemann}, \citenamefont
  {Juzeliunas}, \citenamefont {Kasper}, \citenamefont {Piga}, \citenamefont
  {Ran}, \citenamefont {Rizzi}, \citenamefont {Sierra}, \citenamefont
  {Tagliacozzo}, \citenamefont {Tirrito}, \citenamefont {Zache}, \citenamefont
  {Zakrzewski}, \citenamefont {Zohar},\ and\ \citenamefont
  {Lewenstein}}]{aidelsburger21}%
  \BibitemOpen
  \bibfield  {author} {\bibinfo {author} {\bibfnamefont {M.}~\bibnamefont
  {Aidelsburger}}, \bibinfo {author} {\bibfnamefont {L.}~\bibnamefont
  {Barbiero}}, \bibinfo {author} {\bibfnamefont {A.}~\bibnamefont {Bermudez}},
  \bibinfo {author} {\bibfnamefont {T.}~\bibnamefont {Chanda}}, \bibinfo
  {author} {\bibfnamefont {A.}~\bibnamefont {Dauphin}}, \bibinfo {author}
  {\bibfnamefont {D.}~\bibnamefont {Gonz{\'a}lez-Cuadra}}, \bibinfo {author}
  {\bibfnamefont {P.~R.}\ \bibnamefont {Grzybowski}}, \bibinfo {author}
  {\bibfnamefont {S.}~\bibnamefont {Hands}}, \bibinfo {author} {\bibfnamefont
  {F.}~\bibnamefont {Jendrzejewski}}, \bibinfo {author} {\bibfnamefont
  {J.}~\bibnamefont {J{\"u}nemann}}, \bibinfo {author} {\bibfnamefont
  {G.}~\bibnamefont {Juzeliunas}}, \bibinfo {author} {\bibfnamefont
  {V.}~\bibnamefont {Kasper}}, \bibinfo {author} {\bibfnamefont
  {A.}~\bibnamefont {Piga}}, \bibinfo {author} {\bibfnamefont {S.-J.}\
  \bibnamefont {Ran}}, \bibinfo {author} {\bibfnamefont {M.}~\bibnamefont
  {Rizzi}}, \bibinfo {author} {\bibfnamefont {G.}~\bibnamefont {Sierra}},
  \bibinfo {author} {\bibfnamefont {L.}~\bibnamefont {Tagliacozzo}}, \bibinfo
  {author} {\bibfnamefont {E.}~\bibnamefont {Tirrito}}, \bibinfo {author}
  {\bibfnamefont {T.~V.}\ \bibnamefont {Zache}}, \bibinfo {author}
  {\bibfnamefont {J.}~\bibnamefont {Zakrzewski}}, \bibinfo {author}
  {\bibfnamefont {E.}~\bibnamefont {Zohar}},\ and\ \bibinfo {author}
  {\bibfnamefont {M.}~\bibnamefont {Lewenstein}},\ }\bibfield  {title}
  {\bibinfo {title} {Cold atoms meet lattice gauge theory},\ }\href
  {https://doi.org/10.1098/rsta.2021.0064} {\bibfield  {journal} {\bibinfo
  {journal} {Phil. Trans. R. Soc. A}\ }\textbf {\bibinfo {volume} {380}},\
  \bibinfo {pages} {20210064} (\bibinfo {year} {2021})}\BibitemShut {NoStop}%
\bibitem [{\citenamefont {Oelker}\ \emph {et~al.}(2019)\citenamefont {Oelker},
  \citenamefont {Hutson}, \citenamefont {Kennedy}, \citenamefont {Sonderhouse},
  \citenamefont {Bothwell}, \citenamefont {Goban}, \citenamefont {Kedar},
  \citenamefont {Sanner}, \citenamefont {Robinson}, \citenamefont {Marti},
  \citenamefont {Matei}, \citenamefont {Legero}, \citenamefont {Giunta},
  \citenamefont {Holzwarth}, \citenamefont {Riehle}, \citenamefont {Sterr},\
  and\ \citenamefont {Ye}}]{oelker19}%
  \BibitemOpen
  \bibfield  {author} {\bibinfo {author} {\bibfnamefont {E.}~\bibnamefont
  {Oelker}}, \bibinfo {author} {\bibfnamefont {R.~B.}\ \bibnamefont {Hutson}},
  \bibinfo {author} {\bibfnamefont {C.~J.}\ \bibnamefont {Kennedy}}, \bibinfo
  {author} {\bibfnamefont {L.}~\bibnamefont {Sonderhouse}}, \bibinfo {author}
  {\bibfnamefont {T.}~\bibnamefont {Bothwell}}, \bibinfo {author}
  {\bibfnamefont {A.}~\bibnamefont {Goban}}, \bibinfo {author} {\bibfnamefont
  {D.}~\bibnamefont {Kedar}}, \bibinfo {author} {\bibfnamefont
  {C.}~\bibnamefont {Sanner}}, \bibinfo {author} {\bibfnamefont {J.~M.}\
  \bibnamefont {Robinson}}, \bibinfo {author} {\bibfnamefont {G.~E.}\
  \bibnamefont {Marti}}, \bibinfo {author} {\bibfnamefont {D.~G.}\ \bibnamefont
  {Matei}}, \bibinfo {author} {\bibfnamefont {T.}~\bibnamefont {Legero}},
  \bibinfo {author} {\bibfnamefont {M.}~\bibnamefont {Giunta}}, \bibinfo
  {author} {\bibfnamefont {R.}~\bibnamefont {Holzwarth}}, \bibinfo {author}
  {\bibfnamefont {F.}~\bibnamefont {Riehle}}, \bibinfo {author} {\bibfnamefont
  {U.}~\bibnamefont {Sterr}},\ and\ \bibinfo {author} {\bibfnamefont
  {J.}~\bibnamefont {Ye}},\ }\bibfield  {title} {\bibinfo {title}
  {Demonstration of $4.8\times 10^{-17}$ stability at 1 s for two independent
  optical clocks},\ }\href {https://doi.org/10.1038/s41566-019-0493-4}
  {\bibfield  {journal} {\bibinfo  {journal} {Nature Photonics}\ }\textbf
  {\bibinfo {volume} {13}},\ \bibinfo {pages} {714} (\bibinfo {year}
  {2019})}\BibitemShut {NoStop}%
\bibitem [{\citenamefont {Bothwell}\ \emph {et~al.}(2022)\citenamefont
  {Bothwell}, \citenamefont {Kennedy}, \citenamefont {Aeppli}, \citenamefont
  {Kedar}, \citenamefont {Robinson}, \citenamefont {Oelker}, \citenamefont
  {Staron},\ and\ \citenamefont {Ye}}]{bothwell22}%
  \BibitemOpen
  \bibfield  {author} {\bibinfo {author} {\bibfnamefont {T.}~\bibnamefont
  {Bothwell}}, \bibinfo {author} {\bibfnamefont {C.~J.}\ \bibnamefont
  {Kennedy}}, \bibinfo {author} {\bibfnamefont {A.}~\bibnamefont {Aeppli}},
  \bibinfo {author} {\bibfnamefont {D.}~\bibnamefont {Kedar}}, \bibinfo
  {author} {\bibfnamefont {J.~M.}\ \bibnamefont {Robinson}}, \bibinfo {author}
  {\bibfnamefont {E.}~\bibnamefont {Oelker}}, \bibinfo {author} {\bibfnamefont
  {A.}~\bibnamefont {Staron}},\ and\ \bibinfo {author} {\bibfnamefont
  {J.}~\bibnamefont {Ye}},\ }\bibfield  {title} {\bibinfo {title} {Resolving
  the gravitational redshift across a millimetre-scale atomic sample},\ }\href
  {https://doi.org/10.1038/s41586-021-04349-7} {\bibfield  {journal} {\bibinfo
  {journal} {Nature}\ }\textbf {\bibinfo {volume} {602}},\ \bibinfo {pages}
  {420} (\bibinfo {year} {2022})}\BibitemShut {NoStop}%
\bibitem [{\citenamefont {Bloch}\ \emph {et~al.}(2008)\citenamefont {Bloch},
  \citenamefont {Dalibard},\ and\ \citenamefont {Zwerger}}]{bloch08}%
  \BibitemOpen
  \bibfield  {author} {\bibinfo {author} {\bibfnamefont {I.}~\bibnamefont
  {Bloch}}, \bibinfo {author} {\bibfnamefont {J.}~\bibnamefont {Dalibard}},\
  and\ \bibinfo {author} {\bibfnamefont {W.}~\bibnamefont {Zwerger}},\
  }\bibfield  {title} {\bibinfo {title} {Many-body physics with ultracold
  gases},\ }\href {https://doi.org/10.1103/RevModPhys.80.885} {\bibfield
  {journal} {\bibinfo  {journal} {Rev. Mod. Phys.}\ }\textbf {\bibinfo {volume}
  {80}},\ \bibinfo {pages} {885} (\bibinfo {year} {2008})}\BibitemShut
  {NoStop}%
\bibitem [{\citenamefont {Campbell}\ \emph {et~al.}(2009)\citenamefont
  {Campbell}, \citenamefont {Boyd}, \citenamefont {Thomsen}, \citenamefont
  {Martin}, \citenamefont {Blatt}, \citenamefont {Swallows}, \citenamefont
  {Nicholson}, \citenamefont {Fortier}, \citenamefont {Oates}, \citenamefont
  {Diddams}, \citenamefont {Lemke}, \citenamefont {Naidon}, \citenamefont
  {Julienne}, \citenamefont {Ye},\ and\ \citenamefont {Ludlow}}]{campbell09}%
  \BibitemOpen
  \bibfield  {author} {\bibinfo {author} {\bibfnamefont {G.~K.}\ \bibnamefont
  {Campbell}}, \bibinfo {author} {\bibfnamefont {M.~M.}\ \bibnamefont {Boyd}},
  \bibinfo {author} {\bibfnamefont {J.~W.}\ \bibnamefont {Thomsen}}, \bibinfo
  {author} {\bibfnamefont {M.~J.}\ \bibnamefont {Martin}}, \bibinfo {author}
  {\bibfnamefont {S.}~\bibnamefont {Blatt}}, \bibinfo {author} {\bibfnamefont
  {M.~D.}\ \bibnamefont {Swallows}}, \bibinfo {author} {\bibfnamefont {T.~L.}\
  \bibnamefont {Nicholson}}, \bibinfo {author} {\bibfnamefont {T.}~\bibnamefont
  {Fortier}}, \bibinfo {author} {\bibfnamefont {C.~W.}\ \bibnamefont {Oates}},
  \bibinfo {author} {\bibfnamefont {S.~A.}\ \bibnamefont {Diddams}}, \bibinfo
  {author} {\bibfnamefont {N.~D.}\ \bibnamefont {Lemke}}, \bibinfo {author}
  {\bibfnamefont {P.}~\bibnamefont {Naidon}}, \bibinfo {author} {\bibfnamefont
  {P.}~\bibnamefont {Julienne}}, \bibinfo {author} {\bibfnamefont
  {J.}~\bibnamefont {Ye}},\ and\ \bibinfo {author} {\bibfnamefont {A.~D.}\
  \bibnamefont {Ludlow}},\ }\bibfield  {title} {\bibinfo {title} {Probing
  interactions between ultracold fermions},\ }\href
  {https://doi.org/10.1126/science.1169724} {\bibfield  {journal} {\bibinfo
  {journal} {Science}\ }\textbf {\bibinfo {volume} {324}},\ \bibinfo {pages}
  {360} (\bibinfo {year} {2009})}\BibitemShut {NoStop}%
\bibitem [{\citenamefont {Lemke}\ \emph {et~al.}(2011)\citenamefont {Lemke},
  \citenamefont {von Stecher}, \citenamefont {Sherman}, \citenamefont {Rey},
  \citenamefont {Oates},\ and\ \citenamefont {Ludlow}}]{lemke11}%
  \BibitemOpen
  \bibfield  {author} {\bibinfo {author} {\bibfnamefont {N.~D.}\ \bibnamefont
  {Lemke}}, \bibinfo {author} {\bibfnamefont {J.}~\bibnamefont {von Stecher}},
  \bibinfo {author} {\bibfnamefont {J.~A.}\ \bibnamefont {Sherman}}, \bibinfo
  {author} {\bibfnamefont {A.~M.}\ \bibnamefont {Rey}}, \bibinfo {author}
  {\bibfnamefont {C.~W.}\ \bibnamefont {Oates}},\ and\ \bibinfo {author}
  {\bibfnamefont {A.~D.}\ \bibnamefont {Ludlow}},\ }\bibfield  {title}
  {\bibinfo {title} {$p$-wave cold collisions in an optical lattice clock},\
  }\href {https://doi.org/10.1103/PhysRevLett.107.103902} {\bibfield  {journal}
  {\bibinfo  {journal} {Phys. Rev. Lett.}\ }\textbf {\bibinfo {volume} {107}},\
  \bibinfo {pages} {103902} (\bibinfo {year} {2011})}\BibitemShut {NoStop}%
\bibitem [{\citenamefont {Martin}\ \emph {et~al.}(2013)\citenamefont {Martin},
  \citenamefont {Bishof}, \citenamefont {Swallows}, \citenamefont {Zhang},
  \citenamefont {Benko}, \citenamefont {von Stecher}, \citenamefont {Gorshkov},
  \citenamefont {Rey},\ and\ \citenamefont {Ye}}]{martin13b}%
  \BibitemOpen
  \bibfield  {author} {\bibinfo {author} {\bibfnamefont {M.~J.}\ \bibnamefont
  {Martin}}, \bibinfo {author} {\bibfnamefont {M.}~\bibnamefont {Bishof}},
  \bibinfo {author} {\bibfnamefont {M.~D.}\ \bibnamefont {Swallows}}, \bibinfo
  {author} {\bibfnamefont {X.}~\bibnamefont {Zhang}}, \bibinfo {author}
  {\bibfnamefont {C.}~\bibnamefont {Benko}}, \bibinfo {author} {\bibfnamefont
  {J.}~\bibnamefont {von Stecher}}, \bibinfo {author} {\bibfnamefont {A.~V.}\
  \bibnamefont {Gorshkov}}, \bibinfo {author} {\bibfnamefont {A.~M.}\
  \bibnamefont {Rey}},\ and\ \bibinfo {author} {\bibfnamefont {J.}~\bibnamefont
  {Ye}},\ }\bibfield  {title} {\bibinfo {title} {A quantum many-body spin
  system in an optical lattice clock},\ }\href
  {https://doi.org/10.1126/science.1236929} {\bibfield  {journal} {\bibinfo
  {journal} {Science}\ }\textbf {\bibinfo {volume} {341}},\ \bibinfo {pages}
  {632} (\bibinfo {year} {2013})}\BibitemShut {NoStop}%
\bibitem [{\citenamefont {Swallows}\ \emph {et~al.}(2011)\citenamefont
  {Swallows}, \citenamefont {Bishof}, \citenamefont {Lin}, \citenamefont
  {Blatt}, \citenamefont {Martin}, \citenamefont {Rey},\ and\ \citenamefont
  {Ye}}]{swallows11}%
  \BibitemOpen
  \bibfield  {author} {\bibinfo {author} {\bibfnamefont {M.~D.}\ \bibnamefont
  {Swallows}}, \bibinfo {author} {\bibfnamefont {M.}~\bibnamefont {Bishof}},
  \bibinfo {author} {\bibfnamefont {Y.}~\bibnamefont {Lin}}, \bibinfo {author}
  {\bibfnamefont {S.}~\bibnamefont {Blatt}}, \bibinfo {author} {\bibfnamefont
  {M.~J.}\ \bibnamefont {Martin}}, \bibinfo {author} {\bibfnamefont {A.~M.}\
  \bibnamefont {Rey}},\ and\ \bibinfo {author} {\bibfnamefont {J.}~\bibnamefont
  {Ye}},\ }\bibfield  {title} {\bibinfo {title} {Suppression of collisional
  shifts in a strongly interacting lattice clock},\ }\href
  {https://doi.org/10.1126/science.1196442} {\bibfield  {journal} {\bibinfo
  {journal} {Science}\ }\textbf {\bibinfo {volume} {331}},\ \bibinfo {pages}
  {1043–1046} (\bibinfo {year} {2011})}\BibitemShut {NoStop}%
\bibitem [{\citenamefont {Akatsuka}\ \emph {et~al.}(2010)\citenamefont
  {Akatsuka}, \citenamefont {Takamoto},\ and\ \citenamefont
  {Katori}}]{akatsuka10}%
  \BibitemOpen
  \bibfield  {author} {\bibinfo {author} {\bibfnamefont {T.}~\bibnamefont
  {Akatsuka}}, \bibinfo {author} {\bibfnamefont {M.}~\bibnamefont {Takamoto}},\
  and\ \bibinfo {author} {\bibfnamefont {H.}~\bibnamefont {Katori}},\
  }\bibfield  {title} {\bibinfo {title} {Three-dimensional optical lattice
  clock with bosonic {$^{88}\mathrm{Sr}$} atoms},\ }\href
  {https://doi.org/10.1103/PhysRevA.81.023402} {\bibfield  {journal} {\bibinfo
  {journal} {Phys. Rev. A}\ }\textbf {\bibinfo {volume} {81}},\ \bibinfo
  {pages} {023402} (\bibinfo {year} {2010})}\BibitemShut {NoStop}%
\bibitem [{\citenamefont {Campbell}\ \emph {et~al.}(2017)\citenamefont
  {Campbell}, \citenamefont {Hutson}, \citenamefont {Marti}, \citenamefont
  {Goban}, \citenamefont {Oppong}, \citenamefont {McNally}, \citenamefont
  {Sonderhouse}, \citenamefont {Robinson}, \citenamefont {Zhang}, \citenamefont
  {Bloom},\ and\ \citenamefont {Ye}}]{campbell17}%
  \BibitemOpen
  \bibfield  {author} {\bibinfo {author} {\bibfnamefont {S.~L.}\ \bibnamefont
  {Campbell}}, \bibinfo {author} {\bibfnamefont {R.~B.}\ \bibnamefont
  {Hutson}}, \bibinfo {author} {\bibfnamefont {G.~E.}\ \bibnamefont {Marti}},
  \bibinfo {author} {\bibfnamefont {A.}~\bibnamefont {Goban}}, \bibinfo
  {author} {\bibfnamefont {N.~D.}\ \bibnamefont {Oppong}}, \bibinfo {author}
  {\bibfnamefont {R.~L.}\ \bibnamefont {McNally}}, \bibinfo {author}
  {\bibfnamefont {L.}~\bibnamefont {Sonderhouse}}, \bibinfo {author}
  {\bibfnamefont {J.~M.}\ \bibnamefont {Robinson}}, \bibinfo {author}
  {\bibfnamefont {W.}~\bibnamefont {Zhang}}, \bibinfo {author} {\bibfnamefont
  {B.~J.}\ \bibnamefont {Bloom}},\ and\ \bibinfo {author} {\bibfnamefont
  {J.}~\bibnamefont {Ye}},\ }\bibfield  {title} {\bibinfo {title} {A
  {Fermi}-degenerate three-dimensional optical lattice clock},\ }\href
  {https://doi.org/10.1126/science.aam5538} {\bibfield  {journal} {\bibinfo
  {journal} {Science}\ }\textbf {\bibinfo {volume} {358}},\ \bibinfo {pages}
  {90} (\bibinfo {year} {2017})}\BibitemShut {NoStop}%
\bibitem [{\citenamefont {Nicholson}\ \emph {et~al.}(2012)\citenamefont
  {Nicholson}, \citenamefont {Martin}, \citenamefont {Williams}, \citenamefont
  {Bloom}, \citenamefont {Bishof}, \citenamefont {Swallows}, \citenamefont
  {Campbell},\ and\ \citenamefont {Ye}}]{nicholson12}%
  \BibitemOpen
  \bibfield  {author} {\bibinfo {author} {\bibfnamefont {T.~L.}\ \bibnamefont
  {Nicholson}}, \bibinfo {author} {\bibfnamefont {M.~J.}\ \bibnamefont
  {Martin}}, \bibinfo {author} {\bibfnamefont {J.~R.}\ \bibnamefont
  {Williams}}, \bibinfo {author} {\bibfnamefont {B.~J.}\ \bibnamefont {Bloom}},
  \bibinfo {author} {\bibfnamefont {M.}~\bibnamefont {Bishof}}, \bibinfo
  {author} {\bibfnamefont {M.~D.}\ \bibnamefont {Swallows}}, \bibinfo {author}
  {\bibfnamefont {S.~L.}\ \bibnamefont {Campbell}},\ and\ \bibinfo {author}
  {\bibfnamefont {J.}~\bibnamefont {Ye}},\ }\bibfield  {title} {\bibinfo
  {title} {Comparison of two independent sr optical clocks with
  $1\mathbf{\times}{10}^{-17}$ stability at ${10}^{3}\text{ }$s},\ }\href
  {https://doi.org/10.1103/PhysRevLett.109.230801} {\bibfield  {journal}
  {\bibinfo  {journal} {Phys. Rev. Lett.}\ }\textbf {\bibinfo {volume} {109}},\
  \bibinfo {pages} {230801} (\bibinfo {year} {2012})}\BibitemShut {NoStop}%
\bibitem [{\citenamefont {Hart}\ \emph {et~al.}(2015)\citenamefont {Hart},
  \citenamefont {Duarte}, \citenamefont {Yang}, \citenamefont {Liu},
  \citenamefont {Paiva}, \citenamefont {Khatami}, \citenamefont {Scalettar},
  \citenamefont {Trivedi}, \citenamefont {Huse},\ and\ \citenamefont
  {Hulet}}]{hart15}%
  \BibitemOpen
  \bibfield  {author} {\bibinfo {author} {\bibfnamefont {R.~A.}\ \bibnamefont
  {Hart}}, \bibinfo {author} {\bibfnamefont {P.~M.}\ \bibnamefont {Duarte}},
  \bibinfo {author} {\bibfnamefont {T.-L.}\ \bibnamefont {Yang}}, \bibinfo
  {author} {\bibfnamefont {X.}~\bibnamefont {Liu}}, \bibinfo {author}
  {\bibfnamefont {T.}~\bibnamefont {Paiva}}, \bibinfo {author} {\bibfnamefont
  {E.}~\bibnamefont {Khatami}}, \bibinfo {author} {\bibfnamefont {R.~T.}\
  \bibnamefont {Scalettar}}, \bibinfo {author} {\bibfnamefont {N.}~\bibnamefont
  {Trivedi}}, \bibinfo {author} {\bibfnamefont {D.~A.}\ \bibnamefont {Huse}},\
  and\ \bibinfo {author} {\bibfnamefont {R.~G.}\ \bibnamefont {Hulet}},\
  }\bibfield  {title} {\bibinfo {title} {Observation of antiferromagnetic
  correlations in the hubbard model with ultracold atoms},\ }\href
  {https://doi.org/10.1038/nature14223} {\bibfield  {journal} {\bibinfo
  {journal} {Nature}\ }\textbf {\bibinfo {volume} {519}},\ \bibinfo {pages}
  {211} (\bibinfo {year} {2015})}\BibitemShut {NoStop}%
\bibitem [{\citenamefont {Mazurenko}\ \emph {et~al.}(2017)\citenamefont
  {Mazurenko}, \citenamefont {Chiu}, \citenamefont {Ji}, \citenamefont
  {Parsons}, \citenamefont {Kan{\'a}sz-Nagy}, \citenamefont {Schmidt},
  \citenamefont {Grusdt}, \citenamefont {Demler}, \citenamefont {Greif},\ and\
  \citenamefont {Greiner}}]{mazurenko17}%
  \BibitemOpen
  \bibfield  {author} {\bibinfo {author} {\bibfnamefont {A.}~\bibnamefont
  {Mazurenko}}, \bibinfo {author} {\bibfnamefont {C.~S.}\ \bibnamefont {Chiu}},
  \bibinfo {author} {\bibfnamefont {G.}~\bibnamefont {Ji}}, \bibinfo {author}
  {\bibfnamefont {M.~F.}\ \bibnamefont {Parsons}}, \bibinfo {author}
  {\bibfnamefont {M.}~\bibnamefont {Kan{\'a}sz-Nagy}}, \bibinfo {author}
  {\bibfnamefont {R.}~\bibnamefont {Schmidt}}, \bibinfo {author} {\bibfnamefont
  {F.}~\bibnamefont {Grusdt}}, \bibinfo {author} {\bibfnamefont
  {E.}~\bibnamefont {Demler}}, \bibinfo {author} {\bibfnamefont
  {D.}~\bibnamefont {Greif}},\ and\ \bibinfo {author} {\bibfnamefont
  {M.}~\bibnamefont {Greiner}},\ }\bibfield  {title} {\bibinfo {title} {A
  cold-atom {Fermi--Hubbard} antiferromagnet},\ }\href
  {https://doi.org/10.1038/nature22362} {\bibfield  {journal} {\bibinfo
  {journal} {Nature}\ }\textbf {\bibinfo {volume} {545}},\ \bibinfo {pages}
  {462} (\bibinfo {year} {2017})}\BibitemShut {NoStop}%
\bibitem [{\citenamefont {Chiu}\ \emph {et~al.}(2018)\citenamefont {Chiu},
  \citenamefont {Ji}, \citenamefont {Mazurenko}, \citenamefont {Greif},\ and\
  \citenamefont {Greiner}}]{chiu18}%
  \BibitemOpen
  \bibfield  {author} {\bibinfo {author} {\bibfnamefont {C.~S.}\ \bibnamefont
  {Chiu}}, \bibinfo {author} {\bibfnamefont {G.}~\bibnamefont {Ji}}, \bibinfo
  {author} {\bibfnamefont {A.}~\bibnamefont {Mazurenko}}, \bibinfo {author}
  {\bibfnamefont {D.}~\bibnamefont {Greif}},\ and\ \bibinfo {author}
  {\bibfnamefont {M.}~\bibnamefont {Greiner}},\ }\bibfield  {title} {\bibinfo
  {title} {Quantum state engineering of a {Hubbard} system with ultracold
  fermions},\ }\href {https://doi.org/10.1103/PhysRevLett.120.243201}
  {\bibfield  {journal} {\bibinfo  {journal} {Phys. Rev. Lett.}\ }\textbf
  {\bibinfo {volume} {120}},\ \bibinfo {pages} {243201} (\bibinfo {year}
  {2018})}\BibitemShut {NoStop}%
\bibitem [{\citenamefont {Gall}\ \emph {et~al.}(2021)\citenamefont {Gall},
  \citenamefont {Wurz}, \citenamefont {Samland}, \citenamefont {Chan},\ and\
  \citenamefont {K{\"o}hl}}]{gall21}%
  \BibitemOpen
  \bibfield  {author} {\bibinfo {author} {\bibfnamefont {M.}~\bibnamefont
  {Gall}}, \bibinfo {author} {\bibfnamefont {N.}~\bibnamefont {Wurz}}, \bibinfo
  {author} {\bibfnamefont {J.}~\bibnamefont {Samland}}, \bibinfo {author}
  {\bibfnamefont {C.~F.}\ \bibnamefont {Chan}},\ and\ \bibinfo {author}
  {\bibfnamefont {M.}~\bibnamefont {K{\"o}hl}},\ }\bibfield  {title} {\bibinfo
  {title} {Competing magnetic orders in a bilayer {Hubbard} model with
  ultracold atoms},\ }\href {https://doi.org/10.1038/s41586-020-03058-x}
  {\bibfield  {journal} {\bibinfo  {journal} {Nature}\ }\textbf {\bibinfo
  {volume} {589}},\ \bibinfo {pages} {40} (\bibinfo {year} {2021})}\BibitemShut
  {NoStop}%
\bibitem [{\citenamefont {Wei}\ \emph {et~al.}(2021)\citenamefont {Wei},
  \citenamefont {Rubio-Abadal}, \citenamefont {Ye}, \citenamefont {Machado},
  \citenamefont {Kemp}, \citenamefont {Srakaew}, \citenamefont {Hollerith},
  \citenamefont {Rui}, \citenamefont {Gopalakrishnan}, \citenamefont {Y.~Yao},
  \citenamefont {Bloch},\ and\ \citenamefont {Zeiher}}]{wei21}%
  \BibitemOpen
  \bibfield  {author} {\bibinfo {author} {\bibfnamefont {D.}~\bibnamefont
  {Wei}}, \bibinfo {author} {\bibfnamefont {A.}~\bibnamefont {Rubio-Abadal}},
  \bibinfo {author} {\bibfnamefont {B.}~\bibnamefont {Ye}}, \bibinfo {author}
  {\bibfnamefont {F.}~\bibnamefont {Machado}}, \bibinfo {author} {\bibfnamefont
  {J.}~\bibnamefont {Kemp}}, \bibinfo {author} {\bibfnamefont {K.}~\bibnamefont
  {Srakaew}}, \bibinfo {author} {\bibfnamefont {S.}~\bibnamefont {Hollerith}},
  \bibinfo {author} {\bibfnamefont {J.}~\bibnamefont {Rui}}, \bibinfo {author}
  {\bibfnamefont {S.}~\bibnamefont {Gopalakrishnan}}, \bibinfo {author}
  {\bibfnamefont {N.}~\bibnamefont {Y.~Yao}}, \bibinfo {author} {\bibfnamefont
  {I.}~\bibnamefont {Bloch}},\ and\ \bibinfo {author} {\bibfnamefont
  {J.}~\bibnamefont {Zeiher}},\ }\bibfield  {title} {\bibinfo {title} {Quantum
  gas microscopy of {K}ardar-{P}arisi-{Z}hang superdiffusion},\ }\href@noop {}
  {\bibfield  {journal} {\bibinfo  {journal} {arXiv:2107.00038}\ } (\bibinfo
  {year} {2021})}\BibitemShut {NoStop}%
\bibitem [{\citenamefont {Brusch}\ \emph {et~al.}(2006)\citenamefont {Brusch},
  \citenamefont {Le~Targat}, \citenamefont {Baillard}, \citenamefont
  {Fouch\'e},\ and\ \citenamefont {Lemonde}}]{brusch06}%
  \BibitemOpen
  \bibfield  {author} {\bibinfo {author} {\bibfnamefont {A.}~\bibnamefont
  {Brusch}}, \bibinfo {author} {\bibfnamefont {R.}~\bibnamefont {Le~Targat}},
  \bibinfo {author} {\bibfnamefont {X.}~\bibnamefont {Baillard}}, \bibinfo
  {author} {\bibfnamefont {M.}~\bibnamefont {Fouch\'e}},\ and\ \bibinfo
  {author} {\bibfnamefont {P.}~\bibnamefont {Lemonde}},\ }\bibfield  {title}
  {\bibinfo {title} {Hyperpolarizability effects in a sr optical lattice
  clock},\ }\href {https://doi.org/10.1103/PhysRevLett.96.103003} {\bibfield
  {journal} {\bibinfo  {journal} {Phys. Rev. Lett.}\ }\textbf {\bibinfo
  {volume} {96}},\ \bibinfo {pages} {103003} (\bibinfo {year}
  {2006})}\BibitemShut {NoStop}%
\bibitem [{\citenamefont {Yi}\ \emph {et~al.}(2011)\citenamefont {Yi},
  \citenamefont {Mejri}, \citenamefont {McFerran}, \citenamefont {Coq},\ and\
  \citenamefont {Bize}}]{yi11}%
  \BibitemOpen
  \bibfield  {author} {\bibinfo {author} {\bibfnamefont {L.}~\bibnamefont
  {Yi}}, \bibinfo {author} {\bibfnamefont {S.}~\bibnamefont {Mejri}}, \bibinfo
  {author} {\bibfnamefont {J.~J.}\ \bibnamefont {McFerran}}, \bibinfo {author}
  {\bibfnamefont {Y.~L.}\ \bibnamefont {Coq}},\ and\ \bibinfo {author}
  {\bibfnamefont {S.}~\bibnamefont {Bize}},\ }\bibfield  {title} {\bibinfo
  {title} {Optical lattice trapping of $^{199}\mathrm{Hg}$ and determination of
  the magic wavelength for the ultraviolet
  $^{1}{S}_{0}\ensuremath{\leftrightarrow}^{3}{P}_{0}$ clock transition},\
  }\href {https://doi.org/10.1103/PhysRevLett.106.073005} {\bibfield  {journal}
  {\bibinfo  {journal} {Phys. Rev. Lett.}\ }\textbf {\bibinfo {volume} {106}},\
  \bibinfo {pages} {073005} (\bibinfo {year} {2011})}\BibitemShut {NoStop}%
\bibitem [{\citenamefont {Kulosa}\ \emph {et~al.}(2015)\citenamefont {Kulosa},
  \citenamefont {Fim}, \citenamefont {Zipfel}, \citenamefont {R\"uhmann},
  \citenamefont {Sauer}, \citenamefont {Jha}, \citenamefont {Gibble},
  \citenamefont {Ertmer}, \citenamefont {Rasel}, \citenamefont {Safronova},
  \citenamefont {Safronova},\ and\ \citenamefont {Porsev}}]{kulosa15}%
  \BibitemOpen
  \bibfield  {author} {\bibinfo {author} {\bibfnamefont {A.~P.}\ \bibnamefont
  {Kulosa}}, \bibinfo {author} {\bibfnamefont {D.}~\bibnamefont {Fim}},
  \bibinfo {author} {\bibfnamefont {K.~H.}\ \bibnamefont {Zipfel}}, \bibinfo
  {author} {\bibfnamefont {S.}~\bibnamefont {R\"uhmann}}, \bibinfo {author}
  {\bibfnamefont {S.}~\bibnamefont {Sauer}}, \bibinfo {author} {\bibfnamefont
  {N.}~\bibnamefont {Jha}}, \bibinfo {author} {\bibfnamefont {K.}~\bibnamefont
  {Gibble}}, \bibinfo {author} {\bibfnamefont {W.}~\bibnamefont {Ertmer}},
  \bibinfo {author} {\bibfnamefont {E.~M.}\ \bibnamefont {Rasel}}, \bibinfo
  {author} {\bibfnamefont {M.~S.}\ \bibnamefont {Safronova}}, \bibinfo {author}
  {\bibfnamefont {U.~I.}\ \bibnamefont {Safronova}},\ and\ \bibinfo {author}
  {\bibfnamefont {S.~G.}\ \bibnamefont {Porsev}},\ }\bibfield  {title}
  {\bibinfo {title} {Towards a mg lattice clock: Observation of the
  $^{1}{S}_{0}\text{\ensuremath{-}}^{3}{P}_{0}$ transition and determination of
  the magic wavelength},\ }\href
  {https://doi.org/10.1103/PhysRevLett.115.240801} {\bibfield  {journal}
  {\bibinfo  {journal} {Phys. Rev. Lett.}\ }\textbf {\bibinfo {volume} {115}},\
  \bibinfo {pages} {240801} (\bibinfo {year} {2015})}\BibitemShut {NoStop}%
\bibitem [{\citenamefont {Brown}\ \emph {et~al.}(2017)\citenamefont {Brown},
  \citenamefont {Phillips}, \citenamefont {Beloy}, \citenamefont {McGrew},
  \citenamefont {Schioppo}, \citenamefont {Fasano}, \citenamefont {Milani},
  \citenamefont {Zhang}, \citenamefont {Hinkley}, \citenamefont {Leopardi},
  \citenamefont {Yoon}, \citenamefont {Nicolodi}, \citenamefont {Fortier},\
  and\ \citenamefont {Ludlow}}]{brown17}%
  \BibitemOpen
  \bibfield  {author} {\bibinfo {author} {\bibfnamefont {R.~C.}\ \bibnamefont
  {Brown}}, \bibinfo {author} {\bibfnamefont {N.~B.}\ \bibnamefont {Phillips}},
  \bibinfo {author} {\bibfnamefont {K.}~\bibnamefont {Beloy}}, \bibinfo
  {author} {\bibfnamefont {W.~F.}\ \bibnamefont {McGrew}}, \bibinfo {author}
  {\bibfnamefont {M.}~\bibnamefont {Schioppo}}, \bibinfo {author}
  {\bibfnamefont {R.~J.}\ \bibnamefont {Fasano}}, \bibinfo {author}
  {\bibfnamefont {G.}~\bibnamefont {Milani}}, \bibinfo {author} {\bibfnamefont
  {X.}~\bibnamefont {Zhang}}, \bibinfo {author} {\bibfnamefont
  {N.}~\bibnamefont {Hinkley}}, \bibinfo {author} {\bibfnamefont
  {H.}~\bibnamefont {Leopardi}}, \bibinfo {author} {\bibfnamefont {T.~H.}\
  \bibnamefont {Yoon}}, \bibinfo {author} {\bibfnamefont {D.}~\bibnamefont
  {Nicolodi}}, \bibinfo {author} {\bibfnamefont {T.~M.}\ \bibnamefont
  {Fortier}},\ and\ \bibinfo {author} {\bibfnamefont {A.~D.}\ \bibnamefont
  {Ludlow}},\ }\bibfield  {title} {\bibinfo {title} {Hyperpolarizability and
  operational magic wavelength in an optical lattice clock},\ }\href
  {https://doi.org/10.1103/PhysRevLett.119.253001} {\bibfield  {journal}
  {\bibinfo  {journal} {Phys. Rev. Lett.}\ }\textbf {\bibinfo {volume} {119}},\
  \bibinfo {pages} {253001} (\bibinfo {year} {2017})}\BibitemShut {NoStop}%
\bibitem [{\citenamefont {Yamaguchi}\ \emph {et~al.}(2019)\citenamefont
  {Yamaguchi}, \citenamefont {Safronova}, \citenamefont {Gibble},\ and\
  \citenamefont {Katori}}]{yamaguchi19}%
  \BibitemOpen
  \bibfield  {author} {\bibinfo {author} {\bibfnamefont {A.}~\bibnamefont
  {Yamaguchi}}, \bibinfo {author} {\bibfnamefont {M.~S.}\ \bibnamefont
  {Safronova}}, \bibinfo {author} {\bibfnamefont {K.}~\bibnamefont {Gibble}},\
  and\ \bibinfo {author} {\bibfnamefont {H.}~\bibnamefont {Katori}},\
  }\bibfield  {title} {\bibinfo {title} {Narrow-line cooling and determination
  of the magic wavelength of {C}d},\ }\href
  {https://doi.org/10.1103/PhysRevLett.123.113201} {\bibfield  {journal}
  {\bibinfo  {journal} {Phys. Rev. Lett.}\ }\textbf {\bibinfo {volume} {123}},\
  \bibinfo {pages} {113201} (\bibinfo {year} {2019})}\BibitemShut {NoStop}%
\bibitem [{\citenamefont {Kawasaki}\ \emph {et~al.}(2020)\citenamefont
  {Kawasaki}, \citenamefont {Braverman}, \citenamefont {Pedrozo-Pe\~nafiel},
  \citenamefont {Shu}, \citenamefont {Colombo}, \citenamefont {Li},\ and\
  \citenamefont {Vuleti\ifmmode~\acute{c}\else \'{c}\fi{}}}]{kawasaki20}%
  \BibitemOpen
  \bibfield  {author} {\bibinfo {author} {\bibfnamefont {A.}~\bibnamefont
  {Kawasaki}}, \bibinfo {author} {\bibfnamefont {B.}~\bibnamefont {Braverman}},
  \bibinfo {author} {\bibfnamefont {E.}~\bibnamefont {Pedrozo-Pe\~nafiel}},
  \bibinfo {author} {\bibfnamefont {C.}~\bibnamefont {Shu}}, \bibinfo {author}
  {\bibfnamefont {S.}~\bibnamefont {Colombo}}, \bibinfo {author} {\bibfnamefont
  {Z.}~\bibnamefont {Li}},\ and\ \bibinfo {author} {\bibfnamefont
  {V.}~\bibnamefont {Vuleti\ifmmode~\acute{c}\else \'{c}\fi{}}},\ }\bibfield
  {title} {\bibinfo {title} {Trapping $^{171}\mathrm{Yb}$ atoms into a
  one-dimensional optical lattice with a small waist},\ }\href
  {https://doi.org/10.1103/PhysRevA.102.013114} {\bibfield  {journal} {\bibinfo
   {journal} {Phys. Rev. A}\ }\textbf {\bibinfo {volume} {102}},\ \bibinfo
  {pages} {013114} (\bibinfo {year} {2020})}\BibitemShut {NoStop}%
\bibitem [{\citenamefont {Bowden}\ \emph {et~al.}(2019)\citenamefont {Bowden},
  \citenamefont {Hobson}, \citenamefont {Hill}, \citenamefont {Vianello},
  \citenamefont {Schioppo}, \citenamefont {Silva}, \citenamefont {Margolis},
  \citenamefont {Baird},\ and\ \citenamefont {Gill}}]{bowden19}%
  \BibitemOpen
  \bibfield  {author} {\bibinfo {author} {\bibfnamefont {W.}~\bibnamefont
  {Bowden}}, \bibinfo {author} {\bibfnamefont {R.}~\bibnamefont {Hobson}},
  \bibinfo {author} {\bibfnamefont {I.~R.}\ \bibnamefont {Hill}}, \bibinfo
  {author} {\bibfnamefont {A.}~\bibnamefont {Vianello}}, \bibinfo {author}
  {\bibfnamefont {M.}~\bibnamefont {Schioppo}}, \bibinfo {author}
  {\bibfnamefont {A.}~\bibnamefont {Silva}}, \bibinfo {author} {\bibfnamefont
  {H.~S.}\ \bibnamefont {Margolis}}, \bibinfo {author} {\bibfnamefont
  {P.~E.~G.}\ \bibnamefont {Baird}},\ and\ \bibinfo {author} {\bibfnamefont
  {P.}~\bibnamefont {Gill}},\ }\bibfield  {title} {\bibinfo {title} {A pyramid
  {MOT} with integrated optical cavities as a cold atom platform for an optical
  lattice clock},\ }\href {https://doi.org/10.1038/s41598-019-48168-3}
  {\bibfield  {journal} {\bibinfo  {journal} {Scientific Reports}\ }\textbf
  {\bibinfo {volume} {9}},\ \bibinfo {pages} {11704} (\bibinfo {year}
  {2019})}\BibitemShut {NoStop}%
\bibitem [{\citenamefont {Cai}\ \emph {et~al.}(2020)\citenamefont {Cai},
  \citenamefont {Allman}, \citenamefont {Evans}, \citenamefont {Sabharwal},\
  and\ \citenamefont {Wright}}]{cai20}%
  \BibitemOpen
  \bibfield  {author} {\bibinfo {author} {\bibfnamefont {Y.}~\bibnamefont
  {Cai}}, \bibinfo {author} {\bibfnamefont {D.~G.}\ \bibnamefont {Allman}},
  \bibinfo {author} {\bibfnamefont {J.}~\bibnamefont {Evans}}, \bibinfo
  {author} {\bibfnamefont {P.}~\bibnamefont {Sabharwal}},\ and\ \bibinfo
  {author} {\bibfnamefont {K.~C.}\ \bibnamefont {Wright}},\ }\bibfield  {title}
  {\bibinfo {title} {Monolithic bowtie cavity traps for ultracold gases},\
  }\href {https://doi.org/10.1364/JOSAB.401262} {\bibfield  {journal} {\bibinfo
   {journal} {J. Opt. Soc. Am. B}\ }\textbf {\bibinfo {volume} {37}},\ \bibinfo
  {pages} {3596} (\bibinfo {year} {2020})}\BibitemShut {NoStop}%
\bibitem [{\citenamefont {Heinz}(2020)}]{heinz20b}%
  \BibitemOpen
  \bibfield  {author} {\bibinfo {author} {\bibfnamefont {A.}~\bibnamefont
  {Heinz}},\ }\emph {\bibinfo {title} {Ultracold strontium in state-dependent
  optical lattices}},\ \href {10.5282/edoc.26329} {Ph.D. thesis},\ \bibinfo
  {school} {Ludwig-Maximilians-Universit{\"a}t M{\"u}nchen, Department of
  Physics} (\bibinfo {year} {2020})\BibitemShut {NoStop}%
\bibitem [{\citenamefont {Heinz}\ \emph {et~al.}(2021)\citenamefont {Heinz},
  \citenamefont {Trautmann}, \citenamefont {\v{S}anti\'{c}}, \citenamefont
  {Park}, \citenamefont {Bloch},\ and\ \citenamefont {Blatt}}]{heinz21}%
  \BibitemOpen
  \bibfield  {author} {\bibinfo {author} {\bibfnamefont {A.}~\bibnamefont
  {Heinz}}, \bibinfo {author} {\bibfnamefont {J.}~\bibnamefont {Trautmann}},
  \bibinfo {author} {\bibfnamefont {N.}~\bibnamefont {\v{S}anti\'{c}}},
  \bibinfo {author} {\bibfnamefont {A.~J.}\ \bibnamefont {Park}}, \bibinfo
  {author} {\bibfnamefont {I.}~\bibnamefont {Bloch}},\ and\ \bibinfo {author}
  {\bibfnamefont {S.}~\bibnamefont {Blatt}},\ }\bibfield  {title} {\bibinfo
  {title} {Crossed optical cavities with large mode diameters},\ }\href
  {https://doi.org/10.1364/OL.414076} {\bibfield  {journal} {\bibinfo
  {journal} {Opt. Lett.}\ }\textbf {\bibinfo {volume} {46}},\ \bibinfo {pages}
  {250} (\bibinfo {year} {2021})}\BibitemShut {NoStop}%
\bibitem [{\citenamefont {Gonz\'alez-Tudela}\ and\ \citenamefont
  {Cirac}(2017{\natexlab{a}})}]{tudela17a}%
  \BibitemOpen
  \bibfield  {author} {\bibinfo {author} {\bibfnamefont {A.}~\bibnamefont
  {Gonz\'alez-Tudela}}\ and\ \bibinfo {author} {\bibfnamefont {J.~I.}\
  \bibnamefont {Cirac}},\ }\bibfield  {title} {\bibinfo {title} {Quantum
  emitters in two-dimensional structured reservoirs in the nonperturbative
  regime},\ }\href {https://doi.org/10.1103/PhysRevLett.119.143602} {\bibfield
  {journal} {\bibinfo  {journal} {Phys. Rev. Lett.}\ }\textbf {\bibinfo
  {volume} {119}},\ \bibinfo {pages} {143602} (\bibinfo {year}
  {2017}{\natexlab{a}})}\BibitemShut {NoStop}%
\bibitem [{\citenamefont {Heinz}\ \emph {et~al.}(2020)\citenamefont {Heinz},
  \citenamefont {Park}, \citenamefont {\v{S}anti{\'c}}, \citenamefont
  {Trautmann}, \citenamefont {Porsev}, \citenamefont {Safronova}, \citenamefont
  {Bloch},\ and\ \citenamefont {Blatt}}]{heinz20}%
  \BibitemOpen
  \bibfield  {author} {\bibinfo {author} {\bibfnamefont {A.}~\bibnamefont
  {Heinz}}, \bibinfo {author} {\bibfnamefont {A.~J.}\ \bibnamefont {Park}},
  \bibinfo {author} {\bibfnamefont {N.}~\bibnamefont {\v{S}anti{\'c}}},
  \bibinfo {author} {\bibfnamefont {J.}~\bibnamefont {Trautmann}}, \bibinfo
  {author} {\bibfnamefont {S.~G.}\ \bibnamefont {Porsev}}, \bibinfo {author}
  {\bibfnamefont {M.~S.}\ \bibnamefont {Safronova}}, \bibinfo {author}
  {\bibfnamefont {I.}~\bibnamefont {Bloch}},\ and\ \bibinfo {author}
  {\bibfnamefont {S.}~\bibnamefont {Blatt}},\ }\bibfield  {title} {\bibinfo
  {title} {State-dependent optical lattices for the strontium optical qubit},\
  }\href@noop {} {\bibfield  {journal} {\bibinfo  {journal} {Phys. Rev. Lett.}\
  }\textbf {\bibinfo {volume} {124}},\ \bibinfo {pages} {203201} (\bibinfo
  {year} {2020})}\BibitemShut {NoStop}%
\bibitem [{\citenamefont {Snigirev}\ \emph {et~al.}(2019)\citenamefont
  {Snigirev}, \citenamefont {Park}, \citenamefont {Heinz}, \citenamefont
  {Bloch},\ and\ \citenamefont {Blatt}}]{snigirev19}%
  \BibitemOpen
  \bibfield  {author} {\bibinfo {author} {\bibfnamefont {S.}~\bibnamefont
  {Snigirev}}, \bibinfo {author} {\bibfnamefont {A.~J.}\ \bibnamefont {Park}},
  \bibinfo {author} {\bibfnamefont {A.}~\bibnamefont {Heinz}}, \bibinfo
  {author} {\bibfnamefont {I.}~\bibnamefont {Bloch}},\ and\ \bibinfo {author}
  {\bibfnamefont {S.}~\bibnamefont {Blatt}},\ }\bibfield  {title} {\bibinfo
  {title} {Fast and dense magneto-optical traps for strontium},\ }\href
  {https://doi.org/10.1103/PhysRevA.99.063421} {\bibfield  {journal} {\bibinfo
  {journal} {Phys. Rev. A}\ }\textbf {\bibinfo {volume} {99}},\ \bibinfo
  {pages} {063421} (\bibinfo {year} {2019})}\BibitemShut {NoStop}%
\bibitem [{\citenamefont {L{\'{e}}onard}\ \emph {et~al.}(2014)\citenamefont
  {L{\'{e}}onard}, \citenamefont {Lee}, \citenamefont {Morales}, \citenamefont
  {Karg}, \citenamefont {Esslinger},\ and\ \citenamefont {Donner}}]{leonard14}%
  \BibitemOpen
  \bibfield  {author} {\bibinfo {author} {\bibfnamefont {J.}~\bibnamefont
  {L{\'{e}}onard}}, \bibinfo {author} {\bibfnamefont {M.}~\bibnamefont {Lee}},
  \bibinfo {author} {\bibfnamefont {A.}~\bibnamefont {Morales}}, \bibinfo
  {author} {\bibfnamefont {T.~M.}\ \bibnamefont {Karg}}, \bibinfo {author}
  {\bibfnamefont {T.}~\bibnamefont {Esslinger}},\ and\ \bibinfo {author}
  {\bibfnamefont {T.}~\bibnamefont {Donner}},\ }\bibfield  {title} {\bibinfo
  {title} {Optical transport and manipulation of an ultracold atomic cloud
  using focus-tunable lenses},\ }\href
  {https://doi.org/10.1088/1367-2630/16/9/093028} {\bibfield  {journal}
  {\bibinfo  {journal} {N. J. Phys.}\ }\textbf {\bibinfo {volume} {16}},\
  \bibinfo {pages} {093028} (\bibinfo {year} {2014})}\BibitemShut {NoStop}%
\bibitem [{\citenamefont {Leibfried}\ \emph {et~al.}(2003)\citenamefont
  {Leibfried}, \citenamefont {Blatt}, \citenamefont {Monroe},\ and\
  \citenamefont {Wineland}}]{leibfried03}%
  \BibitemOpen
  \bibfield  {author} {\bibinfo {author} {\bibfnamefont {D.}~\bibnamefont
  {Leibfried}}, \bibinfo {author} {\bibfnamefont {R.}~\bibnamefont {Blatt}},
  \bibinfo {author} {\bibfnamefont {C.}~\bibnamefont {Monroe}},\ and\ \bibinfo
  {author} {\bibfnamefont {D.}~\bibnamefont {Wineland}},\ }\bibfield  {title}
  {\bibinfo {title} {Quantum dynamics of single trapped ions},\ }\href
  {https://doi.org/10.1103/RevModPhys.75.281} {\bibfield  {journal} {\bibinfo
  {journal} {Rev. Mod. Phys.}\ }\textbf {\bibinfo {volume} {75}},\ \bibinfo
  {pages} {281} (\bibinfo {year} {2003})}\BibitemShut {NoStop}%
\bibitem [{\citenamefont {Ido}\ and\ \citenamefont {Katori}(2003)}]{ido03}%
  \BibitemOpen
  \bibfield  {author} {\bibinfo {author} {\bibfnamefont {T.}~\bibnamefont
  {Ido}}\ and\ \bibinfo {author} {\bibfnamefont {H.}~\bibnamefont {Katori}},\
  }\bibfield  {title} {\bibinfo {title} {Recoil-free spectroscopy of neutral
  {Sr} atoms in the {Lamb-Dicke} regime},\ }\href
  {https://doi.org/10.1103/PhysRevLett.91.053001} {\bibfield  {journal}
  {\bibinfo  {journal} {Phys. Rev. Lett.}\ }\textbf {\bibinfo {volume} {91}},\
  \bibinfo {pages} {053001} (\bibinfo {year} {2003})}\BibitemShut {NoStop}%
\bibitem [{\citenamefont {Boyd}\ \emph {et~al.}(2007)\citenamefont {Boyd},
  \citenamefont {Zelevinsky}, \citenamefont {Ludlow}, \citenamefont {Blatt},
  \citenamefont {Zanon-Willette}, \citenamefont {Foreman},\ and\ \citenamefont
  {Ye}}]{boyd07b}%
  \BibitemOpen
  \bibfield  {author} {\bibinfo {author} {\bibfnamefont {M.}~\bibnamefont
  {Boyd}}, \bibinfo {author} {\bibfnamefont {T.}~\bibnamefont {Zelevinsky}},
  \bibinfo {author} {\bibfnamefont {A.}~\bibnamefont {Ludlow}}, \bibinfo
  {author} {\bibfnamefont {S.}~\bibnamefont {Blatt}}, \bibinfo {author}
  {\bibfnamefont {T.}~\bibnamefont {Zanon-Willette}}, \bibinfo {author}
  {\bibfnamefont {S.}~\bibnamefont {Foreman}},\ and\ \bibinfo {author}
  {\bibfnamefont {J.}~\bibnamefont {Ye}},\ }\bibfield  {title} {\bibinfo
  {title} {Nuclear spin effects in optical lattice clocks},\ }\href
  {https://doi.org/10.1103/PhysRevA.76.022510} {\bibfield  {journal} {\bibinfo
  {journal} {Phys. Rev. A}\ }\textbf {\bibinfo {volume} {76}},\ \bibinfo
  {pages} {022510} (\bibinfo {year} {2007})}\BibitemShut {NoStop}%
\bibitem [{\citenamefont {Muniz}\ \emph {et~al.}(2021)\citenamefont {Muniz},
  \citenamefont {Young}, \citenamefont {Cline},\ and\ \citenamefont
  {Thompson}}]{muniz21}%
  \BibitemOpen
  \bibfield  {author} {\bibinfo {author} {\bibfnamefont {J.~A.}\ \bibnamefont
  {Muniz}}, \bibinfo {author} {\bibfnamefont {D.~J.}\ \bibnamefont {Young}},
  \bibinfo {author} {\bibfnamefont {J.~R.~K.}\ \bibnamefont {Cline}},\ and\
  \bibinfo {author} {\bibfnamefont {J.~K.}\ \bibnamefont {Thompson}},\
  }\bibfield  {title} {\bibinfo {title} {Cavity-{QED} measurements of the
  $^{87}\mathrm{Sr}$ millihertz optical clock transition and determination of
  its natural linewidth},\ }\href
  {https://doi.org/10.1103/PhysRevResearch.3.023152} {\bibfield  {journal}
  {\bibinfo  {journal} {Phys. Rev. Research}\ }\textbf {\bibinfo {volume}
  {3}},\ \bibinfo {pages} {023152} (\bibinfo {year} {2021})}\BibitemShut
  {NoStop}%
\bibitem [{\citenamefont {Taichenachev}\ \emph {et~al.}(2006)\citenamefont
  {Taichenachev}, \citenamefont {Yudin}, \citenamefont {Oates}, \citenamefont
  {Hoyt}, \citenamefont {Barber},\ and\ \citenamefont
  {Hollberg}}]{taichenachev06}%
  \BibitemOpen
  \bibfield  {author} {\bibinfo {author} {\bibfnamefont {A.~V.}\ \bibnamefont
  {Taichenachev}}, \bibinfo {author} {\bibfnamefont {V.~I.}\ \bibnamefont
  {Yudin}}, \bibinfo {author} {\bibfnamefont {C.~W.}\ \bibnamefont {Oates}},
  \bibinfo {author} {\bibfnamefont {C.~W.}\ \bibnamefont {Hoyt}}, \bibinfo
  {author} {\bibfnamefont {Z.~W.}\ \bibnamefont {Barber}},\ and\ \bibinfo
  {author} {\bibfnamefont {L.}~\bibnamefont {Hollberg}},\ }\bibfield  {title}
  {\bibinfo {title} {Magnetic field-induced spectroscopy of forbidden optical
  transitions with application to lattice-based optical atomic clocks},\ }\href
  {https://doi.org/10.1103/PhysRevLett.96.083001} {\bibfield  {journal}
  {\bibinfo  {journal} {Phys. Rev. Lett.}\ }\textbf {\bibinfo {volume} {96}},\
  \bibinfo {pages} {083001} (\bibinfo {year} {2006})}\BibitemShut {NoStop}%
\bibitem [{\citenamefont {Lisdat}\ \emph {et~al.}(2009)\citenamefont {Lisdat},
  \citenamefont {{Vellore Winfred}}, \citenamefont {Middelmann}, \citenamefont
  {Riehle},\ and\ \citenamefont {Sterr}}]{lisdat09}%
  \BibitemOpen
  \bibfield  {author} {\bibinfo {author} {\bibfnamefont {C.}~\bibnamefont
  {Lisdat}}, \bibinfo {author} {\bibfnamefont {J.~S.~R.}\ \bibnamefont
  {{Vellore Winfred}}}, \bibinfo {author} {\bibfnamefont {T.}~\bibnamefont
  {Middelmann}}, \bibinfo {author} {\bibfnamefont {F.}~\bibnamefont {Riehle}},\
  and\ \bibinfo {author} {\bibfnamefont {U.}~\bibnamefont {Sterr}},\ }\bibfield
   {title} {\bibinfo {title} {Collisional losses, decoherence, and frequency
  shifts in optical lattice clocks with bosons},\ }\href
  {https://doi.org/10.1103/PhysRevLett.103.090801} {\bibfield  {journal}
  {\bibinfo  {journal} {Phys. Rev. Lett.}\ }\textbf {\bibinfo {volume} {103}},\
  \bibinfo {pages} {090801} (\bibinfo {year} {2009})}\BibitemShut {NoStop}%
\bibitem [{\citenamefont {Bishof}\ \emph {et~al.}(2011)\citenamefont {Bishof},
  \citenamefont {Martin}, \citenamefont {Swallows}, \citenamefont {Benko},
  \citenamefont {Lin}, \citenamefont {Qu{\'e}m{\'e}ner}, \citenamefont {Rey},\
  and\ \citenamefont {Ye}}]{bishof11}%
  \BibitemOpen
  \bibfield  {author} {\bibinfo {author} {\bibfnamefont {M.}~\bibnamefont
  {Bishof}}, \bibinfo {author} {\bibfnamefont {M.}~\bibnamefont {Martin}},
  \bibinfo {author} {\bibfnamefont {M.}~\bibnamefont {Swallows}}, \bibinfo
  {author} {\bibfnamefont {C.}~\bibnamefont {Benko}}, \bibinfo {author}
  {\bibfnamefont {Y.}~\bibnamefont {Lin}}, \bibinfo {author} {\bibfnamefont
  {G.}~\bibnamefont {Qu{\'e}m{\'e}ner}}, \bibinfo {author} {\bibfnamefont
  {A.}~\bibnamefont {Rey}},\ and\ \bibinfo {author} {\bibfnamefont
  {J.}~\bibnamefont {Ye}},\ }\bibfield  {title} {\bibinfo {title} {Inelastic
  collisions and density-dependent excitation suppression in a {$^{87}$Sr}
  optical lattice clock},\ }\href {https://doi.org/10.1103/PhysRevA.84.052716}
  {\bibfield  {journal} {\bibinfo  {journal} {Phys. Rev. A}\ }\textbf {\bibinfo
  {volume} {84}},\ \bibinfo {pages} {052716} (\bibinfo {year}
  {2011})}\BibitemShut {NoStop}%
\bibitem [{\citenamefont {Blatt}\ \emph {et~al.}(2009)\citenamefont {Blatt},
  \citenamefont {Thomsen}, \citenamefont {Campbell}, \citenamefont {Ludlow},
  \citenamefont {Swallows}, \citenamefont {Martin}, \citenamefont {Boyd},\ and\
  \citenamefont {Ye}}]{blatt09}%
  \BibitemOpen
  \bibfield  {author} {\bibinfo {author} {\bibfnamefont {S.}~\bibnamefont
  {Blatt}}, \bibinfo {author} {\bibfnamefont {J.~W.}\ \bibnamefont {Thomsen}},
  \bibinfo {author} {\bibfnamefont {G.~K.}\ \bibnamefont {Campbell}}, \bibinfo
  {author} {\bibfnamefont {A.~D.}\ \bibnamefont {Ludlow}}, \bibinfo {author}
  {\bibfnamefont {M.~D.}\ \bibnamefont {Swallows}}, \bibinfo {author}
  {\bibfnamefont {M.~J.}\ \bibnamefont {Martin}}, \bibinfo {author}
  {\bibfnamefont {M.~M.}\ \bibnamefont {Boyd}},\ and\ \bibinfo {author}
  {\bibfnamefont {J.}~\bibnamefont {Ye}},\ }\bibfield  {title} {\bibinfo
  {title} {Rabi spectroscopy and excitation inhomogeneity in a one-dimensional
  optical lattice clock},\ }\href {https://doi.org/10.1103/PhysRevA.80.052703}
  {\bibfield  {journal} {\bibinfo  {journal} {Phys. Rev. A}\ }\textbf {\bibinfo
  {volume} {80}},\ \bibinfo {pages} {052703} (\bibinfo {year}
  {2009})}\BibitemShut {NoStop}%
\bibitem [{\citenamefont {McDonald}\ \emph {et~al.}(2015)\citenamefont
  {McDonald}, \citenamefont {McGuyer}, \citenamefont {Iwata},\ and\
  \citenamefont {Zelevinsky}}]{mcdonald15}%
  \BibitemOpen
  \bibfield  {author} {\bibinfo {author} {\bibfnamefont {M.}~\bibnamefont
  {McDonald}}, \bibinfo {author} {\bibfnamefont {B.~H.}\ \bibnamefont
  {McGuyer}}, \bibinfo {author} {\bibfnamefont {G.~Z.}\ \bibnamefont {Iwata}},\
  and\ \bibinfo {author} {\bibfnamefont {T.}~\bibnamefont {Zelevinsky}},\
  }\bibfield  {title} {\bibinfo {title} {Thermometry via light shifts in
  optical lattices},\ }\href {https://doi.org/10.1103/PhysRevLett.114.023001}
  {\bibfield  {journal} {\bibinfo  {journal} {Phys. Rev. Lett.}\ }\textbf
  {\bibinfo {volume} {114}},\ \bibinfo {pages} {023001} (\bibinfo {year}
  {2015})}\BibitemShut {NoStop}%
\bibitem [{\citenamefont {Han}\ \emph {et~al.}(2018)\citenamefont {Han},
  \citenamefont {Zhou}, \citenamefont {Zhang}, \citenamefont {Gao},
  \citenamefont {Xu}, \citenamefont {Li}, \citenamefont {Zhang},\ and\
  \citenamefont {Xu}}]{han18}%
  \BibitemOpen
  \bibfield  {author} {\bibinfo {author} {\bibfnamefont {C.}~\bibnamefont
  {Han}}, \bibinfo {author} {\bibfnamefont {M.}~\bibnamefont {Zhou}}, \bibinfo
  {author} {\bibfnamefont {X.}~\bibnamefont {Zhang}}, \bibinfo {author}
  {\bibfnamefont {Q.}~\bibnamefont {Gao}}, \bibinfo {author} {\bibfnamefont
  {Y.}~\bibnamefont {Xu}}, \bibinfo {author} {\bibfnamefont {S.}~\bibnamefont
  {Li}}, \bibinfo {author} {\bibfnamefont {S.}~\bibnamefont {Zhang}},\ and\
  \bibinfo {author} {\bibfnamefont {X.}~\bibnamefont {Xu}},\ }\bibfield
  {title} {\bibinfo {title} {Carrier thermometry of cold ytterbium atoms in an
  optical lattice clock},\ }\href {https://doi.org/10.1038/s41598-018-26367-8}
  {\bibfield  {journal} {\bibinfo  {journal} {Scientific Reports}\ }\textbf
  {\bibinfo {volume} {8}},\ \bibinfo {pages} {7927} (\bibinfo {year}
  {2018})}\BibitemShut {NoStop}%
\bibitem [{\citenamefont {Blatt}\ \emph {et~al.}(2015)\citenamefont {Blatt},
  \citenamefont {Mazurenko}, \citenamefont {Parsons}, \citenamefont {Chiu},
  \citenamefont {Huber},\ and\ \citenamefont {Greiner}}]{blatt15}%
  \BibitemOpen
  \bibfield  {author} {\bibinfo {author} {\bibfnamefont {S.}~\bibnamefont
  {Blatt}}, \bibinfo {author} {\bibfnamefont {A.}~\bibnamefont {Mazurenko}},
  \bibinfo {author} {\bibfnamefont {M.~F.}\ \bibnamefont {Parsons}}, \bibinfo
  {author} {\bibfnamefont {C.~S.}\ \bibnamefont {Chiu}}, \bibinfo {author}
  {\bibfnamefont {F.}~\bibnamefont {Huber}},\ and\ \bibinfo {author}
  {\bibfnamefont {M.}~\bibnamefont {Greiner}},\ }\bibfield  {title} {\bibinfo
  {title} {Low-noise optical lattices for ultracold {$^6$Li}},\ }\href
  {https://doi.org/10.1103/PhysRevA.92.021402} {\bibfield  {journal} {\bibinfo
  {journal} {Phys. Rev. A}\ }\textbf {\bibinfo {volume} {92}},\ \bibinfo
  {pages} {021402(R)} (\bibinfo {year} {2015})}\BibitemShut {NoStop}%
\bibitem [{\citenamefont {Savard}\ \emph {et~al.}(1997)\citenamefont {Savard},
  \citenamefont {O'Hara},\ and\ \citenamefont {Thomas}}]{savard97}%
  \BibitemOpen
  \bibfield  {author} {\bibinfo {author} {\bibfnamefont {T.~A.}\ \bibnamefont
  {Savard}}, \bibinfo {author} {\bibfnamefont {K.~M.}\ \bibnamefont {O'Hara}},\
  and\ \bibinfo {author} {\bibfnamefont {J.~E.}\ \bibnamefont {Thomas}},\
  }\bibfield  {title} {\bibinfo {title} {Laser-noise-induced heating in far-off
  resonance optical traps},\ }\href {https://doi.org/10.1103/PhysRevA.56.R1095}
  {\bibfield  {journal} {\bibinfo  {journal} {Phys. Rev. A}\ }\textbf {\bibinfo
  {volume} {56}},\ \bibinfo {pages} {R1095} (\bibinfo {year}
  {1997})}\BibitemShut {NoStop}%
\bibitem [{\citenamefont {Mosk}\ \emph {et~al.}(2001)\citenamefont {Mosk},
  \citenamefont {Jochim}, \citenamefont {Moritz}, \citenamefont {Els\"{a}sser},
  \citenamefont {Weidem\"{u}ller},\ and\ \citenamefont {Grimm}}]{mosk01}%
  \BibitemOpen
  \bibfield  {author} {\bibinfo {author} {\bibfnamefont {A.}~\bibnamefont
  {Mosk}}, \bibinfo {author} {\bibfnamefont {S.}~\bibnamefont {Jochim}},
  \bibinfo {author} {\bibfnamefont {H.}~\bibnamefont {Moritz}}, \bibinfo
  {author} {\bibfnamefont {T.}~\bibnamefont {Els\"{a}sser}}, \bibinfo {author}
  {\bibfnamefont {M.}~\bibnamefont {Weidem\"{u}ller}},\ and\ \bibinfo {author}
  {\bibfnamefont {R.}~\bibnamefont {Grimm}},\ }\bibfield  {title} {\bibinfo
  {title} {Resonator-enhanced optical dipole trap for fermionic lithium
  atoms},\ }\href {http://ol.osa.org/abstract.cfm?URI=ol-26-23-1837} {\bibfield
   {journal} {\bibinfo  {journal} {Opt. Lett.}\ }\textbf {\bibinfo {volume}
  {26}},\ \bibinfo {pages} {1837} (\bibinfo {year} {2001})}\BibitemShut
  {NoStop}%
\bibitem [{\citenamefont {Lodewyck}\ \emph {et~al.}(2010)\citenamefont
  {Lodewyck}, \citenamefont {Westergaard}, \citenamefont {Lecallier},
  \citenamefont {Lorini},\ and\ \citenamefont {Lemonde}}]{lodewyck10}%
  \BibitemOpen
  \bibfield  {author} {\bibinfo {author} {\bibfnamefont {J.}~\bibnamefont
  {Lodewyck}}, \bibinfo {author} {\bibfnamefont {P.~G.}\ \bibnamefont
  {Westergaard}}, \bibinfo {author} {\bibfnamefont {A.}~\bibnamefont
  {Lecallier}}, \bibinfo {author} {\bibfnamefont {L.}~\bibnamefont {Lorini}},\
  and\ \bibinfo {author} {\bibfnamefont {P.}~\bibnamefont {Lemonde}},\
  }\bibfield  {title} {\bibinfo {title} {Frequency stability of optical lattice
  clocks},\ }\href {https://doi.org/10.1088/1367-2630/12/6/065026} {\bibfield
  {journal} {\bibinfo  {journal} {New Journal of Physics}\ }\textbf {\bibinfo
  {volume} {12}},\ \bibinfo {pages} {065026} (\bibinfo {year}
  {2010})}\BibitemShut {NoStop}%
\bibitem [{\citenamefont {Harry}\ and\ \citenamefont {the LIGO
  Scientific~Collaboration}(2010)}]{harry10}%
  \BibitemOpen
  \bibfield  {author} {\bibinfo {author} {\bibfnamefont {G.~M.}\ \bibnamefont
  {Harry}}\ and\ \bibinfo {author} {\bibnamefont {the LIGO
  Scientific~Collaboration}},\ }\bibfield  {title} {\bibinfo {title} {Advanced
  {LIGO}: the next generation of gravitational wave detectors},\ }\href
  {https://doi.org/10.1088/0264-9381/27/8/084006} {\bibfield  {journal}
  {\bibinfo  {journal} {Classical and Quantum Gravity}\ }\textbf {\bibinfo
  {volume} {27}},\ \bibinfo {pages} {084006} (\bibinfo {year}
  {2010})}\BibitemShut {NoStop}%
\bibitem [{\citenamefont {{Le Targat}}\ \emph {et~al.}(2013)\citenamefont {{Le
  Targat}}, \citenamefont {Lorini}, \citenamefont {{Le Coq}}, \citenamefont
  {Zawada}, \citenamefont {Gu{\'e}na}, \citenamefont {Abgrall}, \citenamefont
  {Gurov}, \citenamefont {Rosenbusch}, \citenamefont {Rovera}, \citenamefont
  {Nag{\'o}rny}, \citenamefont {Gartman}, \citenamefont {Westergaard},
  \citenamefont {Tobar}, \citenamefont {Lours}, \citenamefont {Santarelli},
  \citenamefont {Clairon}, \citenamefont {Bize}, \citenamefont {Laurent},
  \citenamefont {Lemonde},\ and\ \citenamefont {Lodewyck}}]{letargat13}%
  \BibitemOpen
  \bibfield  {author} {\bibinfo {author} {\bibfnamefont {R.}~\bibnamefont {{Le
  Targat}}}, \bibinfo {author} {\bibfnamefont {L.}~\bibnamefont {Lorini}},
  \bibinfo {author} {\bibfnamefont {Y.}~\bibnamefont {{Le Coq}}}, \bibinfo
  {author} {\bibfnamefont {M.}~\bibnamefont {Zawada}}, \bibinfo {author}
  {\bibfnamefont {J.}~\bibnamefont {Gu{\'e}na}}, \bibinfo {author}
  {\bibfnamefont {M.}~\bibnamefont {Abgrall}}, \bibinfo {author} {\bibfnamefont
  {M.}~\bibnamefont {Gurov}}, \bibinfo {author} {\bibfnamefont
  {P.}~\bibnamefont {Rosenbusch}}, \bibinfo {author} {\bibfnamefont {D.~G.}\
  \bibnamefont {Rovera}}, \bibinfo {author} {\bibfnamefont {B.}~\bibnamefont
  {Nag{\'o}rny}}, \bibinfo {author} {\bibfnamefont {R.}~\bibnamefont
  {Gartman}}, \bibinfo {author} {\bibfnamefont {P.~G.}\ \bibnamefont
  {Westergaard}}, \bibinfo {author} {\bibfnamefont {M.~E.}\ \bibnamefont
  {Tobar}}, \bibinfo {author} {\bibfnamefont {M.}~\bibnamefont {Lours}},
  \bibinfo {author} {\bibfnamefont {G.}~\bibnamefont {Santarelli}}, \bibinfo
  {author} {\bibfnamefont {A.}~\bibnamefont {Clairon}}, \bibinfo {author}
  {\bibfnamefont {S.}~\bibnamefont {Bize}}, \bibinfo {author} {\bibfnamefont
  {P.}~\bibnamefont {Laurent}}, \bibinfo {author} {\bibfnamefont
  {P.}~\bibnamefont {Lemonde}},\ and\ \bibinfo {author} {\bibfnamefont
  {J.}~\bibnamefont {Lodewyck}},\ }\bibfield  {title} {\bibinfo {title}
  {Experimental realization of an optical second with strontium lattice
  clocks},\ }\href {https://doi.org/10.1038/ncomms3109} {\bibfield  {journal}
  {\bibinfo  {journal} {Nature communications}\ }\textbf {\bibinfo {volume}
  {4}},\ \bibinfo {pages} {1} (\bibinfo {year} {2013})}\BibitemShut {NoStop}%
\bibitem [{\citenamefont {Schiller}\ \emph {et~al.}(2012)\citenamefont
  {Schiller}, \citenamefont {G{\"o}rlitz}, \citenamefont {Nevsky},
  \citenamefont {Alighanbari}, \citenamefont {Vasilyev}, \citenamefont
  {Abou-Jaoudeh}, \citenamefont {Mura}, \citenamefont {Franzen}, \citenamefont
  {Sterr}, \citenamefont {Falke}, \citenamefont {Lisdat}, \citenamefont
  {Rasel}, \citenamefont {Kulosa}, \citenamefont {Bize}, \citenamefont
  {Lodewyck}, \citenamefont {Tino}, \citenamefont {Poli}, \citenamefont
  {Schioppo}, \citenamefont {Bongs}, \citenamefont {Singh}, \citenamefont
  {Gill}, \citenamefont {Barwood}, \citenamefont {Ovchinnikov}, \citenamefont
  {Stuhler}, \citenamefont {Kaenders}, \citenamefont {Braxmaier}, \citenamefont
  {Holzwarth}, \citenamefont {Donati}, \citenamefont {Lecomte}, \citenamefont
  {Calonico},\ and\ \citenamefont {Levi}}]{schiller12}%
  \BibitemOpen
  \bibfield  {author} {\bibinfo {author} {\bibfnamefont {S.}~\bibnamefont
  {Schiller}}, \bibinfo {author} {\bibfnamefont {A.}~\bibnamefont
  {G{\"o}rlitz}}, \bibinfo {author} {\bibfnamefont {A.}~\bibnamefont {Nevsky}},
  \bibinfo {author} {\bibfnamefont {S.}~\bibnamefont {Alighanbari}}, \bibinfo
  {author} {\bibfnamefont {S.}~\bibnamefont {Vasilyev}}, \bibinfo {author}
  {\bibfnamefont {C.}~\bibnamefont {Abou-Jaoudeh}}, \bibinfo {author}
  {\bibfnamefont {G.}~\bibnamefont {Mura}}, \bibinfo {author} {\bibfnamefont
  {T.}~\bibnamefont {Franzen}}, \bibinfo {author} {\bibfnamefont
  {U.}~\bibnamefont {Sterr}}, \bibinfo {author} {\bibfnamefont
  {S.}~\bibnamefont {Falke}}, \bibinfo {author} {\bibfnamefont
  {C.}~\bibnamefont {Lisdat}}, \bibinfo {author} {\bibfnamefont
  {E.}~\bibnamefont {Rasel}}, \bibinfo {author} {\bibfnamefont
  {A.}~\bibnamefont {Kulosa}}, \bibinfo {author} {\bibfnamefont
  {S.}~\bibnamefont {Bize}}, \bibinfo {author} {\bibfnamefont {J.}~\bibnamefont
  {Lodewyck}}, \bibinfo {author} {\bibfnamefont {G.~M.}\ \bibnamefont {Tino}},
  \bibinfo {author} {\bibfnamefont {N.}~\bibnamefont {Poli}}, \bibinfo {author}
  {\bibfnamefont {M.}~\bibnamefont {Schioppo}}, \bibinfo {author}
  {\bibfnamefont {K.}~\bibnamefont {Bongs}}, \bibinfo {author} {\bibfnamefont
  {Y.}~\bibnamefont {Singh}}, \bibinfo {author} {\bibfnamefont
  {P.}~\bibnamefont {Gill}}, \bibinfo {author} {\bibfnamefont {G.}~\bibnamefont
  {Barwood}}, \bibinfo {author} {\bibfnamefont {Y.}~\bibnamefont
  {Ovchinnikov}}, \bibinfo {author} {\bibfnamefont {J.}~\bibnamefont
  {Stuhler}}, \bibinfo {author} {\bibfnamefont {W.}~\bibnamefont {Kaenders}},
  \bibinfo {author} {\bibfnamefont {C.}~\bibnamefont {Braxmaier}}, \bibinfo
  {author} {\bibfnamefont {R.}~\bibnamefont {Holzwarth}}, \bibinfo {author}
  {\bibfnamefont {A.}~\bibnamefont {Donati}}, \bibinfo {author} {\bibfnamefont
  {S.}~\bibnamefont {Lecomte}}, \bibinfo {author} {\bibfnamefont
  {D.}~\bibnamefont {Calonico}},\ and\ \bibinfo {author} {\bibfnamefont
  {F.}~\bibnamefont {Levi}},\ }\bibfield  {title} {\bibinfo {title} {The space
  optical clocks project: Development of high-performance transportable and
  breadboard optical clocks and advanced subsystems},\ }in\ \href
  {https://doi.org/10.1109/EFTF.2012.6502414} {\emph {\bibinfo {booktitle}
  {2012 European Frequency and Time Forum}}}\ (\bibinfo {organization} {IEEE},\
  \bibinfo {year} {2012})\ p.\ \bibinfo {pages} {412}\BibitemShut {NoStop}%
\bibitem [{\citenamefont {Nicholson}\ \emph {et~al.}(2015)\citenamefont
  {Nicholson}, \citenamefont {Campbell}, \citenamefont {Hutson}, \citenamefont
  {Marti}, \citenamefont {Bloom}, \citenamefont {McNally}, \citenamefont
  {Zhang}, \citenamefont {Barrett}, \citenamefont {Safronova}, \citenamefont
  {Strouse}, \citenamefont {Tew},\ and\ \citenamefont {Ye}}]{nicholson15}%
  \BibitemOpen
  \bibfield  {author} {\bibinfo {author} {\bibfnamefont {T.}~\bibnamefont
  {Nicholson}}, \bibinfo {author} {\bibfnamefont {S.}~\bibnamefont {Campbell}},
  \bibinfo {author} {\bibfnamefont {R.}~\bibnamefont {Hutson}}, \bibinfo
  {author} {\bibfnamefont {G.}~\bibnamefont {Marti}}, \bibinfo {author}
  {\bibfnamefont {B.}~\bibnamefont {Bloom}}, \bibinfo {author} {\bibfnamefont
  {R.}~\bibnamefont {McNally}}, \bibinfo {author} {\bibfnamefont
  {W.}~\bibnamefont {Zhang}}, \bibinfo {author} {\bibfnamefont
  {M.}~\bibnamefont {Barrett}}, \bibinfo {author} {\bibfnamefont
  {M.}~\bibnamefont {Safronova}}, \bibinfo {author} {\bibfnamefont
  {G.}~\bibnamefont {Strouse}}, \bibinfo {author} {\bibfnamefont
  {W.}~\bibnamefont {Tew}},\ and\ \bibinfo {author} {\bibfnamefont
  {J.}~\bibnamefont {Ye}},\ }\bibfield  {title} {\bibinfo {title} {Systematic
  evaluation of an atomic clock at {$2\times 10^{-18}$} total uncertainty},\
  }\href {https://doi.org/10.1038/ncomms7896} {\bibfield  {journal} {\bibinfo
  {journal} {Nat. Commun.}\ }\textbf {\bibinfo {volume} {6}},\ \bibinfo {pages}
  {6896} (\bibinfo {year} {2015})}\BibitemShut {NoStop}%
\bibitem [{\citenamefont {Lemonde}\ and\ \citenamefont
  {Wolf}(2005)}]{lemonde05}%
  \BibitemOpen
  \bibfield  {author} {\bibinfo {author} {\bibfnamefont {P.}~\bibnamefont
  {Lemonde}}\ and\ \bibinfo {author} {\bibfnamefont {P.}~\bibnamefont {Wolf}},\
  }\bibfield  {title} {\bibinfo {title} {Optical lattice clock with atoms
  confined in a shallow trap},\ }\href
  {https://doi.org/10.1103/PhysRevA.72.033409} {\bibfield  {journal} {\bibinfo
  {journal} {Phys. Rev. A}\ }\textbf {\bibinfo {volume} {72}},\ \bibinfo
  {pages} {033409} (\bibinfo {year} {2005})}\BibitemShut {NoStop}%
\bibitem [{\citenamefont {Shi}\ \emph {et~al.}(2015)\citenamefont {Shi},
  \citenamefont {Robyr}, \citenamefont {Eismann}, \citenamefont {Zawada},
  \citenamefont {Lorini}, \citenamefont {{Le Targat}},\ and\ \citenamefont
  {Lodewyck}}]{shi15}%
  \BibitemOpen
  \bibfield  {author} {\bibinfo {author} {\bibfnamefont {C.}~\bibnamefont
  {Shi}}, \bibinfo {author} {\bibfnamefont {J.-L.}\ \bibnamefont {Robyr}},
  \bibinfo {author} {\bibfnamefont {U.}~\bibnamefont {Eismann}}, \bibinfo
  {author} {\bibfnamefont {M.}~\bibnamefont {Zawada}}, \bibinfo {author}
  {\bibfnamefont {L.}~\bibnamefont {Lorini}}, \bibinfo {author} {\bibfnamefont
  {R.}~\bibnamefont {{Le Targat}}},\ and\ \bibinfo {author} {\bibfnamefont
  {J.}~\bibnamefont {Lodewyck}},\ }\bibfield  {title} {\bibinfo {title}
  {Polarizabilities of the {$^{87}\mathrm{Sr}$} clock transition},\ }\href
  {https://doi.org/10.1103/PhysRevA.92.012516} {\bibfield  {journal} {\bibinfo
  {journal} {Phys. Rev. A}\ }\textbf {\bibinfo {volume} {92}},\ \bibinfo
  {pages} {012516} (\bibinfo {year} {2015})}\BibitemShut {NoStop}%
\bibitem [{\citenamefont {Ushijima}\ \emph {et~al.}(2015)\citenamefont
  {Ushijima}, \citenamefont {Takamoto}, \citenamefont {Das}, \citenamefont
  {Ohkubo},\ and\ \citenamefont {Katori}}]{ushijima15}%
  \BibitemOpen
  \bibfield  {author} {\bibinfo {author} {\bibfnamefont {I.}~\bibnamefont
  {Ushijima}}, \bibinfo {author} {\bibfnamefont {M.}~\bibnamefont {Takamoto}},
  \bibinfo {author} {\bibfnamefont {M.}~\bibnamefont {Das}}, \bibinfo {author}
  {\bibfnamefont {T.}~\bibnamefont {Ohkubo}},\ and\ \bibinfo {author}
  {\bibfnamefont {H.}~\bibnamefont {Katori}},\ }\bibfield  {title} {\bibinfo
  {title} {Cryogenic optical lattice clocks},\ }\href
  {https://doi.org/10.1038/nphoton.2015.5} {\bibfield  {journal} {\bibinfo
  {journal} {Nature Photonics}\ }\textbf {\bibinfo {volume} {9}},\ \bibinfo
  {pages} {185} (\bibinfo {year} {2015})}\BibitemShut {NoStop}%
\bibitem [{\citenamefont {Schymik}\ \emph {et~al.}(2021)\citenamefont
  {Schymik}, \citenamefont {Pancaldi}, \citenamefont {Nogrette}, \citenamefont
  {Barredo}, \citenamefont {Paris}, \citenamefont {Browaeys},\ and\
  \citenamefont {Lahaye}}]{schymik21}%
  \BibitemOpen
  \bibfield  {author} {\bibinfo {author} {\bibfnamefont {K.-N.}\ \bibnamefont
  {Schymik}}, \bibinfo {author} {\bibfnamefont {S.}~\bibnamefont {Pancaldi}},
  \bibinfo {author} {\bibfnamefont {F.}~\bibnamefont {Nogrette}}, \bibinfo
  {author} {\bibfnamefont {D.}~\bibnamefont {Barredo}}, \bibinfo {author}
  {\bibfnamefont {J.}~\bibnamefont {Paris}}, \bibinfo {author} {\bibfnamefont
  {A.}~\bibnamefont {Browaeys}},\ and\ \bibinfo {author} {\bibfnamefont
  {T.}~\bibnamefont {Lahaye}},\ }\bibfield  {title} {\bibinfo {title} {Single
  atoms with 6000-second trapping lifetimes in optical-tweezer arrays at
  cryogenic temperatures},\ }\href
  {https://doi.org/10.1103/PhysRevApplied.16.034013} {\bibfield  {journal}
  {\bibinfo  {journal} {Phys. Rev. Applied}\ }\textbf {\bibinfo {volume}
  {16}},\ \bibinfo {pages} {034013} (\bibinfo {year} {2021})}\BibitemShut
  {NoStop}%
\bibitem [{\citenamefont {Shibata}\ \emph {et~al.}(2014)\citenamefont
  {Shibata}, \citenamefont {Yamamoto}, \citenamefont {Seki},\ and\
  \citenamefont {Takahashi}}]{shibata14}%
  \BibitemOpen
  \bibfield  {author} {\bibinfo {author} {\bibfnamefont {K.}~\bibnamefont
  {Shibata}}, \bibinfo {author} {\bibfnamefont {R.}~\bibnamefont {Yamamoto}},
  \bibinfo {author} {\bibfnamefont {Y.}~\bibnamefont {Seki}},\ and\ \bibinfo
  {author} {\bibfnamefont {Y.}~\bibnamefont {Takahashi}},\ }\bibfield  {title}
  {\bibinfo {title} {Optical spectral imaging of a single layer of a quantum
  gas with an ultranarrow optical transition},\ }\href
  {https://doi.org/10.1103/PhysRevA.89.031601} {\bibfield  {journal} {\bibinfo
  {journal} {Phys. Rev. A}\ }\textbf {\bibinfo {volume} {89}},\ \bibinfo
  {pages} {031601(R)} (\bibinfo {year} {2014})}\BibitemShut {NoStop}%
\bibitem [{\citenamefont {Marti}\ \emph {et~al.}(2018)\citenamefont {Marti},
  \citenamefont {Hutson}, \citenamefont {Goban}, \citenamefont {Campbell},
  \citenamefont {Poli},\ and\ \citenamefont {Ye}}]{marti18}%
  \BibitemOpen
  \bibfield  {author} {\bibinfo {author} {\bibfnamefont {G.~E.}\ \bibnamefont
  {Marti}}, \bibinfo {author} {\bibfnamefont {R.~B.}\ \bibnamefont {Hutson}},
  \bibinfo {author} {\bibfnamefont {A.}~\bibnamefont {Goban}}, \bibinfo
  {author} {\bibfnamefont {S.~L.}\ \bibnamefont {Campbell}}, \bibinfo {author}
  {\bibfnamefont {N.}~\bibnamefont {Poli}},\ and\ \bibinfo {author}
  {\bibfnamefont {J.}~\bibnamefont {Ye}},\ }\bibfield  {title} {\bibinfo
  {title} {Imaging optical frequencies with 100~$\mu$hz precision and
  1.1~$\mu$m resolution},\ }\href
  {https://doi.org/10.1103/PhysRevLett.120.103201} {\bibfield  {journal}
  {\bibinfo  {journal} {Phys. Rev. Lett.}\ }\textbf {\bibinfo {volume} {120}},\
  \bibinfo {pages} {103201} (\bibinfo {year} {2018})}\BibitemShut {NoStop}%
\bibitem [{\citenamefont {Daley}\ \emph {et~al.}(2008)\citenamefont {Daley},
  \citenamefont {Boyd}, \citenamefont {Ye},\ and\ \citenamefont
  {Zoller}}]{daley08}%
  \BibitemOpen
  \bibfield  {author} {\bibinfo {author} {\bibfnamefont {A.}~\bibnamefont
  {Daley}}, \bibinfo {author} {\bibfnamefont {M.}~\bibnamefont {Boyd}},
  \bibinfo {author} {\bibfnamefont {J.}~\bibnamefont {Ye}},\ and\ \bibinfo
  {author} {\bibfnamefont {P.}~\bibnamefont {Zoller}},\ }\bibfield  {title}
  {\bibinfo {title} {Quantum computing with alkaline-earth-metal atoms},\
  }\href {https://doi.org/10.1103/PhysRevLett.101.170504} {\bibfield  {journal}
  {\bibinfo  {journal} {Phys. Rev. Lett.}\ }\textbf {\bibinfo {volume} {101}},\
  \bibinfo {pages} {170504} (\bibinfo {year} {2008})}\BibitemShut {NoStop}%
\bibitem [{\citenamefont {Daley}\ \emph {et~al.}(2011)\citenamefont {Daley},
  \citenamefont {Ye},\ and\ \citenamefont {Zoller}}]{daley11}%
  \BibitemOpen
  \bibfield  {author} {\bibinfo {author} {\bibfnamefont {A.~J.}\ \bibnamefont
  {Daley}}, \bibinfo {author} {\bibfnamefont {J.}~\bibnamefont {Ye}},\ and\
  \bibinfo {author} {\bibfnamefont {P.}~\bibnamefont {Zoller}},\ }\bibfield
  {title} {\bibinfo {title} {State-dependent lattices for quantum computing
  with alkaline-earth-metal atoms},\ }\href
  {https://doi.org/10.1140/epjd/e2011-20095-2} {\bibfield  {journal} {\bibinfo
  {journal} {Eur. Phys. J. D}\ }\textbf {\bibinfo {volume} {65}},\ \bibinfo
  {pages} {207} (\bibinfo {year} {2011})}\BibitemShut {NoStop}%
\bibitem [{\citenamefont {Daley}(2011)}]{daley11b}%
  \BibitemOpen
  \bibfield  {author} {\bibinfo {author} {\bibfnamefont {A.~J.}\ \bibnamefont
  {Daley}},\ }\bibfield  {title} {\bibinfo {title} {Quantum computing and
  quantum simulation with {group-II} atoms},\ }\href
  {https://doi.org/10.1007/s11128-011-0293-3} {\bibfield  {journal} {\bibinfo
  {journal} {Quantum Inf. Process.}\ }\textbf {\bibinfo {volume} {10}},\
  \bibinfo {pages} {865} (\bibinfo {year} {2011})}\BibitemShut {NoStop}%
\bibitem [{\citenamefont {Gonz\'alez-Tudela}\ and\ \citenamefont
  {Cirac}(2017{\natexlab{b}})}]{tudela17b}%
  \BibitemOpen
  \bibfield  {author} {\bibinfo {author} {\bibfnamefont {A.}~\bibnamefont
  {Gonz\'alez-Tudela}}\ and\ \bibinfo {author} {\bibfnamefont {J.~I.}\
  \bibnamefont {Cirac}},\ }\bibfield  {title} {\bibinfo {title} {{Markovian}
  and non-{Markovian} dynamics of quantum emitters coupled to two-dimensional
  structured reservoirs},\ }\href {https://doi.org/10.1103/PhysRevA.96.043811}
  {\bibfield  {journal} {\bibinfo  {journal} {Phys. Rev. A}\ }\textbf {\bibinfo
  {volume} {96}},\ \bibinfo {pages} {043811} (\bibinfo {year}
  {2017}{\natexlab{b}})}\BibitemShut {NoStop}%
\bibitem [{\citenamefont {Koller}\ \emph {et~al.}(2017)\citenamefont {Koller},
  \citenamefont {Grotti}, \citenamefont {Vogt}, \citenamefont {Al-Masoudi},
  \citenamefont {D{\"o}rscher}, \citenamefont {H{\"a}fner}, \citenamefont
  {Sterr},\ and\ \citenamefont {Lisdat}}]{koller17}%
  \BibitemOpen
  \bibfield  {author} {\bibinfo {author} {\bibfnamefont {S.~B.}\ \bibnamefont
  {Koller}}, \bibinfo {author} {\bibfnamefont {J.}~\bibnamefont {Grotti}},
  \bibinfo {author} {\bibfnamefont {S.}~\bibnamefont {Vogt}}, \bibinfo {author}
  {\bibfnamefont {A.}~\bibnamefont {Al-Masoudi}}, \bibinfo {author}
  {\bibfnamefont {S.}~\bibnamefont {D{\"o}rscher}}, \bibinfo {author}
  {\bibfnamefont {S.}~\bibnamefont {H{\"a}fner}}, \bibinfo {author}
  {\bibfnamefont {U.}~\bibnamefont {Sterr}},\ and\ \bibinfo {author}
  {\bibfnamefont {C.}~\bibnamefont {Lisdat}},\ }\bibfield  {title} {\bibinfo
  {title} {Transportable optical lattice clock with {$7\times 10^{-17}$}
  uncertainty},\ }\href@noop {} {\bibfield  {journal} {\bibinfo  {journal}
  {Physical Review Letters}\ }\textbf {\bibinfo {volume} {118}},\ \bibinfo
  {pages} {073601} (\bibinfo {year} {2017})}\BibitemShut {NoStop}%
\bibitem [{\citenamefont {Grotti}\ \emph {et~al.}(2018)\citenamefont {Grotti},
  \citenamefont {Koller}, \citenamefont {Vogt}, \citenamefont {H{\"a}fner},
  \citenamefont {Sterr}, \citenamefont {Lisdat}, \citenamefont {Denker},
  \citenamefont {Voigt}, \citenamefont {Timmen}, \citenamefont {Rolland},
  \citenamefont {Baynes}, \citenamefont {Margolis}, \citenamefont {Zampaolo},
  \citenamefont {Thoumany}, \citenamefont {Pizzocaro}, \citenamefont {Rauf},
  \citenamefont {Bregolin}, \citenamefont {Tampellini}, \citenamefont
  {Barbieri}, \citenamefont {Zucco}, \citenamefont {Costanzo}, \citenamefont
  {Clivati}, \citenamefont {Levi},\ and\ \citenamefont {Calonico}}]{grotti18}%
  \BibitemOpen
  \bibfield  {author} {\bibinfo {author} {\bibfnamefont {J.}~\bibnamefont
  {Grotti}}, \bibinfo {author} {\bibfnamefont {S.}~\bibnamefont {Koller}},
  \bibinfo {author} {\bibfnamefont {S.}~\bibnamefont {Vogt}}, \bibinfo {author}
  {\bibfnamefont {S.}~\bibnamefont {H{\"a}fner}}, \bibinfo {author}
  {\bibfnamefont {U.}~\bibnamefont {Sterr}}, \bibinfo {author} {\bibfnamefont
  {C.}~\bibnamefont {Lisdat}}, \bibinfo {author} {\bibfnamefont
  {H.}~\bibnamefont {Denker}}, \bibinfo {author} {\bibfnamefont
  {C.}~\bibnamefont {Voigt}}, \bibinfo {author} {\bibfnamefont
  {L.}~\bibnamefont {Timmen}}, \bibinfo {author} {\bibfnamefont
  {A.}~\bibnamefont {Rolland}}, \bibinfo {author} {\bibfnamefont {F.~N.}\
  \bibnamefont {Baynes}}, \bibinfo {author} {\bibfnamefont {H.~S.}\
  \bibnamefont {Margolis}}, \bibinfo {author} {\bibfnamefont {M.}~\bibnamefont
  {Zampaolo}}, \bibinfo {author} {\bibfnamefont {P.}~\bibnamefont {Thoumany}},
  \bibinfo {author} {\bibfnamefont {M.}~\bibnamefont {Pizzocaro}}, \bibinfo
  {author} {\bibfnamefont {B.}~\bibnamefont {Rauf}}, \bibinfo {author}
  {\bibfnamefont {F.}~\bibnamefont {Bregolin}}, \bibinfo {author}
  {\bibfnamefont {A.}~\bibnamefont {Tampellini}}, \bibinfo {author}
  {\bibfnamefont {P.}~\bibnamefont {Barbieri}}, \bibinfo {author}
  {\bibfnamefont {M.}~\bibnamefont {Zucco}}, \bibinfo {author} {\bibfnamefont
  {G.~A.}\ \bibnamefont {Costanzo}}, \bibinfo {author} {\bibfnamefont
  {C.}~\bibnamefont {Clivati}}, \bibinfo {author} {\bibfnamefont
  {F.}~\bibnamefont {Levi}},\ and\ \bibinfo {author} {\bibfnamefont
  {D.}~\bibnamefont {Calonico}},\ }\bibfield  {title} {\bibinfo {title}
  {Geodesy and metrology with a transportable optical clock},\ }\href@noop {}
  {\bibfield  {journal} {\bibinfo  {journal} {Nature Physics}\ }\textbf
  {\bibinfo {volume} {14}},\ \bibinfo {pages} {437} (\bibinfo {year}
  {2018})}\BibitemShut {NoStop}%
\bibitem [{\citenamefont {Wolf}\ \emph {et~al.}(2009)\citenamefont {Wolf},
  \citenamefont {Bord{\'e}}, \citenamefont {Clairon}, \citenamefont {Duchayne},
  \citenamefont {Landragin}, \citenamefont {Lemonde}, \citenamefont
  {Santarelli}, \citenamefont {Ertmer}, \citenamefont {Rasel}, \citenamefont
  {Cataliotti}, \citenamefont {Inguscio}, \citenamefont {Tino}, \citenamefont
  {Gill}, \citenamefont {Klein}, \citenamefont {Reynaud}, \citenamefont
  {Salomon}, \citenamefont {Peik}, \citenamefont {Bertolami}, \citenamefont
  {Gil}, \citenamefont {P{\'a}ramos}, \citenamefont {Jentsch}, \citenamefont
  {Johann}, \citenamefont {Rathke}, \citenamefont {Bouyer}, \citenamefont
  {Cacciapuoti}, \citenamefont {Izzo}, \citenamefont {Natale}, \citenamefont
  {Christophe}, \citenamefont {Touboul}, \citenamefont {Turyshev},
  \citenamefont {Anderson}, \citenamefont {Tobar}, \citenamefont
  {Schmidt-Kaler}, \citenamefont {Vigu{\'e}}, \citenamefont {Madej},
  \citenamefont {Marmet}, \citenamefont {Angonin}, \citenamefont {Delva},
  \citenamefont {Tourrenc}, \citenamefont {Metris}, \citenamefont {M{\"u}ller},
  \citenamefont {Walsworth}, \citenamefont {Lu}, \citenamefont {Wang},
  \citenamefont {Bongs}, \citenamefont {Toncelli}, \citenamefont {Tonelli},
  \citenamefont {Dittus}, \citenamefont {L{\"a}mmerzahl}, \citenamefont
  {Galzerano}, \citenamefont {Laporta}, \citenamefont {Laskar}, \citenamefont
  {Fienga}, \citenamefont {Roques},\ and\ \citenamefont {Sengstock}}]{wolf09}%
  \BibitemOpen
  \bibfield  {author} {\bibinfo {author} {\bibfnamefont {P.}~\bibnamefont
  {Wolf}}, \bibinfo {author} {\bibfnamefont {C.~J.}\ \bibnamefont {Bord{\'e}}},
  \bibinfo {author} {\bibfnamefont {A.}~\bibnamefont {Clairon}}, \bibinfo
  {author} {\bibfnamefont {L.}~\bibnamefont {Duchayne}}, \bibinfo {author}
  {\bibfnamefont {A.}~\bibnamefont {Landragin}}, \bibinfo {author}
  {\bibfnamefont {P.}~\bibnamefont {Lemonde}}, \bibinfo {author} {\bibfnamefont
  {G.}~\bibnamefont {Santarelli}}, \bibinfo {author} {\bibfnamefont
  {W.}~\bibnamefont {Ertmer}}, \bibinfo {author} {\bibfnamefont
  {E.}~\bibnamefont {Rasel}}, \bibinfo {author} {\bibfnamefont {F.~S.}\
  \bibnamefont {Cataliotti}}, \bibinfo {author} {\bibfnamefont
  {M.}~\bibnamefont {Inguscio}}, \bibinfo {author} {\bibfnamefont {G.~M.}\
  \bibnamefont {Tino}}, \bibinfo {author} {\bibfnamefont {P.}~\bibnamefont
  {Gill}}, \bibinfo {author} {\bibfnamefont {H.}~\bibnamefont {Klein}},
  \bibinfo {author} {\bibfnamefont {S.}~\bibnamefont {Reynaud}}, \bibinfo
  {author} {\bibfnamefont {C.}~\bibnamefont {Salomon}}, \bibinfo {author}
  {\bibfnamefont {E.}~\bibnamefont {Peik}}, \bibinfo {author} {\bibfnamefont
  {O.}~\bibnamefont {Bertolami}}, \bibinfo {author} {\bibfnamefont
  {P.}~\bibnamefont {Gil}}, \bibinfo {author} {\bibfnamefont {J.}~\bibnamefont
  {P{\'a}ramos}}, \bibinfo {author} {\bibfnamefont {C.}~\bibnamefont
  {Jentsch}}, \bibinfo {author} {\bibfnamefont {U.}~\bibnamefont {Johann}},
  \bibinfo {author} {\bibfnamefont {A.}~\bibnamefont {Rathke}}, \bibinfo
  {author} {\bibfnamefont {P.}~\bibnamefont {Bouyer}}, \bibinfo {author}
  {\bibfnamefont {L.}~\bibnamefont {Cacciapuoti}}, \bibinfo {author}
  {\bibfnamefont {D.}~\bibnamefont {Izzo}}, \bibinfo {author} {\bibfnamefont
  {P.~D.}\ \bibnamefont {Natale}}, \bibinfo {author} {\bibfnamefont
  {B.}~\bibnamefont {Christophe}}, \bibinfo {author} {\bibfnamefont
  {P.}~\bibnamefont {Touboul}}, \bibinfo {author} {\bibfnamefont {S.~G.}\
  \bibnamefont {Turyshev}}, \bibinfo {author} {\bibfnamefont {J.}~\bibnamefont
  {Anderson}}, \bibinfo {author} {\bibfnamefont {M.~E.}\ \bibnamefont {Tobar}},
  \bibinfo {author} {\bibfnamefont {F.}~\bibnamefont {Schmidt-Kaler}}, \bibinfo
  {author} {\bibfnamefont {J.}~\bibnamefont {Vigu{\'e}}}, \bibinfo {author}
  {\bibfnamefont {A.~A.}\ \bibnamefont {Madej}}, \bibinfo {author}
  {\bibfnamefont {L.}~\bibnamefont {Marmet}}, \bibinfo {author} {\bibfnamefont
  {M.-C.}\ \bibnamefont {Angonin}}, \bibinfo {author} {\bibfnamefont
  {P.}~\bibnamefont {Delva}}, \bibinfo {author} {\bibfnamefont
  {P.}~\bibnamefont {Tourrenc}}, \bibinfo {author} {\bibfnamefont
  {G.}~\bibnamefont {Metris}}, \bibinfo {author} {\bibfnamefont
  {H.}~\bibnamefont {M{\"u}ller}}, \bibinfo {author} {\bibfnamefont
  {R.}~\bibnamefont {Walsworth}}, \bibinfo {author} {\bibfnamefont {Z.~H.}\
  \bibnamefont {Lu}}, \bibinfo {author} {\bibfnamefont {L.~J.}\ \bibnamefont
  {Wang}}, \bibinfo {author} {\bibfnamefont {K.}~\bibnamefont {Bongs}},
  \bibinfo {author} {\bibfnamefont {A.}~\bibnamefont {Toncelli}}, \bibinfo
  {author} {\bibfnamefont {M.}~\bibnamefont {Tonelli}}, \bibinfo {author}
  {\bibfnamefont {H.}~\bibnamefont {Dittus}}, \bibinfo {author} {\bibfnamefont
  {C.}~\bibnamefont {L{\"a}mmerzahl}}, \bibinfo {author} {\bibfnamefont
  {G.}~\bibnamefont {Galzerano}}, \bibinfo {author} {\bibfnamefont
  {P.}~\bibnamefont {Laporta}}, \bibinfo {author} {\bibfnamefont
  {J.}~\bibnamefont {Laskar}}, \bibinfo {author} {\bibfnamefont
  {A.}~\bibnamefont {Fienga}}, \bibinfo {author} {\bibfnamefont
  {F.}~\bibnamefont {Roques}},\ and\ \bibinfo {author} {\bibfnamefont
  {K.}~\bibnamefont {Sengstock}},\ }\bibfield  {title} {\bibinfo {title}
  {Quantum physics exploring gravity in the outer solar system: the {SAGAS}
  project},\ }\href@noop {} {\bibfield  {journal} {\bibinfo  {journal}
  {Experimental Astronomy}\ }\textbf {\bibinfo {volume} {23}},\ \bibinfo
  {pages} {651} (\bibinfo {year} {2009})}\BibitemShut {NoStop}%
\bibitem [{\citenamefont {Origlia}\ \emph {et~al.}(2018)\citenamefont
  {Origlia}, \citenamefont {Pramod}, \citenamefont {Schiller}, \citenamefont
  {Singh}, \citenamefont {Bongs}, \citenamefont {Schwarz}, \citenamefont
  {Al-Masoudi}, \citenamefont {D{\"o}rscher}, \citenamefont {Herbers},
  \citenamefont {H{\"a}fner}, \citenamefont {Sterr},\ and\ \citenamefont
  {Lisdat}}]{origlia18}%
  \BibitemOpen
  \bibfield  {author} {\bibinfo {author} {\bibfnamefont {S.}~\bibnamefont
  {Origlia}}, \bibinfo {author} {\bibfnamefont {M.~S.}\ \bibnamefont {Pramod}},
  \bibinfo {author} {\bibfnamefont {S.}~\bibnamefont {Schiller}}, \bibinfo
  {author} {\bibfnamefont {Y.}~\bibnamefont {Singh}}, \bibinfo {author}
  {\bibfnamefont {K.}~\bibnamefont {Bongs}}, \bibinfo {author} {\bibfnamefont
  {R.}~\bibnamefont {Schwarz}}, \bibinfo {author} {\bibfnamefont
  {A.}~\bibnamefont {Al-Masoudi}}, \bibinfo {author} {\bibfnamefont
  {S.}~\bibnamefont {D{\"o}rscher}}, \bibinfo {author} {\bibfnamefont
  {S.}~\bibnamefont {Herbers}}, \bibinfo {author} {\bibfnamefont
  {S.}~\bibnamefont {H{\"a}fner}}, \bibinfo {author} {\bibfnamefont
  {U.}~\bibnamefont {Sterr}},\ and\ \bibinfo {author} {\bibfnamefont
  {C.}~\bibnamefont {Lisdat}},\ }\bibfield  {title} {\bibinfo {title} {Towards
  an optical clock for space: Compact, high-performance optical lattice clock
  based on bosonic atoms},\ }\href@noop {} {\bibfield  {journal} {\bibinfo
  {journal} {Physical Review A}\ }\textbf {\bibinfo {volume} {98}},\ \bibinfo
  {pages} {053443} (\bibinfo {year} {2018})}\BibitemShut {NoStop}%
\bibitem [{\citenamefont {Wang}\ \emph {et~al.}(2020)\citenamefont {Wang},
  \citenamefont {Shevate}, \citenamefont {Wintermantel}, \citenamefont
  {Morgado}, \citenamefont {Lochead},\ and\ \citenamefont {Whitlock}}]{wang20}%
  \BibitemOpen
  \bibfield  {author} {\bibinfo {author} {\bibfnamefont {Y.}~\bibnamefont
  {Wang}}, \bibinfo {author} {\bibfnamefont {S.}~\bibnamefont {Shevate}},
  \bibinfo {author} {\bibfnamefont {T.~M.}\ \bibnamefont {Wintermantel}},
  \bibinfo {author} {\bibfnamefont {M.}~\bibnamefont {Morgado}}, \bibinfo
  {author} {\bibfnamefont {G.}~\bibnamefont {Lochead}},\ and\ \bibinfo {author}
  {\bibfnamefont {S.}~\bibnamefont {Whitlock}},\ }\bibfield  {title} {\bibinfo
  {title} {Preparation of hundreds of microscopic atomic ensembles in optical
  tweezer arrays},\ }\href {https://doi.org/10.1038/s41534-020-0285-1}
  {\bibfield  {journal} {\bibinfo  {journal} {npj Quantum Information}\
  }\textbf {\bibinfo {volume} {6}},\ \bibinfo {pages} {54} (\bibinfo {year}
  {2020})}\BibitemShut {NoStop}%
\bibitem [{\citenamefont {Safronova}\ \emph {et~al.}(1999)\citenamefont
  {Safronova}, \citenamefont {Johnson},\ and\ \citenamefont
  {Derevianko}}]{safronova99}%
  \BibitemOpen
  \bibfield  {author} {\bibinfo {author} {\bibfnamefont {M.~S.}\ \bibnamefont
  {Safronova}}, \bibinfo {author} {\bibfnamefont {W.~R.}\ \bibnamefont
  {Johnson}},\ and\ \bibinfo {author} {\bibfnamefont {A.}~\bibnamefont
  {Derevianko}},\ }\bibfield  {title} {\bibinfo {title} {Relativistic many-body
  calculations of energy levels, hyperfine constants, electric-dipole matrix
  elements, and static polarizabilities for alkali-metal atoms},\ }\href
  {https://doi.org/10.1103/PhysRevA.60.4476} {\bibfield  {journal} {\bibinfo
  {journal} {Phys. Rev. A}\ }\textbf {\bibinfo {volume} {60}},\ \bibinfo
  {pages} {4476} (\bibinfo {year} {1999})}\BibitemShut {NoStop}%
\bibitem [{\citenamefont {Safronova}\ \emph {et~al.}(2015)\citenamefont
  {Safronova}, \citenamefont {Zuhrianda}, \citenamefont {Safronova},\ and\
  \citenamefont {Clark}}]{safronova15}%
  \BibitemOpen
  \bibfield  {author} {\bibinfo {author} {\bibfnamefont {M.~S.}\ \bibnamefont
  {Safronova}}, \bibinfo {author} {\bibfnamefont {Z.}~\bibnamefont
  {Zuhrianda}}, \bibinfo {author} {\bibfnamefont {U.~I.}\ \bibnamefont
  {Safronova}},\ and\ \bibinfo {author} {\bibfnamefont {C.~W.}\ \bibnamefont
  {Clark}},\ }\bibfield  {title} {\bibinfo {title} {Extracting transition rates
  from zero-polarizability spectroscopy},\ }\href
  {https://doi.org/10.1103/PhysRevA.92.040501} {\bibfield  {journal} {\bibinfo
  {journal} {Phys. Rev. A}\ }\textbf {\bibinfo {volume} {92}},\ \bibinfo
  {pages} {040501(R)} (\bibinfo {year} {2015})}\BibitemShut {NoStop}%
\bibitem [{\citenamefont {Kramida}\ \emph {et~al.}(2020)\citenamefont
  {Kramida}, \citenamefont {{Yu.~Ralchenko}}, \citenamefont {Reader},\ and\
  \citenamefont {{and NIST ASD Team}}}]{nistasd}%
  \BibitemOpen
  \bibfield  {author} {\bibinfo {author} {\bibfnamefont {A.}~\bibnamefont
  {Kramida}}, \bibinfo {author} {\bibnamefont {{Yu.~Ralchenko}}}, \bibinfo
  {author} {\bibfnamefont {J.}~\bibnamefont {Reader}},\ and\ \bibinfo {author}
  {\bibnamefont {{and NIST ASD Team}}},\ }\href
  {https://doi.org/10.18434/T4W30F} {}\bibinfo {howpublished} {{NIST Atomic
  Spectra Database (ver. 5.8), [Online]. Available:
  {\tt{https://physics.nist.gov/asd}} [10/30/2020]. National Institute of
  Standards and Technology, Gaithersburg, MD.}} (\bibinfo {year}
  {2020})\BibitemShut {NoStop}%
\bibitem [{\citenamefont {Safronova}\ \emph {et~al.}(2013)\citenamefont
  {Safronova}, \citenamefont {Porsev}, \citenamefont {Safronova}, \citenamefont
  {Kozlov},\ and\ \citenamefont {Clark}}]{safronova13}%
  \BibitemOpen
  \bibfield  {author} {\bibinfo {author} {\bibfnamefont {M.~S.}\ \bibnamefont
  {Safronova}}, \bibinfo {author} {\bibfnamefont {S.~G.}\ \bibnamefont
  {Porsev}}, \bibinfo {author} {\bibfnamefont {U.~I.}\ \bibnamefont
  {Safronova}}, \bibinfo {author} {\bibfnamefont {M.~G.}\ \bibnamefont
  {Kozlov}},\ and\ \bibinfo {author} {\bibfnamefont {C.~W.}\ \bibnamefont
  {Clark}},\ }\bibfield  {title} {\bibinfo {title} {Blackbody-radiation shift
  in the {Sr} optical atomic clock},\ }\href
  {https://doi.org/10.1103/PhysRevA.87.012509} {\bibfield  {journal} {\bibinfo
  {journal} {Phys. Rev. A}\ }\textbf {\bibinfo {volume} {87}},\ \bibinfo
  {pages} {012509} (\bibinfo {year} {2013})}\BibitemShut {NoStop}%
\bibitem [{\citenamefont {Safronova}()}]{safronovapriv}%
  \BibitemOpen
  \bibfield  {author} {\bibinfo {author} {\bibfnamefont {M.~S.}\ \bibnamefont
  {Safronova}},\ }\href@noop {} {\bibinfo {title} {private
  communication}}\BibitemShut {NoStop}%
\end{thebibliography}

%

\end{document}